\documentclass[3p,times,english]{elsarticle}
\usepackage{ecrc}
\volume{00}

\firstpage{1}

\journalname{Annals of Physics}

\runauth{S.~M. Kopeikin and A.~N. Petrov}

\jid{annphys}

\jnltitlelogo{Annals of Physics}

\CopyrightLine{2011}{Published by Elsevier Ltd.}

\usepackage{german}
\usepackage{amssymb}
\usepackage{amsmath}
\usepackage{amsthm}
\usepackage{mathtools}
\usepackage{bm}
\usepackage{emerald}
\usepackage{babel}
\usepackage{slantsc}
\usepackage{array}
\usepackage{url}

\biboptions{sort&compress}

\def\disp{\displaystyle}
\def\st{\scriptstyle}

\def\be{\begin{equation}}
\def\ee{\end{equation}}
\def\ba{\begin{eqnarray}}
\def\ea{\end{eqnarray}}
\def\bit{\begin{itemize}}
\def\eit{\end{itemize}}
\def\eq{\equiv}
\def\bsu{\begin{subequations}}
\def\esu{\end{subequations}}

\def\F{{\mathcal F}}
\def\H{{\mathcal H}}
\def\Q{{\mathcal  Q}}
\def\L{{\mathcal L}}
\def\M{{\mathcal M}}

\def\Tc{{\mathcal T}}
\def\Y{{\mathcal Y}}

\def\gag{\mathfrak{g}}

\def\a{\alpha}
\def\a {\alpha}
\def\b{\beta}
\def\g{\gamma}     \def\G{\Gamma}
\def\d{\delta}
\def\de{\delta}
\def\k{\kappa}
\def\e{\epsilon}

\def\l{\lambda}
\def\m{\mu}
\def\n{\nu}
   
\def\p{\phi}

\def\r{\rho}
\def\s{\sigma}
\def\t{\tau}

\def\la{\label}
\def\pd{\partial}
\def\lef{\left}
\def\ri{\right}

\def\lag{{\L}}                                
\def\mt{{\mathfrak{t}}}
\def\mm{{\mathtt f}}
\def\T{{\mathfrak{T}}}
\def\hatl{\mathfrak{h}}
\def\hatg{\bar\gag}

\begin{document}
\begin{frontmatter}
\title{Dynamic Field Theory and Equations of Motion in Cosmology}
\author{Sergei M. Kopeikin\corref{cor1}}
\address{Department of Physics \& Astronomy, University of Missouri, 322 Physics Bldg., Columbia, MO 65211, USA}
\ead{kopeikins@missouri.edu}
\cortext[cor1]{Corresponding author.}
\author{Alexander N. Petrov}
\address{Sternberg Astronomical Institute, Lomonosov Moscow State University, Universitetskij Prospect 13, Moscow 119992, Russia}
\ead{alex.petrov55@gmail.com}

\begin{abstract}
We discuss a field-theoretical approach based on general-relativistic variational principle to derive the covariant field equations and hydrodynamic equations of motion of baryonic matter governed by cosmological perturbations of dark matter and dark energy. The action depends on the gravitational and matter Lagrangian. The gravitational Lagrangian depends on the metric tensor and its first and second derivatives. The matter Lagrangian includes dark matter, dark energy and the ordinary baryonic matter which plays a role of a bare perturbation. The total Lagrangian is expanded in an asymptotic Taylor series around the background cosmological manifold defined as a solution of Einstein's equations in the form of the Friedmann-Lema\^itre-Robertson-Walker (FLRW) metric tensor. The small parameter of the decomposition is the magnitude of the metric tensor perturbation. Each term of the series expansion is gauge-invariant and all of them together form a basis for the successive post-Friedmannian approximations around the background metric. The approximation scheme is covariant and the asymptotic nature of the Lagrangian decomposition does not require the post-Friedmannian perturbations to be small though computationally it works the most effectively when the perturbed metric is close enough to the background FLRW metric. The temporal evolution of the background metric is governed by dark matter and dark energy and we associate the large scale inhomogeneities in these two components as those generated by the primordial cosmological perturbations with an effective matter density contrast $\d\r/\r\le 1$. The small scale inhomogeneities are generated by the condensations of baryonic matter considered as the bare perturbations of the background manifold that admits $\d\r/\r\gg 1$. Mathematically, the large scale perturbations are given by the homogeneous solution of the linearized field equations while the small scale perturbations are described by a particular solution of these equations with the bare stress-energy tensor of the baryonic matter. We explicitly work out the covariant field equations of the successive post-Friedmannian approximations of Einstein's equations in cosmology and derive equations of motion of large and small scale inhomogeneities of dark matter and dark energy. We apply these equations to derive the post-Friedmannian equations of motion of baryonic matter comprising stars, galaxies and their clusters.  
\end{abstract}
 
\end{frontmatter}
\newpage
\tableofcontents
\newpage
\section{Introduction}\la{sec1}
\subsection{Perturbation techniques in cosmology}
The multiwavelength satellite observations of cosmic microwave background (CMB) radiation have opened a new chapter in cosmology \citep{Verkhodanov_2012}. The standard cosmological model \citep{Peacock_2002,mukh_book,weinberg_2008,rubak_2011} has been worked out to fit the model parameters to the CMB map  and  with the decisive confidence level of 95\% \citep{2004PhRvD..69j3501T}. The Planck satellite observations provide further evidences in robustness of the standard model \citep{Planck_cosmological_parameters,Planck_primordial_non-Gaussianity} (see \url{http://www.sciops.esa.int/index.php?project=PLANCK} for a comprehensive list of Planck collaboration papers) though it might be still difficult to discern between various scenarios of the early universe \citep{2013PhRvD..87l3533L}. 

Study of the formation and evolution of the large scale structure in the universe is a key for understanding the present state of the universe and for predicting its uttermost fate \citep{LSS_2000}. It is extensively researched but, as of today, remains yet unsolved problem in physical cosmology. It is dark matter which plays a key role in the large scale structure formation. The dark matter consists of weakly interacting massive particles whose true nature is not known so far except that they interact with baryons mainly by the force of gravity. The baryonic matter forms galaxies which, at early stage of structure formation, simply follow the evolution of dark matter condensations. Therefore, it is supposed that the observed large scale distribution of galaxies trace the distribution of dark matter. 

At the linear stage the effective matter density contrast, $\d\r=\r-\bar\r$, is much smaller than the background (mean) density $\bar\r$ of the universe: $\d\r/\bar\r\ll 1$. At later stages of cosmological evolution the structure formation enters a non-linear regime where $\d\r/\bar\r\simeq 1$, and caustics are formed. Further growth of the perturbations leads to the development of small scale structures like nuclei of galaxies, dwarf galaxies, globular clusters, stars and more compact relativistic objects which have $\d\r/\bar\r\gg 1$. Gravitational field and matter of these super-dense baryonic objects counteract with the gravitational potential of dark matter and dark energy but details of this process are still unclear because it involves rather complicated physics of fluid's magnetohydrodynamics, turbulence and the strong gravity field that implies a general-relativistic approach taking into account the non-linear interaction of gravitational field with itself and the surrounding matter. Presumably, some insight to the solution of this problem can be gained by exploring exact Lema\^itre-Tolman cosmological solution of Einstein's equations admitting spatially inhomogeneity along radial coordinate \citep{Krasinski_2002,Krasinski_2004,Jacewicz_2012}. This purely geometric approach is mathematically sound but not very realistic as it describes a pressureless, spherically-symmetric accretion of dust to a single point of a cosmological manifold while the real early universe has a continuous set of the accretion points (seeds of future galaxies) determined by the initial spectrum of the primordial density fluctuations \citep{mukh_book,weinberg_2008,rubak_2011}. In addition, the baryonic fluid pressure cannot be ignored at the non-linear regime.

Exact solution of Einstein's equations is unavailable for the general case of perturbed universe. Therefore, we have to resort to approximations in order to treat non-linear gravitational effects in the structure formation. Two approximation schemes of solving Einstein's equations are known in asymptotically-flat spacetime - post-Newtonian and post-Minkowskian approximations \citep[pp. 340-344]{kopeikin_2011book}. Post-Minkowskian approximations (PMA) rely on the assumptions that gravitational field is weak everywhere without any limitation on velocity of matter besides that it must be smaller than the fundamental speed, $c$.  Post-Newtonian approximations (PNA) are made under assumption that the field is weak and velocity of matter is much smaller than the speed of light. The PMA formalism has been basically developed in a series of papers by Damour and Blanchet for studying the mechanism of emission and propagation of gravitational waves emitted by isolated astronomical systems \citep{bd1,bd2,bd3,2006LRR.....9....4B}. The PNA formalism has been developed by a number of independent researchers \citep{fock_book,Papapetrou_1951,Chandra_1969,1970ApJ...160..153C,andec} for describing non-linear gravitational effects in fluids, for deriving equations of motion of binary stars \citep{Spyrou_1981,Kopeikin_1985,Breuer_1982,Breuer_1981}, for calculating equilibrium models of rapidly rotating neutron stars \citep{Shapiro_1976,Ciufolini_1983,Shapiro_1998ApJS}, etc. Both PMA and PNA expand explicitly only the metric tensor of the manifold by making use either Landau-Lifshitz pseudotensor or 3+1 ADM decomposition. However, the metric tensor PMA/PNA expansion is not sufficient in cosmology because the background spacetime is not asymptotically flat and we have to take into account not only the perturbations of the metric tensor but also those of the background stress-energy tensor of the cosmological matter that governs evolution of the cosmological spacetime. Additional non-trivial problem of the perturbation technique in cosmology is to separate the contribution of bare perturbations of the baryonic matter of small-scale inhomogeneities  from the large-scale perturbations of the background matter and the metric tensor of spacetime manifold. Thus, we have to generalize PMA/PNA schemes of finding solutions of Einstein's equations to include the case of more general background manifold. We follow Tegmark \citep{tegmark_2002PhRvD} and use the name of post-Friedmannian approximations for such a more general, iterative procedure.

A number of theoretical attempts was undertaken to work out the first post-Friedmannian approximation and equations of motion of perfect fluid on cosmological manifold \citep{Takada_1999MNRAS,Petry_2000,Szekeres_2000GReGr,Hwang_2008JCAP}. These works provide a good insight to the possible solution of the problem but are insufficiently consistent in separation of perturbations from their background values. They do not suggest a systematic approach for extending the calculations of the linear perturbation theory to the second, and higher, post-Friedmannian approximations either. We also point out a result by Oliynyk \citep{Oliynyk_2010CMaPh,Oliynyk_2012ATMP} who analyzed the general structure of the post-Friedmannian expansions on cosmological manifold and arrived to the conclusion that the post-Friedmannian series are differentiable, but not analytic, with respect to the small parameter  $\varepsilon =v/c$, where $v$ is a peculiar velocity of fluid with respect to the Hubble flow and $c$ is the constant fundamental speed \citep{2004CQGra..21.3251K}. The Oliynyk’s conclusion differs with the results obtained by making use of the post-Newtonian expansions in
asymptotically-ﬂat spacetime \citep{PhysRevD.25.2038,Blanchet_2000LNP,Schaefer_2011mmgr} due to the compactness of the spatial slices in the cosmological manifold.

Recently, a new interest for developing a self-consistent theory of post-Friedmannian approximations in precision cosmology was triggered by a lively discussion \citep{Buchert_2008GReGr,Kolb_2008PhRvD,Kolb_2010GReGr,wald_2011,wald_2012} on whether the small-scale structure of the universe affects its Hubble expansion rate and, thus, can explain the cosmic acceleration of the universe discovered in 1998-99 \citep{Riess_1998AJ,Perlmutter_1999ApJ} without invoking a dark energy. This is a, so-called, backreaction problem which intimately relates to the procedure of averaging the small-scale matter perturbations on a curved cosmological manifold \citep{Zalaletdinov_2008IJMPA,Futamase_1996PhRvD}.  A certain progress in this direction was achieved but many mathematical aspects of the backreaction problem are still poorly understood \citep{Clarkson:2011zq}. The task is to build a rigorous mathematical formalism being able to describe on equal footing both the large-scale perturbations of the background matter of cosmological manifold with the density contrast $\d\r/\bar\r\ll 1$ and the small-scale perturbations at present epoch caused by small-scale structures (galaxy, globular cluster, star) having the density contrast $\d\r/\bar\r\gg 1$. Einstein's equations tell us that the density perturbation, $\d\r/\bar\r$, is proportional to the second derivatives of the metric tensor perturbation which can be very large if $\d\r/\bar\r\gg 1$. At the same time, the metric tensor perturbation, $\varkappa_{\a\b}=g_{\a\b}-\bar g_{\a\b}$, and its first derivatives, $\varkappa_{\a\b,\g}$, can still remain small enough in order to apply a perturbation technique for solving the Einstein equations. This is similar to the situation in the solar system where the matter density contrast is huge but, nonetheless, the gravitational weak-field approximation for solving Einstein's equations is fully applicable \citep{kopeikin_2011book}. It supports the idea that the perturbation technique in cosmology (under above assumptions) is valid for calculating physical effects of inhomogeneities on both the large and small scales  \citep{Ellis_2011cqg,wald_2011,kopetr}. The question is what mathematical technique is the most adequate to deal with physical applications. 

Historically, the very first perturbation scheme in cosmology was worked out by Lifshitz \citep{1964SvPhU...6..495L,lif}. It is technically convenient  for calculating time evolution of matter's large scale structure inhomogeneities and gravitational field perturbations  \citep{weinberg_1972,weinberg_2008} but is unsuitable for discussing the process of formation and time evolution of the small scale structures in universe. This is because the Lifshitz approach uses the synchronous gauge where the time-time component of the metric tensor is fixed, $g_{00}=-1$, at any order of approximation. The small scale structure in cosmology corresponds to a localized astronomical system having a large density contrast, $\d\r/\bar\r$, and governed by the Newtonian law of gravity which demands the presence of the Newtonian potential, $U$, making $g_{00}=-1+2U/c^2$. It breaks down the synchronous gauge condition at the regime of $\d\r/\bar\r\gg 1$.   

Bardeen's perturbation approach \citep{1980PhRvD..22.1882B} (see also \citep{kodama_1984}) is more flexible as it admits a rather large freedom in choosing a particular gauge condition for solving cosmological problems \citep{mukh_book,rubak_2011}. In the framework of this approach, the longitudinal (conformal-Newtonian) gauge is the most appropriate for discussing the small scale structure formation and its physical effects \citep{ishibashi_2006CQG}. Some mathematical disadvantage of Bardeen's approach is in imposing a scalar-vector-tensor decomposition on the metric tensor. It requires application of the Helmholtz theorem \citep{baierlein_1995} that demands to foliate spacetime by a family of spacelike hypersurfaces and to integrate the metric tensor over these hypersurfaces. It makes Bardeen's approach non-local as contrasted to Lifshitz's perturbation scheme. Moreover, in order to preserve the gauge-invariance of the Bardeen perturbation scheme one has to decompose the gauge functions in the same fashion as the metric tensor. As the universe evolves the gauge-invariance of the overall Bardeen's scheme can be preserved, if and only if, one maintains the mapping of spatial points on the foliations along the vector field of time coordinate world lines. Evidently, this approximation scheme differs significantly from the post-Newtonian approximations in asymptotically flat spacetime \citep{fock_book,1991ercm.book.....B,kopeikin_2011book} which are more similar to Lifshitz's approach but use a different gauge condition (harmonic gauge).

A gauge-invariant alternative to Bardeen's approach was suggested by Ellis and Bruni \citep{ellis1} (see also \citep{ellis2,Maartens_2008}), who developed a perturbation scheme based on a reduction of full Einstein's equations down to a system of field equations that are linear around a particular background. The Ellis-Bruni approach uses gauge-invariant variables which are spatial projections on the local comoving-observer frame threading the entire space-time of a real universe. Thus, the Bardeen's foliation has been replaced in Ellis-Bruni approach with the frame threading and, thus, observer dependent. This is not convenient, and is not used, for developing the post-Newtonian approximations in asymptotically-flat spacetime. 

At the epoch of precise cosmology we need more transparent theoretical scheme of the post-Friedmannian approximations for handling the iterative calculation of cosmological perturbations and derivation of their equations of motion. This iterative scheme must satisfy a number of well-established criteria like to be covariant, gauge-invariant, operate with locally defined quantities, be systematic and self-consistent in improving the order of approximations, clearly separate the large-scale from small-scale matter perturbations, be independent of the mathematical ambiguities introduced by the averaging procedures, etc. Some steps in developing such a scheme were done by Green and Wald \citep{wald_2011,wald_2012}. However, their work was focused mainly on the discussion of the averaging procedure in cosmology, on the proof that the small-scale  inhomogeneities do not produce a noticeable backreaction and on finding mathematical evidences that the Newtonian approximation is sufficient in numerical N-body simulations of large scale structure formation \citep{shandarin_2012PhRvD}.

No doubt, theoretical questions about how to perform the averaging in cosmology and whether it produces any backreaction at all, are important for understanding the mathematics of averaging of differential operators in non-linear equations and for clarification of the true nature of dark matter and dark energy. However, the post-Friedmannian approximation scheme in cosmology has broader implications that are going beyond the discussion of averaging and backreaction problems and relates to the problem of interpretation of precise measurement of cosmological parameters by the advanced gravitational wave detector's technique \citep{2012PhRvD..86d3011D,2012PhRvD..85b3535T} and formation of small-scale structures in the universe at the non-linear regime. The formalism of the post-Friedmannian approximations can be also helpful in better understanding of the influence of cosmological expansion on celestial mechanics of isolated astronomical systems like binary pulsars which are currently the best laboratories for testing non-linear regime of general relativity \citep{krawex_2009,2010ApJ...722.1030W}. These tests will be made significantly more precise with advent of gravitational-wave astronomy and  Square Kilometer Array (SKA) radio telescope \citep{kramer_2012IAUS}. 

Recently, we have started a systematic investigation of the dynamics of small-scale inhomogeneities moving on the FLRW background manifold. We have set up a Lagrangian formalism to derive the post-Friedmannian field equations for linearised cosmological perturbations \citep{kopetr} and analysed the Newtonian limit of these equations \citep{Kopeikin_2012ephemerid}. The present paper goes beyond the linear regime and explores some non-linear effects. In particular, we derive the post-Friedmannian hydrodynamic equations of motion of the background matter (dark matter and dark energy) along with the equations of motion of the baryonic matter forming a small-scale structure with high-density contrast like a star, or galaxy or a cluster of galaxies.

We explain the idea of manifold and underlying geometric objects in \ref{sec2}. The concept of the covariant and Lie derivatives on manifold are explained in section \ref{vg34v}. This section also defines the variational derivatives on manifold in the context of the dynamic field theory. Geometric theory of Euler-type perturbations of arbitrary background manifold is set up in section \ref{sec3}. This theory is applied to the FLRW universe, governed by dark matter and dark energy, in section \ref{sec4}. Section \ref{sec5} derives the stress-energy tensors for perturbations of the gravitational field, dark matter and dark energy. Finally, we derive equations of motion of the small-scale (bare) perturbations in section \ref{sec6} and compare our framework against other theoretical approaches in section \ref{diss56}. Appendix outlines some particular mathematical aspects of our derivation.

Before going into details of our presentation we explain the notations adopted in the present paper.
\subsection{Notations}

We use $G$ to denote the universal gravitational constant and $c$ for the ultimate speed in Minkowski spacetime. Every time, when there is no confusion about the system of units, we use a geometrized system of units where $G=c=1$.
We put a bar over any function that belongs to the background manifold of the FLRW cosmological model. Any function without such a bar belongs to the perturbed manifold. The other notations used in the present paper are as follows:
\begin{itemize}
\item $T$ and $X^i=\{X,Y,Z\}$ are the coordinate time and isotropic spatial coordinates on the background manifold;
\item $X^\a=\{X^0,X^i\}=\{c\eta,X^i\}$ are the conformal coordinates with $\eta$ being a conformal time;
\item $x^\a=\{x^0,x^i\}=\{ct,x^i\}$ is an arbitrary coordinate chart on the background manifold;
\item Greek indices $\a,\b,\g,\ldots$ run through values $0,1,2,3$, and label spacetime coordinates;
\item Roman indices $i,j,k,\ldots$ take values $1,2,3$, and label spatial coordinates;
\item Einstein summation rule is applied for repeated (dummy) indices, for example,  $P^\a Q_\a\equiv P^0 Q_0+P^1 Q_1+P^2 Q_2 + P^3 Q_3$, and $P^i Q_i\equiv P^1 Q_1+P^2 Q_2 + P^3 Q_3$;
\item $g_{\a\b}$ is a full metric on the cosmological spacetime manifold;
\item $\bar g_{\a\b}$ is the FLRW metric on the background spacetime manifold;
\item ${\gag}^{\mu\nu}=\sqrt{-g}g^{\mu\nu}$ -- the metric tensor density of weight +1 where we accept the standard definition of tensor density explained, for example, in \citep[page 501]{mitowh}. The reader should be warned that the definition of tensor density chosen in the book by S. Weinberg \citep[Chapter 4, \S 4]{weinberg_1972} differs by sign from the standard definition, and is not commonly accepted;
\item $\bar {\gag}^{\mu\nu}=\sqrt{-\bar g}\bar g^{\mu\nu}$ -- the  background metric tensor density of  weight +1;
\item $\mm_{\a\b}$ is the metric on the conformal spacetime manifold;
\item $\eta_{\a\b}={\rm diag}\{-1,+1,+1,+1\}$ is the Minkowski metric;
\item the scale factor of the FLRW metric is denoted as $R=R(T)$, or as $a=a(\eta)=R[T(\eta)]$;
\item the Hubble parameter, $H=R^{-1}dR/dT$;
\item the conformal Hubble parameter, $\H= a^{-1}da/d\eta$;
\item $\F$ denotes a geometric object on the manifold. It can be either a scalar, or a vector, or a tensor field, or a corresponding tensor density;
\item a bar, $\bar\F$ above a geometric object $\F$, denotes the unperturbed value of $\F$ on the background manifold;
\item the tensor indices of geometric objects on the background manifold are raised and lowered with the background metric $\bar g_{\a\b}$, for example $\F_{\a\b}=\bar g_{\a\m}\bar g_{\b\n}\F^{\m\n}$;
\item the tensor indices of geometric objects on the conformal spacetime are raised and lowered with the conformal metric $\mm_{\a\b}$;
\item symmetry of a geometric object with respect to two indices is denoted with round parenthesis, $\F_{(\a\b)}\eq (1/2)\lef(\F_{\a\b}+\F_{\b\a}\ri)$;
\item antisymmetry of a geometric object with respect to two indices is denoted with square parenthesis, $\F_{[\a\b]}\eq (1/2)\lef(\F_{\a\b}-\F_{\b\a}\ri)$;
\item a prime $\F'=d\F/d\eta$ denotes a total derivative with respect to the conformal time $\eta$;
\item a dot $\dot\F=d\F/dT$ denotes a total derivative with respect to the coordinate time $T$;
\item $\pd_\a=\pd/\pd x^\a$ is a partial derivative with respect to the coordinate $x^\a$;
\item a comma with a following index $\F_{,\a}\eq\pd_\a \F$ is an other designation of a partial derivative with respect to a coordinate $x^\a$ which is more convenient in some cases. In some cases which may not cause confusion, the comma as a symbol of the partial derivative is omitted. For example, we denote the partial derivatives of the perturbations of matter variables as $\p_\a\eq\p_{,\a}$, $\psi_\a\eq\psi_{,\a}$, etc.;
\item a vertical bar, $\F_{|\a}$ denotes a covariant derivative of a geometric object $\F$ with respect to the background metric $\bar g_{\a\b}$. Covariant derivatives of scalar fields coincide with their partial derivatives;
\item a semicolon, $\F_{;\a}$ denotes a covariant derivative of a geometric object $\F$ with respect to the conformal metric $\mm_{\a\b}$;
\item $\Phi^A$ -- a multiplet of $A=\{1,2,\ldots,a\}$ matter fields, $\Phi^A=\{\Phi^1,\Phi^2,\ldots,\Phi^a\}$. These fields generate the full metric $g_{\mu\nu}$ of FLRW universe via the Einstein equations;
\item an operator $\nabla_\a$ denotes the covariant derivative with respect to the full metric $g_{\a\b}$;
\item $\bar \Phi^A$ -- the background value of the fields $\Phi^A$. These fields generate the background metric $\bar g_{\mu\nu}$ of FLRW universe via the Einstein equations;
\item $\Theta^B$ -- a multiplet of $B=\{1,2,\ldots,b\}$ matter fields, $\Theta^B=\{\Theta^1,\Theta^2,\ldots,\Theta^b\}$. They generate the stress energy tensor of the bare perturbation of the metric tensor $g_{\m\n}$ and that of the fields $\Phi^A$;
\item $\bar\Theta^B$ -- the background value of the fields $\Theta^B$. 
\item $\phi^A \equiv \Phi^A - \bar \Phi^A$ -- the perturbation of the field $\Phi^A$. Fields $\Phi^A$ and $\bar\Phi^A$ refer to the same point on the manifold;
\item $\t^B \equiv \Theta^B - \bar \Theta^B$ -- the perturbation of the field $\Theta^B$ caused by the counteraction of the metric tensor perturbations $l_{\m\n}$ and those of the dynamic fields $\p^A$ on the stress-energy tensor of the bare perturbations;
\item $\varkappa_{\mu\nu}\equiv g_{\mu\nu}-\bar g_{\mu\nu}$ -- the metric tensor perturbation. Fields $g_{\mu\nu}$ and $\bar g_{\mu\nu}$ refer to the same point on the manifold;
\item $\mathfrak{h}^{\mu\nu} \equiv {\gag}^{\mu\nu} - \bar {\gag}^{\mu\nu}$ --
the perturbation of the metric density caused by $\Theta^B$;
\item $l^{\mu\nu}\equiv\mathfrak{h}^{\mu\nu}/\sqrt{-\bar g}$. In a linear approximation, $l^{\mu\nu}=-\varkappa^{\mu\nu}+\frac12\bar g^{\mu\nu} \varkappa^\a{}_\a$, where $\varkappa^\a{}_\a=\bar g^{\a\b}\varkappa_{\a\b}$;
\item the Christoffel symbols, $\G^\a{}_{\b\g}=\disp\frac12 g^{\a\n}\lef( g_{\n\b,\g}+ g_{\n\g,\b}- g_{\b\g,\n}\ri)\;;$
\item the Riemann tensor, $R^\a{}_{\b\m\n}=\Gamma^\a{}_{\b\n,\m}-\Gamma^\a{}_{\b\m,\n}+\Gamma^\a{}_{\m\g}\Gamma^\g{}_{\b\n}-\Gamma^\a{}_{\n\g}\Gamma^\g{}_{\b\m}\;;$
\item the Ricci tensor, $R_{\a\b}=R^\m{}_{\a\m\b}$\;;
\item the Ricci scalar, $R=g^{\a\b}R_{\a\b}$\;.
\end{itemize}
We shall often employ the term {\it on-shell}. By {\it on-shell} we mean {\it satisfying the equations of motion}. For instance, Noether's theorem links conserved quantities to symmetries of the system on-shell. It is invalid off-shell. We shall introduce and explain other notations as they appear in the main text of the paper.

\section{Geometric manifold}\la{sec2}
Manifold ${\cal M}$ is a geometric arena for gravitational physics. Topologically manifold is a set of points endowed with some particular differential structure giving the manifold certain rigidity and physical properties. The basic element of this structure is the metric tensor $g_{\a\b}$ that allows to measure the distance between infinitesimally close points on the manifold and the angles between two vectors attached to the same point on the manifold. In general relativity, the metric tensor represents gravitational field which is a tensor field of rank two. Alternative theories of gravity either introduce other fields (scalar, vector, etc.) which yield additional contribution to the overall gravitational field besides the metric tensor or operate with the Lagrangian which is more complicated than the Hilbert Lagrangian of general theory of relativity like $F(R)$ theories of gravity \citep{Lima_2013PhRvD,Chakrab_2013GReGr}. The present paper deals exclusively with general theory of relativity and does not discuss the alternative theories in order to treat either gravitational field or dark matter and dark energy.

\subsection{Affine connection}\la{affc7}

The first level of differential structure of manifold is associated with the affine connection allowing us to differentiate geometric objects and to transport them from one point of manifold to another. The affine connection consists of three algebraically-irreducible components which are the Christoffel symbols, torsion and non-metricity \citep{kopeikin_2011book}. Torsion and non-metricity do not appear in general relativity and we do not mention them from now on.  

We define the Christoffel symbols of the second kind as usual \citep{kopeikin_2011book}
\ba\la{aa1}
\G^\a{}_{\b\g}&\eq&\frac12 g^{\a\d}\lef(g_{\d\b,\g}+g_{\d\g,\b}-g_{\b\g,\d}\ri)\;.\ea
The Christoffel symbols of the first kind
\ba
\la{aa2}
\G_{\a\b\g}&=&g_{\a\s}\G^\s{}_{\b\g}=\frac12\lef(g_{\a\b,\g}+g_{\a\g,\b}-g_{\b\g,\a}\ri)\;,\ea
We notice the symmetry with respect to the last two indices $\G_{\a\b\g}=\G_{\a(\b\g)}$. There is no any symmetry with respect to the first two indices. In general,
\ba\la{aa2a}
\G_{\a\b\g}&=&\G_{(\a\b)\g}+\G_{[\a\b]\g}\;,
\ea
where
\be
\la{aa10}
\G_{(\a\b)\g}=\frac12 g_{\a\b,\g}\qquad,\qquad
\G_{[\a\b]\g}=\frac12 \lef(g_{\g\a,\b}-g_{\g\b,\a}\ri)\;,
\ee
There are two, particularly useful symbols that are obtained by contracting indices of the Christoffel symbols of the first kind. They are denoted as
\be
\la{aa3}
{\cal Y}_\a\eq \G^\b{}_{\a\b}\qquad,\qquad{\cal Y}^\a=g^{\a\b}{\cal Y}_\b\;,\ee
and
\be
\la{aa5}
\G^\a\eq g^{\b\g}\G^\a{}_{\b\g}\qquad,\qquad\G_\a=g_{\a\b}\G^\b\;,\ee
Direct inspection shows that
\ba
\la{aa4a}
{\cal Y}_\a&=&=\frac12 g^{\b\g}g_{\b\g,\a}=\lef(\ln{\sqrt{-g}}\ri)_{,\a}\;.\ea
The two symbols are interrelated
\ba
\la{aa7}
\G_\a&=&-{\cal Y}_a+g^{\b\g}g_{\a\b,\g}\;,\\
\la{aa8}
\G^\a&=&-{\cal Y}^a-g^{\a\b}{}_{,\b}\;,\ea

\subsection{Curvature}

The second level of differential structure of manifold is its curvature defined in terms of the Riemann tensor. We define the Riemann tensor as follows \citep{kopeikin_2011book}
\ba\la{ri1}
R^\a{}_{\m\b\n}&=&\Gamma^\a{}_{\m\n,\b}-\Gamma^\a{}_{\m\b,\n}+\Gamma^\a{}_{\b\g}\Gamma^\g{}_{\m\n}-\Gamma^\a{}_{\n\g}\Gamma^\g{}_{\m\b}\;.
\ea
Riemann tensor can be also expressed in terms of the second partial derivatives of the metric tensor and the Christoffel symbols
\be\la{ri1c}
R_{\a\m\b\n}=\frac{1}{2}\lef(g_{\m\b,\a\n}+g_{\n\a,\b\m}-g_{\a\b,\m\n}-g_{\m\n,\a\b} \ri)+\Gamma_{\r\m\b}\Gamma^\r{}_{\a\n}-\Gamma_{\r\m\n}\Gamma^\r{}_{\a\b}\;.
\ee
Contraction of two indices in the Riemann tensor yields the Ricci tensor
\ba
\la{ri2}
R_{\m\n}&=&\Gamma^\a{}_{\m\n,\a}-{\cal Y}_{\m,\n}+{\cal Y}_\g\Gamma^\g{}_{\m\n}-\Gamma^\a{}_{\n\g}\Gamma^\g{}_{\m\a}\;,\ea
or, in terms of the second derivatives from the metric tensor and the Christoffel symbols,
\be\la{ri2a}
R_{\m\n}=\frac{1}{2}g^{\k\e}\lef(g_{\m\k,\e\n}+g_{\n\k,\e\m}-g_{\k\e,\m\n}-g_{\m\n,\k\e}\ri) +g^{\k\e}\Gamma_{\r\m\e}\Gamma^\r{}_{\k\n}-\Gamma_{\r\m\n}\Gamma^\r\;.
\ee

One more contraction of indices in the Ricci tensor brings about the Ricci scalar which we shall write down in the form suggested by Fock \citep[Appendix B]{fock_book}
\ba
\la{ri4}
R&=&L+\Y_\a\G^\a-\Y_\a\Y^\a+\G^\a{}_{,\a}-\Y^\a{}_{,\a}\;,
\ea
where
\ba
\la{ri5}
L&=&g^{\m\n}\lef(\Gamma^\a{}_{\n\g}\Gamma^\g{}_{\m\a}-\Y_\a\G^\a{}_{\m\n}\ri)\;,
\ea
is (up to a constant factor) the gravitational Lagrangian introduced by Einstein \citep{LanLif} as an alternative to the gravitational Lagrangian, $R$, of Hilbert. The Hilbert Lagrangian is the Ricci scalar which depends on the second derivatives of the metric tensor while the Einstein Lagrangian does not.

The two Lagrangians are interrelated
\be
\la{ri9}
R=L+\lef(-g\ri)^{-1/2}{\cal A}^\a{}_{,\a}\;,
\ee
where
\be
\la{ri10}
{\cal A}^\a=\sqrt{-g}\lef(\G^\a-\Y^\a\ri)\;,
\ee
is a vector density of weight $+1$. After performing differentiation in (\ref{ri9}), and accounting for (\ref{aa3}) we can easily prove that (\ref{ri9}) reproduces (\ref{ri4}).

One more form of relation between $R$ and $L$ will be useful for calculating the variational derivative in Appendix \ref{cvdhl}. To this end we introduce a new notation
\ba
\la{ri6}
\G&\eq&\G^\a{}_{,\a}+\Y_\a\G^\a\;,\ea
and notice that
\ba
\la{ri7}
g^{\a\b}\Y_{\a,\b}&=&\Y^\a{}_{,\a}+\Y_\a\G^\a+\Y_\a\Y^\a\;,
\ea
Equations (\ref{ri6}), (\ref{ri7}) allows us to cast (\ref{ri4}) to the following form
\ba
\la{ri8}
R&=&L+\G+\Y_\a\G^\a-g^{\a\b}\Y_{\a,\b}\;,
\ea
that was found by Fock \citep[appendix B]{fock_book}.

\section{Derivatives on manifold}\la{vg34v}
\subsection{Covariant derivative}

Covariant derivative on manifold is a rule of transportation of geometric objects from one point of the manifold to another. If the geometric object is a tensor density $\F=\F^{\mu_1\ldots\mu_p}_{\nu_1\ldots\nu_q}$ of type $(p,q)$ and weight $m$, the covariant derivative is defined by the following rule
\ba\la{cd4z2}
\F^{\mu_1\ldots\mu_p}_{\nu_1\ldots\nu_q;\a}&=&\F^{\mu_1\ldots\mu_p}_{\nu_1\ldots\nu_q,\a}+\G^{\mu_1}{}_{\a\b}\F^{\b\ldots\mu_p}_{\nu_1\ldots\nu_q}+\ldots+\G^{\mu_p}{}_{\a\b}\F^{\mu_1\ldots\b}_{\nu_1\ldots\nu_q}-\G^\b{}_{\a\nu_1}\F^{\mu_1\ldots\mu_p}_{\b\ldots\nu_q}-\ldots-\G^\b{}_{\a\nu_q}\F^{\mu_1\ldots\mu_p}_{\nu_1\ldots\b}-m{\cal Y}_\a\F^{\mu_1\ldots\mu_p}_{\nu_1\ldots\nu_q}\;.
\ea

Second covariant derivatives of tensors do not commute due to the curvature of spacetime. For example, for a covector field $\F_\a$ and a covariant tensor field of second rank, $\F^\a_\b$ the following commutation relations are hold
\ba\la{ri1a}
\F_{\a;\b\g}&=&\F_{\a;\g\b}+R^\m{}_{\a\b\g}\F_\m\;,\\
\la{ri1b}
\F^\a{}_{\b;\g\d}&=&\F^\a{}_{\b;\d\g}-R^\a{}_{\m\g\d}\F^\m{}_{\b}+R^\m{}_{\b\g\d}\F^\a{}_{\m}\;. 
\ea
It is straightforward to extend these commutation relations to tensors and tensor densities of higher rank.

\subsection{Variational derivative}\la{thld}

Theory of perturbations of physical fields on manifolds rely upon the principle of the least action of a functional $S$ called action. Variational derivative arises in the problem of finding solutions of the gravitational field equation that extremize the action 
\be\la{u0}
S=\int \F d^4x\;,
\ee
where $\F\equiv \sqrt{-g}f=\sqrt{-g}f\lef(Q,Q_{\a},Q_{\a\b}\ri)$, is a scalar density of weight $+1$. Let $\F= \F\lef(Q,Q_{\a},Q_{\a\b}\ri)$ depend on the field variable $Q$, its first - $Q_\a\equiv Q_{,\a}$ and second - $Q_{\a\b}\equiv Q_{,\a\b}$ partial derivatives that play here a similar role as velocity and acceleration in the Lagrangian mechanics of point-like particles. The field variable $Q$ can be a tensor field of an arbitrary type with the covariant and/or contravariant indices. For the time being, we suppress the tensor indices of $Q$ as it may not lead to a confusion. Function $\F$ depends on the determinant $g$ of the metric tensor and can also depend on its derivatives. We shall discuss this case in the sections that follow.

A certain care should be taken in choosing the dynamic variables of the Lagrangian formalism in case when the variable $Q$ is a tensor field. For example, if we choose a covariant vector field $A_{\m}$ as an independent variable, the corresponding ``velocity'' and ``acceleration'' variables must be chosen as $A_{\m,\a}$ and $A_{\m,\a\b}$ respectively. On the other hand, if the independent variable is chosen as a contravariant vector $A^{\m}$, the corresponding ``velocity'' and ``acceleration'' variables must be chosen as $A^\m{}_{,\a}$ and $A^\m{}_{,\a\b}$. The same remark is applied to any other tensor field. The reason behind is that $A_\m$ and $A^\m$ are interrelated via the metric tensor, $A^\m=g^{\m\n}A_\n$. Therefore, derivative of $A^\m$ differs from that of $A_\m$ by an additional term involving the derivative of the metric tensor which, if being improperly introduced, can bring about spurious terms to the field equations derived from the principle of the least action.

Variational derivative, $\d\F/\d Q$, taken with respect to the variable $Q$ relates a change, $\d S$, in the functional $S$ to a change, $\d\F$, in the function $\F$ that the functional depends on,
\be\la{u1}
\d S=\int\d\F d^4x\;,
\ee
where
\be\la{u2}
\d\F=\frac{\pd \F}{\pd Q}\d Q+\frac{\pd \F}{\pd Q_\a}\d Q_\a+\frac{\pd \F}{\pd Q_{\a\b}}\d Q_{\a\b}\;.
\ee
This is a functional increment of $\F$. The variational derivative is obtained after we single out a total divergence in the right side of (\ref{u2}) by making use of the commutation relations, $\d Q_\a=(\d Q)_{,\a}$ and $\d Q_{\a\b}=(\d Q)_{,\a\b}$. The total divergence is reduced to a surface term in the integral (\ref{u1}) which vanishes on the boundary of the volume of integration. Thus, the variation of $S$ with respect to $Q$ is given by
\be\la{u3}
\d S=\int\frac{\d\F}{\d Q}\d Q d^4x\;,
\ee
where
\ba\la{lagder}
\frac{\delta\F}{\delta Q}&\equiv&\frac{\pd\F}{\pd Q}-\frac{\pd}{\pd x^\a}\frac{\pd \F}{\pd Q_{\a}}+\frac{\pd^2}{\pd x^\a\pd x^\b}\frac{\pd \F}{\pd Q_{\a\b}}\;.\ea
Similar procedure can be applied to $S$ by varying it with respect to $Q_\a$ and $Q_{\a\b}$. In such a case we get the variational derivatives of $\F$ with respect to $Q_\a$
\be
\la{lagder1}
\frac{\delta \F}{\delta Q_{\a}}\equiv\frac{\pd \F}{\pd Q_{\a}}-\frac{\pd}{\pd x^\b}\frac{\pd \F}{\pd Q_{\a\b}}\;,\ee
and that of $\F$ with respect to $Q_{\a\b}$,
\be
\la{lagder2}
\frac{\delta \F}{\delta Q_{\a\b}}\equiv\frac{\pd \F}{\pd Q_{\a\b}}\;.
\ee

Let us assume that there is another geometric object, $\Tc\lef(Q,Q_{\a},Q_{\a\b}\ri)$, which differs from the original one $\F\lef(Q,Q_{\a},Q_{\a\b}\ri)$ by a total divergence
\be
\la{gd}
\Tc\lef(Q,Q_{\a},Q_{\a\b}\ri)=\F\lef(Q,Q_{\a},Q_{\a\b}\ri)+\partial_\b H^\b{}\lef(Q,Q_{\a}\ri)\;.
\ee
It is well-known \cite{PRD27,1969fpoo.book..326M} that taking the variational derivative (\ref{lagder}) from $\Tc$ and $\F$ yields the same result
\be
\la{gde}
\frac{\delta \Tc}{\delta Q}\equiv\frac{\delta \F}{\delta Q}\;,
\ee
because the variational derivative from the divergence is zero identically.
In fact, it is straightforward to prove a more general result, namely, that the variational derivative (\ref{lagder}), after it applies to a partial derivative of an arbitrary smooth function, vanishes identically
\be
\la{gd+}
\frac{\delta}{\delta Q}\lef(\frac{\pd \F}{\pd x^\a}\ri)\equiv 0\;.
\ee
However, this property does not hold for a covariant derivative in the most general case \cite{1969fpoo.book..326M}.

The variational derivatives are covariant geometric object that is they do not depend on the choice of a particular coordinates on manifold \citep{1969fpoo.book..326M,kopeikin_2011book}. In case, when the dynamic variable $Q$ is not a metric tensor, this statement can be proved by taking the first, $\Q_\a\eq Q_{;\a}$, and second, $\Q_{\a\b}\eq Q_{;\a\b}$, covariant derivatives of $Q$ as independent dynamic variables instead of its partial derivatives, $Q_\a$ and $Q_{\a\b}$. In this case the procedure of derivation of variational derivatives (\ref{lagder}), (\ref{lagder1}) remains the same and the result is
\ba\la{uu4}
\frac{\delta\F}{\delta Q}&=&\frac{\pd\F}{\pd Q}-\lef[\frac{\pd \F}{\pd \Q_{\a}}\ri]_{;\a}+\lef[\frac{\pd \F}{\pd \Q_{\a\b}}\ri]_{;\b\a}\;.\ea 
The order, in which the covariant derivatives are taken, is imposed by the procedure of the extracting the total divergence from the variation of the action in (\ref{u1}). The order of the derivatives is important because the covariant derivatives do not commute.

Variational derivative of $\F$ with respect to the metric tensor $g_{\m\n}$ is defined by the same equations (\ref{lagder})--(\ref{lagder2}) where we identify $Q\eq g_{\m\n}$, $Q_\a\eq g_{\m\n,\a}$, and $Q_{\m\n}\eq g_{\m\n,\a\b}$. It yields
\ba\la{er}
\frac{\delta\F}{\delta g_{\m\n}}&\equiv&\frac{\pd\F}{\pd g_{\m\n}}-\frac{\pd}{\pd x^\a}\frac{\pd \F}{\pd g_{\m\n,\a}}+\frac{\pd^2}{\pd x^\a\pd x^\b}\frac{\pd \F}{\pd g_{\m\n,\a\b}}\;\\
\la{er1}
\frac{\delta \F}{\delta g_{\m\n,\a}}&\equiv&\frac{\pd \F}{\pd g_{\m\n,\a}}-\frac{\pd}{\pd x^\b}\frac{\pd \F}{\pd g_{\m\n,\a\b}}\;,\\
\la{er2}
\frac{\delta \F}{\delta g_{\m\n,\a\b}}&\equiv&\frac{\pd \F}{\pd g_{\m\n,\a\b}}\;.
\ea

If the geometric object $\F$ depends on the contravariant components of the metric tensor, $g^{\a\b}$, and/or its derivatives the partial derivatives in (\ref{er})--(\ref{er2}) are calculated after making use of relations
\ba
\la{ay1}
\frac{\pd{g}^{\a\beta}}{\pd{g}_{\mu\nu}}&=&-g^{\a(\m}g^{\n)\b} \;,\\
\la{ay2}
\frac{\pd{g}^{\a\beta}{}_{,\gamma}}{\pd{g}_{\mu\nu,\sigma}}&=&-g^{\a(\m}g^{\n)\b}\delta^\sigma_\gamma\;,\\
\la{ay3}
\frac{\pd{g}^{\a\beta}{}_{,\gamma\k}}{\pd{g}_{\mu\nu,\r\s}}&=&-g^{\a(\m}g^{\n)\b}
\delta^{(\r}_\g\delta^{\s)}_\k\;.
\ea

Variational derivatives (\ref{er})--(\ref{er2}) preserve covariance. The most simple way to prove it would be to express (\ref{er})--(\ref{er2}) in terms of the covariant derivatives like we did in transformation of variational derivative (\ref{u2}) to its covariant analogue (\ref{uu4}).  However, in case of variational derivative with respect to the metric tensor this procedure is not so straightforward because the covariant derivative of the metric tensor, $g_{\m\n;\a}\equiv 0$, and we cannot use it as a covariant dynamic variable being conjugated to the metric tensor. In this case, we consider a set of the metric tensor, $g_{\m\n}$, the Christoffel symbols $\G^\a{}_{\m\n}$, and the Riemann tensor $R^\a{}_{\b\m\n}$ as independent dynamic variables. The action is given by (\ref{u0}) where $\F\eq \sqrt{-g}f\lef(g_{\m\n},\G^\a{}_{\m\n},R^\a{}_{\b\m\n}\ri)$ is a scalar density of weight $+1$. Variation of $\F$ is
\be\la{er5}
\d\F=\frac{\pd \F}{\pd g_{\m\n}}\d g_{\m\n}+\frac{\pd \F}{\pd \G^\a{}_{\m\n}}\d \G^\a{}_{\m\n}+\frac{\pd \F}{\pd R^\a{}_{\b\m\n}}\d R^\a{}_{\b\m\n}\;,
\ee  
where variations of the Christoffel symbols and the Riemann tensor are tensors that can be expressed in terms of the variation $\d g_{\m\n}$ of the metric tensor \citep{weinberg_1972}
\ba\la{er6}
\d \G^\a{}_{\m\n}&=&\frac12 g^{\a\s}\lef[(\d g_{\s\m})_{;\n}+(\d g_{\s\n})_{;\m}-(\d g_{\m\n})_{;\s}\ri]\;\;,\\
\d R^\a{}_{\b\m\n}&=&(\d \G^\a{}_{\b\n})_{;\m}-(\d \G^\a{}_{\b\m})_{;\n}\;\;.
\ea
Now, we replace variations of the Christoffel symbols and the Riemann tensor in (\ref{er5}) with (\ref{er6}), (\ref{er7}) and single out a total divergence \footnote{The fact that $\F$ is a scalar density is essential for the transformation of covariant derivatives to the total divergence. The total divergences can be converted to surface integrals which vanish on the boundary of integration and, hence, can be dropped off the calculations.}. It yields
\be\la{er7}
\d\F=\frac{\d\F}{\d g_{\m\n}}\d g_{\m\n}+{\cal B}^\a{}_{,\a}\;\;,
\ee
where the total divergence vanishes on the boundary of integration of the action, and the covariant variational derivative is
\ba\la{er8}
\frac{\d\F}{\d g_{\m\n}}&=&\frac{\pd\F}{\pd g_{\m\n}}
-\frac12\lef(g^{\s\m}\frac{\pd\F}{\pd \G^\s{}_{\n\a}}+g^{\s\n}\frac{\pd\F}{\pd \G^\s{}_{\m\a}}-g^{\s\a}\frac{\pd\F}{\pd \G^\s{}_{\m\n}}\ri)_{;\a}\\&&\phantom{\frac{\pd\F}{\pd g_{\m\n}}}\nonumber+
\lef(g^{\s\m}\frac{\pd\F}{\pd R^\s{}_{\a\b\n}}+g^{\s\n}\frac{\pd\F}{\pd R^\s{}_{\m\b\a}}-g^{\s\a}\frac{\pd\F}{\pd R^\s{}_{\m\b\n}}\ri)_{;\b\a}
\ea

This equation can be further simplified if we shall make use of the Christoffel symbols of the first kind, $\G_{\a\m\n}=g_{\a\s}\G^\s{}_{\m\n}$, and $R_{\a\b\m\n}=g_{\a\s}R^\s{}_{\b\m\n}$. The partial derivatives
\be\la{er8a}
\frac{\pd\F}{\pd \G^\s{}_{\m\n}}=g_{\s\r}\frac{\pd\F}{\pd \G_{\r\m\n}}\qquad,\qquad \frac{\pd\F}{\pd R^\s{}_{\l\m\n}}=g_{\s\r}\frac{\pd\F}{\pd R_{\r\l\m\n}}\;.
\ee
Moreover, the cyclic permutation property of the Riemann tensor tells us that
\be\la{er8b} 
\frac{\pd\F}{\pd R_{\r\l\m\n}}=-\frac{\pd\F}{\pd R_{\r\m\n\l}}-\frac{\pd\F}{\pd R_{\r\n\l\m}}\;.
\ee
Employing (\ref{er8a}), (\ref{er8b}) in (\ref{er8}) transforms, and making use of antisymmetry $R_{\s\b\m\n}=-R_{\s\b\n\m}$ of the Riemann tensor, reduces (\ref{er8}) to a more compact form
\be\la{er8a1}
\frac{\d\F}{\d g_{\m\n}}=\frac{\pd\F}{\pd g_{\m\n}}-\frac12\lef(\frac{\pd\F}{\pd \G_{\m\n\a}}+\frac{\pd\F}{\pd \G_{\n\m\a}}-\frac{\pd\F}{\pd \G_{\a\m\n}}\ri)_{;\a}+
\lef(\frac{\pd\F}{\pd R_{\m\a\b\n}}+\frac{\pd\F}{\pd R_{\m\b\a\n}}\ri)_{;\b\a}
\ee

Calculation of variational derivatives requires calculation of partial derivatives with respect to the metric tensor
and other geometric objects like the Christoffel symbols, the Riemann tensor, etc. An example is the partial derivatives from the determinant of the metric tensor
\be
\la{ay4}
\frac{\pd\sqrt{-g}}{\pd{g}^{\mu\nu}}=-\frac12\sqrt{-g}g_{\mu\nu}\qquad,\qquad \frac{\pd\sqrt{-g}}{\pd{g}_{\mu\nu}}=\frac12\sqrt{-g}g^{\mu\nu}  \;,
\ee
where $g$ is the determinant of the metric tensor. Taking partial derivatives from $\F=\F\left(g_{\m\n},\G^\a_{\m\n},R^\a_{\b\m\n}\right)$ with respect to $g_{\m\n}$, $\G^\a_{\m\n}$ and $R^\a_{\b\m\n}$ is performed with the help of the following formulas
\ba\la{er9}
\frac{\pd g_{\a\b}}{\pd g_{\m\n}}&=&\d^{(\m}_{\a}\d^{\n)}_{\b}\;,\\
\la{er10}
\frac{\pd \G^\s{}_{\a\b}}{\pd \G^\r{}_{\m\n}}&=&\d^{\s}_\r\d^{(\m}_{\a}\d^{\n)}_{\b}\;,\\
\la{er11}
\frac{\pd R^\s{}_{\g\a\b}}{\pd R^\r{}_{\k\m\n}}&=&\d^\s_\r\d^\k_\g\d^{[\m}_{\a}\d^{\n]}_{\b}\;,
\ea
where we have accounted for the symmetry of the Christoffel symbols and the antisymmetry of the Riemann tensor.

In case when $\F$ is a function of $g^{\m\n}$ the variational derivative with respect to the contravariant metric tensor is
\be\la{err8}
\frac{\d\F}{\d g^{\m\n}}=\frac{\pd g_{\a\b}}{\pd g^{\m\n}}\frac{\d\F}{\d g_{\a\b}}=-g_{\a\m}g_{\b\n}\frac{\d\F}{\d g_{\a\b}}\;.
\ee
We will also need to calculate the variational derivative with respect to the density of the metric tensor, $\gag^{\mu\nu}$. It relates to the variational derivative of the metric tensor as follows,
\ba
\la{ia6}
 \frac{\delta}{\delta \gag^{\mu\nu}}&=&
\frac{\pd{g}^{\r\s}}{\pd
\gag^{\mu\nu}}\frac{\delta}{\delta{g}^{\r\s}}=A^{\r\s}_{\mu\nu}\frac{1}{\sqrt{-
g}}\frac{\delta}{\delta {g}^{\r\s}}\;,
\ea
where
\be\la{nqv}
A^{\rho\sigma}_{\mu\nu}=\frac{1}{2}\lef(\delta^\rho_\mu\delta^\s_\nu+\delta^\rho_\nu\delta^\s_\mu-g_{\mu\nu} g^{\rho\sigma}\ri)\;.
\ee

The variational derivatives are not linear operators. For example, they do not obey Leibniz's rule \citep[Section 2.3]{MR1383589}. More specifically, for any geometric object, $\H=\F\Tc$, that is a corresponding product of two other geometric objects, $\F=\F\lef(Q,Q_{\a},Q_{\a\b}\ri)$ and $\Tc=\Tc\lef(Q,Q_{\a},Q_{\a\b}\ri)$, the variational derivative
\be\la{xos6}
\frac{\d\lef(\F\Tc\ri)}{\d Q}\not=\frac{\d\F}{\d Q}\Tc+\F\frac{\d\Tc}{\d Q}\;,
\ee
in the most general case.
The chain rule with regard to the variational derivative is preserved in a limited sense. More specifically, let us consider a geometric object $\F=\F\lef(Q,Q_{\a},Q_{\a\b}\ri)$ where $Q$ is a function of a variable $P$, that is $Q=Q(P)$. Then, the variational derivative
\be\la{xcr1a}
\frac{\d\F}{\d P}=\frac{\d\F}{\d Q}\frac{\pd Q}{\pd P}\;,
\ee
that can be confirmed by inspection \cite{1988IJMPA...3.2651P}. On the other hand, if we have a singled-valued function $\H=\H(Q)$, and $Q=Q\lef(P,P_{\a},P_{\a\b}\ri)$, the chain rule
\be\la{xcr1}
\frac{\d\H}{\d P}=\frac{\pd\F}{\pd Q}\frac{\d Q}{\d P}\;,
\ee
is also valid.
The chain rule (\ref{xcr1}) will be often used in calculations of the present paper.

\subsection{Lie derivative}\la{lider}
Lie derivative on the manifold can be viewed as being induced by a diffeomorphism
\be
\la{ct}
x'^\a=x^\a+\xi^\a(x)\;,
\ee
such that a vector field $\xi^\a$ has no self-intersections, thus, defining a congruence of curves which provides a natural mapping of the manifold into itself.
Lie derivative of a geometric object $\F$ is denoted as $\pounds_{\bm\xi}\F$. It is defined by a standard rule
\be
\la{li1}
\pounds_{\bm\xi}\F=\F'(x)-\F(x)\;,
\ee
where $\F'$ is calculated by doing its coordinate transformation induced by the change of the coordinates (\ref{ct}) with subsequent pulling back the transformed object from the point $x'^\alpha$ to $x^\alpha$ along the congruence $\xi^\a$ \citep{kopeikin_2011book}. In particular, for any tensor density $\F=\F^{\mu_1\ldots\mu_p}_{\nu_1\ldots\nu_q}$ of type $(p,q)$ and weight $m$ one has
\ba
\la{gde1}
\pounds_{\bm\xi}\F^{\mu_1\ldots\mu_p}_{\nu_1\ldots\nu_q}&=&\xi^\a \F^{\mu_1\ldots\mu_p}_{\nu_1\ldots\nu_q}{}_{,\a}+m\xi^\a{}_{,\a}\F^{\mu_1\ldots\mu_p}_{\nu_1\ldots\nu_q}\\\nonumber
&&+\,\F^{\mu_1\ldots\mu_p}_{\a\ldots\nu_q}\xi^\a{}_{,\nu_1}+\ldots+\F^{\mu_1\ldots\mu_p}_{\nu_1\ldots\a}\xi^\a{}_{,\nu_q}\\\nonumber&&  -\,\F^{\a\ldots\mu_p}_{\nu_1\ldots\nu_q}\xi^{\mu_1}{}_{,\a}-\ldots-\F^{\mu_1\ldots\a}_{\nu_1\ldots\nu_q}\xi^{\mu_p}{}_{,\a}\;.
\ea
We notice that {\it all} partial derivatives in the right side of equation (\ref{gde1}) can be simultaneously replaced with the covariant derivatives because the terms containing the Christoffel symbols cancel each other.

The Lie derivative commutes with a partial (but not a covariant) derivative
\be
\la{li2}
\pd_\a\lef(\pounds_{\bm\xi}\F\ri)=\pounds_{\bm\xi}\lef(\pd_\a \F\ri)\;,
\ee
where $\F$ is actually an arbitrary geometric object rather than merely a tensor density.
This property allows us to prove that a Lie derivative from a geometric object $\F\lef(Q,Q_{\a},Q_{\a\beta}\ri)$ can be calculated in terms of its variational derivative. Indeed,
\be
\la{li3}
\pounds_{\bm\xi}\F=\frac{\pd\F}{\pd Q}\pounds_{\bm\xi}Q+\frac{\pd \F}{\pd Q_{\a}}\pounds_{\bm\xi}Q_{\a}+\frac{\pd \F}{\pd Q_{\a\b}}\pounds_{\bm\xi}Q_{\a \b}\;.
\ee
Now, after using the commutation property (\ref{li2}) and changing the order of partial derivatives in $\pounds_{\bm\xi}Q_{\a }$ and $\pounds_{\bm\xi}Q_{\a\b}$, one can express (\ref{li3}) as an algebraic sum of the variational derivative and a total divergence
\be
\la{li4}
\pounds_{\bm\xi}\F=\frac{\d\F}{\d Q}\pounds_{\bm\xi}Q+\frac{\pd}{\pd x^\a }\lef(\frac{\d \F}{\d Q_{\a }}\pounds_{\bm\xi}Q
+\frac{\d \F}{\d Q_{\a\b}}\pounds_{\bm\xi}Q_\b 
\ri)\;.
\ee
This property of the Lie derivative indicates its close relation to the variational derivative on manifold and will be used in the calculations that follow this section. 

It is also worth pointing out that (\ref{li4}) is used in derivation of Noether's theorem of conservation of the canonical stress-energy tensor of the field $Q$ in case when $\F=\lag$ is the Lagrangian density of weight $m=+1$ of the field $Q$ which variational derivative vanishes on-shell, $\d\F/\d Q=\d\lag/\d Q=0$. The Lagrangian density has the Lie derivative in the form of total divergence, $\pounds_{\bm\xi}\lag=\pd_\a\lef(\xi^\a\lag\ri)$, and (\ref{li4}) yields the conserved Noether current 
\be\la{q2w3v}
J^\a\eq \xi^\a\lag-\frac{\d \F}{\d Q_{\a }}\pounds_{\bm\xi}Q
-\frac{\d \F}{\d Q_{\a\b}}\pounds_{\bm\xi}Q_\b
\;,
\ee
where $\xi^\a$ is a vector field defining the change of coordinates (\ref{li2}). This field should not be confused with the Killing vector defining isometry of the metric tensor. The Noether current is conserved, $\nabla_\a J^\a=0$, independently of whether the manifold admit isometries or not \citep{2011CQGra..28u5021P}.

\section{Field Perturbation Theory on Spacetime Manifold}\la{sec3}

\subsection{Physical fields and their perturbations}

Let us consider a field theory on a background pseudo-Riemannian manifold $\bar\M$ having the metric tensor $\bar g_{\m\n}$ that is a solution of Einstein's equations 
\be\la{p9m5}
\bar G_{\m\n}-8\pi\bar T^{\rm\st M}_{\m\n}=0\;,
\ee
where $\bar G_{\m\n}=\bar R_{\m\n}-1/2 \bar g_{\m\n}\bar R$ is the Einstein tensor, and $\bar T^{\rm\st M}_{\m\n}$ is the stress-energy tensor of the matter fields $\bar\Phi^A$, where the index $A$ numerates the fields and takes the values $A=1,2,\ldots,a$. We assume that the solution of the background Einstein's equations (\ref{p9m5}) is known. 

In the simplest case the background manifold $\M$ is considered as Minkowski spacetime with the background metric $\bar g_{\m\n}=\eta_{\m\n}={\rm diag}[-1,+1,+1,+1]$ being the Minkowski metric. In this case, both tensors $\bar G_{\m\n}$ and $\bar T^{\rm\st M}_{\m\n}$ vanish identically, and the background Einstein's equations (\ref{p9m5}) satisfy automatically. The use of the Minkowski background is typical in solving the problems of post-Newtonian celestial mechanics \citep{kopeikin_2011book, kopeikin_2014} and in the branch of gravitational wave astronomy dealing with calculation of templates of gravitational waves emitted by coalescing binary systems \citep{Blanchet_2000LNP,futamase_2007LRR,Schaefer_2011mmgr}. 
Minkowski background is not appropriate in cosmology which operates with conformally-flat Friedman-Lema\^itre-Robertson-Walker (FLRW) metric tensor
\be\la{frm1}
d\bar s^2=-dT^2+R^2(T)\lef(1+\frac14 kr^2\ri)^{-2}\d_{ij}dX^i dX^j\;,
\ee
where $T$ is the cosmic time, $X^\a=(T,X^i)$ are the global coordinates associated with the Hubble flow, $r=\sqrt{\d_{ij}X^iX^j}$, $k$ is the curvature of space taking one of the three values $k=-1,0,+1$, and $R(T)$ is the scale factor which temporal evolution is governed by the solution of Einstein's equations (\ref{p9m5}) with the background stress-energy tensor $\bar T^{\rm\st M}_{\m\n}$ determined by the matter fields $\bar\Phi^A$ \citep{weinberg_1972,weinberg_2008,rubak_2011}. FLRW metric (\ref{frm1}) is conformally flat and is not reduced to Minkowski metric globally.
The conformal nature of the cosmological metric does not allow us to apply the post-Newtonian approximations which should be generalized to take into account the curvature of the background spacetime to solve the Einstein equations for the field perturbations. 

Let us perturb the background manifold $\M$ so that the geometry of the perturbed manifold $\M$ is now described by the metric tensor $g_{\m\n}$ that is a solution of perturbed Einstein's equations 
\be\la{n7x4}
G_{\m\n}-8\pi T^{\rm\st M}_{\m\n}=8\pi T_{\m\n}\;,
\ee
with the perturbed values of the Einstein tensor $G_{\m\n}=R_{\m\n}-1/2 g_{\m\n}R$, and the stress-energy tensor $T^{\rm\st M}_{\m\n}$ of the same physical fields $\Phi^A$. Besides the background matter we admit the presence of the stress-energy tensor $T_{\m\n}$ of the other matter fields $\Theta^B$ where the index $B$ numerates the bare fields and takes values $B=1,2,\ldots,b$ in the right side of Einstein's equations. These fields represent the source of the bare perturbation of the background manifold which can be associated in cosmology with a small-scale inhomogeneities having rather large density contrast created by the presence of baryonic matter making stars, planets, etc. or, even, black holes.

The perturbed metric $g_{\m\n}$ and the perturbed matter fields $\Phi^A$ can be always represented as linear combinations $g_{\m\n}=\bar g_{\m\n}+\varkappa_{\m\n}$, and $\Phi^A=\bar\Phi^A+\phi^A$, where $\varkappa_{\m\n}$ and $\phi^A$ are the perturbations of the metric and the matter fields respectively. The fields $\Theta^B$ are not present on the background manifold $\bar\M$, and appear only as bare perturbations ``injected'' to the background manifold from ``outside''. In fact, the only physical system we are dealing with, is the perturbed manifold. Mathematical formalism of the perturbation theory is based on separation of the physical manifold into two parts -- the background and the perturbations. This is done merely for mathematical convenience since it allows us to set up a consistent and rigorous mathematical framework for adequate description of gravitational physics on perturbed manifold $\M$. In what follows, we assume that the fields $\Phi^A$ and $\Theta^B$ are both minimally coupled to the curvature of spacetime in the sense of the strong equivalence principle \citep[Section 3.8.2]{kopeikin_2011book}, \citep[Section 6.13]{1980gmmp.book.....S}. We also assume, for the sake of simplicity, that the fields $\Phi^A$ and $\Theta^B$ do not directly interact one with another.  This assumption can be easily relaxed but the calculations will be longer. We postpone consideration of this problem to a future publication.

In the first approximation the field perturbations $\varkappa_{\m\n}$ and $\phi^A$ can be split in two categories: 
\begin{enumerate}
\item the free perturbations, which are solutions of the homogeneous Einstein's equations (\ref{n7x4}) with the right side $T_{\m\n}=0$;
\item the induced perturbations, which are particular solutions of the inhomogeneous equations (\ref{n7x4}) with the right side $T_{\m\n}\not=0$.
\end{enumerate} 
If the background manifold is Minkowski-flat, only the induced perturbations make physical sense as the free perturbations represent merely coordinate effects which can be removed by a corresponding choice of coordinates on the background manifold $\bar\M$. This is no longer true if the background manifold is curved. For example,  
in cosmology the free perturbations are basically primordial perturbations which are relics of the boundary conditions imposed at the epoch of Big Bang without presence of any particular physical source. There are several alternative explanations of the formation of the primordial (free) perturbations which are discussed, for example, in textbooks \citep{mukh_book,weinberg_2008,Amendola_2010}. The free perturbations of matter fields grow to form the large-scale structure of the universe. The source of the induced perturbations in cosmology is the stress-energy tensor $\T_{\m\n}$ of the bare perturbations $\Theta^B$. Due to the non-linearity of Einstein's equations the perturbations interact between themselves and one with another through the gravitational coupling in non-linear Einstein's equations. It makes the geometric structure of the perturbed manifold $\M$ rather entangled as we go from the first to higher-order approximation theory. 

The present paper describes how to find out the perturbed geometric structure of the manifold $\M$, how to formulate the field equations for the perturbations, and how to derive the equations of motion of perturbations on the basis of the Lagrangian-based variational principle \citep{1984CMaPh..94..379G,1988IJMPA...3.2651P}. The Lagrangian formalism allows us to set up the perturbation theory in a covariant and gauge-invariant representation directly, without splitting the metric tensor perturbations in the scalar, vector and tensor modes which is a rather popular approach in the studies of cosmological perturbations \citep{mukh_book,weinberg_2008}. However, such splitting is usually accompanied with a specific foliation of comsological spacetime which disguises the covariant nature of the perturbation theory and makes the entire perturbation approach look quite different from the post-Newtonian approximation scheme elaborated on the Minkowski-flat background manifold. The Lagrangian-based variational formalism allows us to reconcile the theory of cosmological perturbations with the post-Newtonian approximations in a natural and self-consistent way.

\subsection{The Lagrangian-based variational principle}

The Lagrangian formulation of the dynamic theory of field perturbations in general relativity starts off the Hilbert action defined on the unperturbed background manifold $\bar\M$,
\be\la{una1}
\bar S = \int d^4x\, \bar{\lag}\lef(\bar g_{\mu\nu},\bar\Phi^A\ri)\;,
\ee
where $\bar\lag=\sqrt{-\bar g}\bar L$ is a scalar density of weight $+1$ (because of $\sqrt{-\bar g}$), $\bar L=\bar L\lef(\bar g_{\mu\nu},\bar\Phi^A\ri)$ is the Lagrangian which is a scalar function depending on the metric tensor, the matter fields $\bar \Phi^A=\{\bar\Phi^1,\bar\Phi^2,\ldots,\bar\Phi^a\}$, and their partial derivatives. To avoid superfluous notations we did not show explicitly in (\ref{una1}) but keep in mind, the dependence of the Lagrangian on the partial derivatives of the field. The matter fields $\bar\Phi^A$ determine the dynamic and geometric structure of the background manifold $\bar\M$ via Einstein's equations. For short, we shall call the Lagrangian density, $\bar\lag$, simply the Lagrangian.

The Lagrangian ${\lag}$ is split in two parts
\be\la{y6}
\bar\lag\lef(\bar g_{\mu\nu},\bar\Phi^A\ri)=\bar\lag^{\rm\st G}\lef(\bar g_{\mu\nu}\ri)+\bar\lag^{\rm\st M}\lef(\bar g_{\mu\nu},\bar\Phi^A\ri)\;,
\ee
where the gravitational (Hilbert) Lagrangian
\be\la{gzup}
\bar{\lag}^{\rm\st G}\lef(\bar g_{\alpha\beta}\ri)\equiv-\frac{1}{16\pi}\sqrt{-\bar g}\bar R \;,
\ee
depends on the background metric tensor $\bar g_{\mu\nu}$ as well as on its first and second derivatives. The matter Lagrangian, $\bar\lag^{\rm\st M}$, depends on the matter fields $\bar\Phi^A$ and their derivatives. It also depends on the metric tensor and (for instance, in the case of Yang-Mills fields) on its first derivatives. 

Dynamic equations of the gravitational field and matter are derived from the principle of the least action by varying the action (\ref{una1}) and equating its variation to zero. This procedure is well-known and we shall not repeat it over here (see, for example, \citep[section 3.9]{kopeikin_2011book}). It is equivalent to taking the variational derivatives (\ref{lagder}) from the Lagrangian (\ref{y6}) with the variable $Q=\{\bar g_{\m\n},\bar\Phi^A\}$. It yields the Euler-Lagrange equations
\ba
\la{aq3}
\frac{\de\bar\lag^{\rm\st M}}{\de \bar\Phi^A}&=&0\;,\\
\la{(2.7)}
\frac{\de \bar\lag^{\rm\st G}}{\de \bar\gag^{\mu\nu}}+\frac{\de \bar\lag^{\rm\st M}}{\de \bar\gag^{\mu\nu}}&=&
0\;.
\ea
Equation (\ref{aq3}) describes a dynamic evolution of the matter fields $\bar\Phi^A$.
Equation (\ref{(2.7)}) can be recognized as the Einstein equations (\ref{p9m5}) for the background gravitational field (the metric tensor) after noticing that the variational derivatives
\ba\la{aq1}
\frac{\de \bar\lag^{\rm\st G}}{\de \bar{\gag}^{\mu\nu}}&=&-\frac{1}{16\pi}\bar R_{\mu\nu}\;,\\
\la{aq2}
\frac{\de \bar{\lag}^{\rm\st M}}{\de \bar{\gag}^{\mu\nu}}&=&\frac{1}{2}\lef(\bar T^{\rm\st M}_{\mu\nu} - \frac12 \bar g_{\mu\nu}\bar T^{\rm\st M}\ri)\;, \ea
where $\bar R_{\m\n}$ is the background value of the Ricci tensor calculated with the help of the background metric $\bar g_{\m\n}$, and $\bar T^{\rm\st M}_{\m\n}$ is the stress-energy tensor of the fields $\bar\Phi^A$. Equation (\ref{aq2}) is just a definition of the metrical stress-energy tensor of matter \citep[Section 3.9.5]{kopeikin_2011book}. Equation (\ref{aq1}) is usually derived by varying the gravitational action (see, for instance, \citep[page 310]{kopeikin_2011book}, \citep[page 364]{weinberg_1972}) and extracting the total derivative that vanishes on the boundary of the volume of integration. 
The same result is obtained if we take the variational derivative (\ref{aq1}) directly by making use of (\ref{ia6}), (\ref{err8}) and (\ref{er}). Calculations are straightforward but tedious, and are given in Appendix (\ref{appB2}) of the present paper.
The same result can be achieved in much shorter and attractive way if we use the covariant definition (\ref{er8}) of the variational derivative. Indeed, the Lagrangian (\ref{gzup}) is equivalent to
\be
\la{aq2a}
\bar\lag^{\rm\st G}=-\frac{1}{16\pi}\bar\gag^{\k\l}\d^\s_\r\bar R^\r{}_{\k\s\l}\; 
\ee
This expression depends only on the metric tensor and the Riemann tensor as a linear algebraic function. Hence, partial derivatives with respect to the Christoffel symbols are automatically nil. Moreover, the covariant derivatives from the metric tensor vanish identically. Hence, the variational derivative (\ref{er8}) (along with (\ref{err8})) is reduced to a simple partial derivative
\be\la{aq2b}
\frac{\de \bar\lag^{\rm\st G}}{\de \bar{\gag}^{\mu\nu}}=-\frac{1}{16\pi}\frac{\pd \bar\gag^{\k\l}}{\pd\bar{\gag}^{\m\n}}\bar R_{\k\l}\;,
\ee
that immediately results in (\ref{aq1}). 

Physical perturbations of the background manifold $\bar\M$ are caused either by imposing the initial spectrum of primordial perturbations on the metric tensor $g_{\m\n}$ and fields $\Phi^A$ or by ``injecting'' on the manifold the bare matter field $\bar\Theta^B$ with the Lagrangian $\bar\lag^{\rm\st P}=\sqrt{-\bar g}\bar L^{\rm\st P}$ where $\bar L^{\rm\st P}\equiv \bar L^{\rm\st P}(\bar g_{\m\n},\bar\Theta^B)$ is a scalar function. The present paper assumes that $\bar\Theta^B$ is minimally coupled with gravity but does not interact directly with the fields $\bar\Phi^A$. We postulate that the absolute value of $\bar{\lag}^{\rm\st P}$ is much smaller than the Lagrangian $\bar\lag$ of the background manifold, that is $\bar{\lag}^{\rm\st P}\ll\bar\lag$.
The fields $\bar\Theta^B$ can be conceived, for example, as a baryonic matter composing an isolated astronomical system like the solar system or a galaxy, or a cluster of galaxies. However, it is also admissible to consider $\bar\Theta^B$ as a seed perturbations of the fields $\bar\Phi^A$, for example, in discussion of the formation of the small-scale structure of the universe at the latest stages of its evolution. The assumptions imposed on the Lagrangian of the fields $\bar\Theta^B$ presume that in order to describe the dynamic evolution of the perturbed manifold $\M$ we should add algebraically the Lagrangian $\bar\lag^{\rm\st P}$ of the bare perturbations to the unperturbed Lagrangian $\bar\lag$ of the background manifold, write down the perturbed Einstein equations for the metric tensor perturbations $l_{\m\n}$ along with the equations for the perturbations $\phi^A$ of the fields $\bar\Phi^A$, solve them, and proceed to the second, third, etc. iterations if necessary. The iterative theory of the Lagrangian perturbations of the manifold is described in the following sections.

\subsection{The Lagrangian perturbations of dynamic fields}

Lagrangian-based formulation of the dynamic theory of physical perturbations of a manifold starts off the Hilbert action
\be\la{(2.0)}
S = \int d^4x\, {\lag}(g_{\mu\nu},\Phi^A,\Theta^B)\;,
\ee
where $\lag=\sqrt{-g}L$ is the scalar density of weight $+1$, and $L=L(g_{\mu\nu},\Phi^A,\Theta^B)$ is the Lagrangian depending on the metric tensor, the matter fields $\Phi^A=\{\Phi^1,\Phi^2,\ldots,\Phi^a\}$, and the fields $\Theta^B=\{\Theta^1,\Theta^2,\ldots,\Theta^b\}$ representing the bare perturbation of the manifold.

The Lagrangian ${\lag}$ consists of three parts
\be\la{(2.1)}
\lag\lef(g_{\mu\nu},\Phi^A,\Theta^B\ri)=\lag^{\rm\st G}\lef(g_{\mu\nu}\ri)+\lag^{\rm\st M}\lef(g_{\mu\nu},\Phi^A\ri)+\lag^{\rm\st P}\lef(g_{\m\n},\Theta^B\ri)\;,
\ee
where the gravitational (Hilbert) Lagrangian
\be\la{gzu}
{\lag}^{\rm\st G}\lef(g_{\alpha\beta}\ri)\equiv-\frac{1}{16\pi}\sqrt{- g}R \;,
\ee
depends on the metric tensor $g_{\mu\nu}$, its first and second derivatives. The Lagrangians of matter, $\lag^{\rm\st M}$ and $\lag^{\rm\st P}$, depend solely on the metric tensor and its first derivatives. They also depend directly on the matter fields $\Phi^A$ and $\Theta^B$ and their partial derivatives but we did not show it explicitly to avoid tedious notations. The matter fields $\Phi^A$ and $\Theta^B$ are minimally coupled to gravity but we assume that they are not directly coupled to each other. Hence, the Lagrangian of the interaction between these fields does not appear explicitly in (\ref{(2.1)}). This assumption can be relaxed and successfully handled with the formalism of the present paper but the computational aspects become more intricate and will be considered somewhere else. 

It is worth noticing that ${\lag}^{\rm\st M}$ and $\lag^{\rm\st P}$ depend on the metric tensor $g_{\mu\nu}$ both explicitly and implicitly through the mathematical definition of the matter fields $\Phi^A$ and $\Theta^B$. For example, consider the Lagrangian of the perfect fluid ${\lag}^{\rm\st M}=\rho\lef(1+\Pi\ri)\sqrt{- g}$, where $\Pi$ is the specific internal energy of the fluid and $\rho$ is the energy density. The metric tensor appears explicitly as $\sqrt{-g}$ and implicitly in $\rho$ that is defined as the ratio of the rest energy of the fluid's element to its comoving volume which depends on the determinant of the metric tensor \citep{kopeikin_2011book}. Since $\Pi=\Pi(\rho)$ is a thermodynamic function of $\rho$, it also depends implicitly on the metric tensor. It means that the variational derivatives of $\rho$ and $\Pi$ with respect to the metric tensor have certain values which we shall discuss in the sections which follow and in Appendix \ref{appvdm}. 

We define perturbations of the gravitational and matter fields residing on the background manifold by the following equations,
\ba
\la{(2.4)}
\gag^{\mu\nu}(x)&=& \bar\gag^{\mu\nu}(x) + \hatl^{\mu\nu}(x)\;,\\
\la{(2.5)}
\Phi^A(x)&=& \bar\Phi^A(x) +\phi^A(x)\;,\\\la{2.5as}
\Theta^B(x)&=&\bar\Theta^B(x)+\theta^B(x)\;,
\ea
where all functions are taken at one and the same point $x\eq x^\a$ of the unperturbed manifold $\bar\M$.
The perturbed values of the  fields are assumed to be sufficiently small compared with their background counterparts: $|{\hatl}^{\mu\nu}|<|\bar\gag^{\mu\nu}|$, $|\phi^A|<|\bar\Phi^A|$ and $|\theta^B|< |{\bar\Theta^B}|$. There are no specific limitations on the rate of change of the perturbations that is we assume a slow-motion approximation and do not assume that the time derivatives are much smaller than spatial partial derivatives. The second partial derivatives of the fields are  comparable (due to the field equations) with the magnitude of the stress-energy tensor, $T^{\m\n}$, of the bare perturbations that is $|{\hatl}^{\mu\nu}{}_{,\m\n}|\sim|\phi^A{}_{,\m\n}|\sim|T^{\m\n}|$. Similar assumptions are used in the method of solution of Einstein's equations in asymptotically-flat space time called the post-Minkowskian approximations \citep{bd1,bd2,damour_1987}. The present paper extends the post-Minkowskian approximations to a more sophisticated realm of curved background manifolds.

We consider the conjugated pairs of perturbations and its first partial derivatives $\left\{{\hatl}^{\mu\nu}, {\hatl}^{\mu\nu}{}_{|\a}\right\}$, $\left\{\phi^A,\phi^A{}_{|\a}\right\}$ along with the bare perturbing field $\left\{\theta^B,\theta^B{}_{|\a}\right\}$, where the vertical bar denotes a covariant derivative on the background manifold, as a set of independent dynamic variables which propagate on the background manifold $\bar\M$ with the metric $\bar g_{\mu\nu}$. In order to derive the differential equations governing the evolution of the perturbations we substitute the field decompositions (\ref{(2.4)})--(\ref{2.5as}) to the Lagrangian $\lag$ defined by equations (\ref{(2.1)}) which yields
\be
\la{(2.9)}
\lag=\lag^{\rm\st G}(\bar{\gag}^{\mu\nu} +
{\hatl}^{\mu\nu}) +  \lag^{\rm\st M} (\bar\Phi^A +
\phi^A,\bar\gag^{\mu\nu} + \hatl^{\mu\nu}) + \lag^{\rm\st P}
(\bar\Theta^B+\theta^B,\bar\gag^{\mu\nu} + \hatl^{\mu\nu})\;.
\ee
Because the perturbations $\hatl^{\mu\nu}$, $\phi^A$, $\theta^B$ are linearly superimposed on the background values of the metric tensor $\bar g_{\mu\nu}$ and fields $\bar\Phi$, $\bar\Theta^B$ respectively, the perturbed (total) Lagrangian (\ref{(2.9)}) admits the following property of the variational derivatives,
\be
\la{qq1}
\frac{\de\lag}{\de\hatl^{\mu\nu}}=\frac{\de\lag}{\de\bar\gag^{\mu\nu}} 
\;,\qquad\qquad
\frac{\de\lag}{\de\phi^A}=\frac{\de\lag}{\de\bar\Phi^A} \;,\qquad\qquad
\frac{\de\lag}{\de\theta^B}=\frac{\de\lag}{\de\bar\Theta^B}
\;.
\ee
These relations allow us to replace the variational derivatives of the total Lagrangian taken with respect to the dynamic perturbation of the field for those taken with respect to the background value of the corresponding field. It turns out to be a very useful device in calculations of the variational derivatives and in building the iterative scheme of solving the Einstein equations by successive approximations.

\subsection{The Lagrangian series decomposition}\la{pm4v7}

The perturbative theory of the dynamic fields on the background spacetime manifold $\bar\M$ is based on the Taylor series decomposition of the total Lagrangian with respect to the field perturbations which magnitude plays the role of a small parameter of the theory. The formal procedure is straightforward and has been described by B. DeWitt \citep{dewitt_book}. More specifically, we take the total Lagrangian (\ref{(2.9)}) and expand it in a Taylor series by making use of the variational derivatives of $\lag$ with respect to the dynamic variables $\hatl^{\mu\nu}$ and $\phi^A$. The Taylor expansion of $\lag$ with respect to $\theta$ can be also performed but we prefer to avoid it because physical measurements yield access to the total value of the bare perturbation $\Theta$. The reader should keep in mind that the expansion of the Lagrangian is performed under the sign of the integral in the action functional (\ref{(2.0)}). Therefore, all terms in this expansion which are reduced to a total divergence can be discarded as they do not contribute to the value of the action integral. 

We assume at the beginning of the calculation that the perturbations and their derivatives are sufficiently small to ensure the convergence of the Taylor expansion of the Lagrangian. For the Lagrangian is a function of several variables, the Taylor series has terms with the mixed derivatives starting from the second order. At the first glance, the presence of the mixed derivatives causes mathematical complication in ordering the higher-order terms. It is remarkable that this problem can be nicely handled after taking into account the following property of the commutator of two variational derivatives \citep{1988IJMPA...3.2651P}
\be
\la{ia7}
\hatl^{\a \beta} \frac{\de}{\de
\bar\gag^{\a \beta}}\lef( \phi^A \frac{\delta\bar \lag}
{\delta\bar\Phi^A}\ri)- \phi^A \frac{\de}{\de
\bar\Phi^A}\lef(\hatl^{\a \beta}  \frac{\delta\bar \lag}
{\delta\bar\gag^{\a \beta}}\ri)=\pd_\a{\mathcal H}^\a\;,
\ee
where ${\mathcal H}^\a$ denotes a vector density of weight $+1$ made of the partial derivatives from the background Lagrangian $\bar\lag$, and the repeated field label $A$ denotes Einstein's summation over all fields $\Phi^A$. This commutation rule is also valid for any two fields from the field multiplets $\Phi^A$, $\Theta^B$, etc. Equation (\ref{ia7}) allows us to change the order of the variational derivatives to reshuffle terms with the mixed derivatives in the Taylor expansion of the perturbed Lagrangian $\lag$. In doing this, all terms representing the total divergence can be omitted from the Taylor expansion since the variational derivative from them vanishes identically, and they do not contribute to the field equations according to (\ref{gd}), (\ref{gde}). Using this procedure we can put all terms with the mixed derivatives in a specific order so that the Taylor expansion of the Lagrangian takes the following elegant form
\be\la{dgk}
\lag=\lag^{\rm\st P}+\sum_{n=0}^{\infty}\lag_n\;.
\ee
Here, $\lag^{\rm\st P}$ is the Lagrangian of the bare perturbation, $\lag_0\equiv\bar\lag$ is the Lagrangian (\ref{y6}) describing dynamic properties of the background manifold, and for any $n\geq 1$,
\be\la{ia0}
\lag_n=\frac1n\lef(\hatl^{\mu\nu}
 \frac{\de\lag_{n-1}} {\de \bar\gag^{\mu\nu}} +
\phi^A \frac{\delta\lag_{n-1}}{\delta \bar\Phi^A}\ri)\;,
\ee
represents a collection of terms of the power $n$ with respect to the perturbations $\hatl^{\m\n}$ and $\phi^A$. In particular, the linear and quadratic terms of the expansion (\ref{dgk}) read
\ba
\la{ia1}
\lag_1&=&\hatl^{\mu\nu}
 \frac{\de \bar\lag} {\de \bar\gag^{\mu\nu}} +
\phi^A \frac{\delta \bar\lag}{\delta \bar\Phi^A}\;,\\
\la{ia2}
\lag_2&=&\frac{1}{2}\lef(\hatl^{\mu\nu}
 \frac{\de\lag_1} {\de \bar\gag^{\mu\nu}} +
\phi^A \frac{\delta\lag_1}{\delta \bar\Phi^A}\ri)\;,
\ea
and so on. We conclude that each subsequent term in the Taylor expansion of the Lagrangian (\ref{dgk}) can be obtained from the previous approximation by taking the variational derivative. The entire analytic procedure is easily computerized.

Equation (\ref{ia0}) can be proved by induction starting from the value of $\lag_1$ in (\ref{ia1}) which is apparently true, and operating with the commutation rule (\ref{ia7}) in higher orders in order to confirm that the result is reduced to the original Taylor series. The commutation property (\ref{ia7}) of the variational derivatives allows us to write down the Taylor expansion (\ref{dgk}) as follows
\be
\la{ia11}
\lag=\exp\lef(\hatl^{\mu\nu}
 \frac{\de} {\de \bar\gag^{\mu\nu}} +
\phi^A \frac{\delta}{\delta \bar\Phi^A}\ri)\bar\lag+\lag^{\rm\st P}\;,
\ee
that establishes a mapping relation between the perturbed, $\lag$, and unperturbed, $\bar\lag$, Lagrangians in the most succinct, exponential form.  

By applying equation (\ref{qq1}) to the Taylor series (\ref{dgk}), and making use of $\d\lag^{\rm\st P}/\d \hatl^{\mu\nu}=\d\lag^{\rm\st P}/\d \bar\gag^{\mu\nu}$, we get an important relation between the variational derivatives of the consecutive terms $\lag_n$ and $\lag_{n-1}$ in the series decomposition of the Lagrangian,
\be
\la{qq1o}
\frac{\de\lag_{n}}{\de\hatl^{\mu\nu}}=\frac{\de\lag_{n-1}}{\de\bar\gag^{\mu\nu}}\;,\qquad
\frac{\de\lag_{n}}{\de\phi^A}=\frac{\de\lag_{n-1}}{\de\bar\Phi^A}\;.
\ee
These relations can be confirmed directly by making use of (\ref{ia0}) that establishes a relation between the adjacent orders of the Lagrangian expansion (\ref{dgk}). In doing so, we have to keep in mind that the total divergences can be always discarded.

If necessary, the Lagrangian of the bare perturbation can be also expanded in the Taylor series with respect to $\hatl^{\m\n}$, 
\ba
\la{ia3a}
\lag^{\rm\st P}&=&\lag^\Theta+\hatl^{\mu\nu} \frac{\delta\lag^\Theta}{\delta
\bar\gag^{\mu\nu}}+\frac{1}{2!}\hatl^{\a \beta} \frac{\de}{\de
\bar\gag^{\a \beta}}\lef( \hatl^{\mu\nu} \frac{\de\lag^\Theta}
{\de \bar\gag^{\mu\nu}}\ri)+\ldots\\\nonumber&=&\exp\lef(\hatl^{\mu\nu}
 \frac{\de} {\de \bar\gag^{\mu\nu}}\ri)\lag^\Theta\;,
\ea
where we have defined $\lag^\Theta\equiv\lag^{\rm\st P}(\Theta^B,\bar\gag^{\mu\nu})$. However, in practical calculations it is more convenient to keep $\lag^{\rm\st P}$ unexpanded, remembering that at each iteration the metric tensor $g_{\m\n}$ and the field $\Theta$ entering $\lag^{\rm\st P}$ are known up to the order of the approximation under consideration.

\subsection{The Dynamic and Effective Lagrangians}
In order to build the field perturbation theory on a curved background manifold $\bar\M$ we have to single out the first order terms which represent the linear differential equations for the dynamic field variables. The entire theory is built under assumption that the background field equations are valid exactly. In other words, the perturbation theory is valid {\it on-shell}.

The principle of the least action tells us that the Lagrangian (\ref{(2.9)}) must be stationary with respect to variations of the metric tensor $g_{\m\n}$ and the field variables $\Phi^A$,
\be\la{nbw5}
\frac{\d\lag}{\d\gag^{\m\n}}=0\qquad,\qquad\frac{\d\lag}{\d\Phi^A}=0\;.
\ee
We also assume that the background Lagrangian (\ref{y6}) is stationary with respect to the variations of the background variables $\bar\gag^{\m\n}$ and $\bar\Phi^A$, and the field equations (\ref{aq3}), (\ref{(2.7)}) are valid. It means that the variational derivatives with respect to $\hatl^{\mu\nu}$ and $\phi^A$ from the background Lagrangian $\lag_0\eq\bar\lag$ vanish identically. Therefore, applying equations (\ref{qq1o}) to the terms of the linear order, $n=1$, yields
\ba
\la{lar4}
\frac{\de\lag_{1}}{\de\hatl^{\mu\nu}}&=&\frac{\de\bar\lag}{\de\bar\gag^{\mu\nu}}=0\;,\\\la{lar5}
\frac{\de\lag_{1}}{\de\phi^A}&=&\frac{\de\bar\lag}{\de\bar\Phi^A}=0\;,
\ea
due to the background field equations (\ref{aq3}), (\ref{(2.7)}). 

Equations (\ref{lar4}), (\ref{lar5}) point out that the dynamics of physical field perturbations is governed solely by the quadratic, cubic and higher-order polynomial terms in the Lagrangian decomposition (\ref{dgk}). We define the {\it dynamic} Lagrangian of the dynamic perturbations as follows \cite{1984CMaPh..94..379G,1988IJMPA...3.2651P}
\be
\la{ia4}
\lag^{\rm dyn}\equiv\lag_2+\lag_3+\ldots\;,
\ee
so that the total Lagrangian (\ref{dgk}) can be written down in the following form
\be\la{lar6}
\lag=\bar\lag+\lag_1+\lag^{\rm dyn}+\lag^{\rm\st P}\;.
\ee

The background Lagrangian, $\bar\lag$, does not depend on the dynamic variables, $\hatl^{\m\n}$, $\phi^A$ and $\theta^B$ which represent the field perturbations. Hence, the variational derivative from $\bar\lag$ taken with respect to any of these variables is identically zero,
\be\la{nw5c1}
\frac{\de\bar\lag}{\de\hatl^{\mu\nu}}\eq 0\;,\qquad\frac{\de\bar\lag}{\de\phi^A}\eq 0\;.
\ee
On the other hand, the variational derivative from $\lag_1$ taken with respect to $\hatl^{\m\n}$ or $\phi^A$ vanishes on-shell due to the background field equations, as evident in (\ref{lar4}), (\ref{lar5}). Hence, the Lagrangian perturbation theory of dynamic fields residing on the background manifold $\bar\M$ can be built on-shell with the help of the {\it effective} Lagrangian
\be\la{efla3}
\lag^{\rm eff}\equiv \lag^{\rm dyn}+\lag^{\rm\st P}\;.
\ee
The effective Lagrangian is convenient for deriving the field equations of the physical perturbations and equations of motion of matter which are discussed in the rest of the present paper.

\subsection{Field equations for gravitational perturbations}\la{v5x8n}

By definition, the dynamic perturbations of gravitational field are the perturbations $\hatl^{\m\n}$ of the contravariant components of the metric tensor density. The field equations for the metric perturbations are obtained after taking the variational derivative from the total Lagrangian $\lag$ with respect to $\hatl^{\mu\nu}$, and equating it to zero. Due to equations (\ref{lar4}), (\ref{lar5}) and (\ref{nw5c1}) this derivative is reduced to that taken from the effective Lagrangian $\lag^{\rm eff}$,
\be\la{tyu7}
\frac{\d\lag^{\rm eff}}{\d\hatl^{\m\n}}=0\;.
\ee
Because of (\ref{lar4}), it is equivalent to equation $\de\lef(\L-\bar\L\ri)/\de\hatl^{\m\n}=0$ or, after applying (\ref{qq1}), to $\de\lef(\L-\bar\L\ri)/\de\bar\gag^{\m\n}=0$. Replacing $\L$ in this equation with expansion (\ref{lar6}) and accounting for the background Einstein equations, $\de\bar\L/\de\bar\gag^{\m\n}=0$, we recast (\ref{tyu7}) into the following form
\be
\la{nm3}
-\frac{\de\lag_1}{\de\bar g^{\mu\nu}}=\frac{\delta\lag^{\rm eff}}{\delta\bar{g}^{\mu\nu}}
\;,
\ee
where we have used (\ref{ia6}) in order to replace the variational derivative with respect to $\bar\gag^{\m\n}$ for that with respect to $\bar g^{\m\n}$. The Euler-Lagrangian equation (\ref{nm3}) is equivalent to (\ref{tyu7}) but more convenient to work with. It is worth emphasizing that is equivalent on-shell to the first variational equation (\ref{nbw5}).

By taking the variational derivatives one can reduce equation (\ref{nm3}) to a more tractable tensor form
\be
\la{(2.17)}
F^{\rm\st G}_{\mu\nu} +F^{\rm\st M}_{\mu\nu} = 8\pi \Lambda_{\m\n}\;,
\ee
where
\be\la{kwn8}
\Lambda_{\m\n}\eq\frac2{\sqrt{-\bar g}}\frac{\delta\lag^{\rm eff}}{\delta\bar{g}^{\mu\nu}}\;,
\ee
is the effective stress-energy tensor and the left side of (\ref{(2.17)}) is a Laplace-Beltrami operator for tensor field $\hatl^{\m\n}$ on the background manifold \citep{kopetr} that consists of two parts \citep{1984CMaPh..94..379G,1988IJMPA...3.2651P}
\ba
\la{(2.18)}
F^{\rm\st G}_{\mu\nu} &\equiv& -\frac{16\pi}{\sqrt{-\bar g}}\frac{\delta}
{\delta\bar{g}^{\mu\nu}} \lef(\hatl^{\rho\s}
\frac{\delta\bar\L^{\rm\st G}}{\delta\bar\gag^{\rho\s}}\ri)\;,\\
\la{(2.19)}
F^{\rm\st M}_{\mu\nu}&\equiv& -\frac{16\pi}{\sqrt{-\bar g}} \frac{\delta}{\delta\bar{g}^{\mu\nu}}
\lef(\hatl^{\rho\s}  \frac{\delta\bar\lag^{\rm\st M}}{\delta\bar
\gag^{\rho\s}} + \phi^A \frac{\delta\bar\lag^{\rm\st M}}{\delta\bar\Phi^A}\ri)\;.
\ea

Operator $F^{\rm\st G}_{\mu\nu}$ describes the linearized perturbation of the Ricci tensor and can be easily calculated on any background manifold. Indeed, taking into account (\ref{aq1}), we immediately get
\be\la{azw1}
F^{\rm\st G}_{\mu\nu}=\frac{1}{\sqrt{-\bar g}}\frac{\delta}
{\delta\bar{g}^{\mu\nu}} \lef(\hatl^{\rho\s}\bar R_{\r\s}\ri)\;.
\ee
Now, according to the rule of rising and lowering indices of the variational derivatives, we can recast (\ref{azw1}) to
\be\la{ss3}
F^{\rm\st G}_{\m\n}=-\frac{1}{\sqrt{-\bar g}}\bar g_{\m\chi}g_{\n\e}\frac{\delta}
{\delta\bar{g}_{\chi\e}}\lef(\hatl^{\rho\g}\d^\k_\l\bar R^\l{}_{\r\k\g}\ri)\;.
\ee
Variational derivative in (\ref{ss3}) is calculated with the help of the covariant definition (\ref{er8}) where the covariant derivatives are taken on the background manifold $\bar\M$ and are denoted with a vertical bar. We recall that $\hatl^{\r\g}$ is an independent dynamic variable while the term under the sign of the variational derivative in (\ref{ss3}) depends merely on the background Riemann tensor without explicit appearance of the Christoffel symbols. Therefore, the variational derivative in (\ref{ss3}) is taken only with respect to the Riemann tensor in accordance with (\ref{er8}). It yields
\ba
\la{ss2}
\frac{\delta}
{\delta\bar{g}_{\chi\e}}\lef(\hatl^{\rho\g}\d^\k_\l\bar R^\l{}_{\r\k\g}\ri)
&=&\lef[\hatl^{\rho\g}\d^\k_\l\lef(\bar g^{\s\chi}\d^\l_\s\d^\a_\r\d^{[\b}_\k\d^{\e]}_\g+\bar g^{\s\e}\d^\l_\s\d^\chi_\r\d^{[\b}_\k\d^{\a]}_\g-\bar g^{\s\a}\d^\l_\s\d^\chi_\r\d^{[\b}_\k\d^{\e]}_\g\ri)\ri]_{|\b\a}\nonumber\\\nonumber
&=&\lef(\hatl^{\a[\e}\bar g^{\b]\chi}+\hatl^{\chi[\a}\bar g^{\b]\e}-\hatl^{\chi[\e}\bar g^{\b]\a}\ri)_{|\b\a}\\
&=&\frac12\lef(\hatl^{\a\chi}\bar g^{\b\e}+\hatl^{\a\e}\bar g^{\b\chi}-\hatl^{\chi\e}\bar g^{\a\b}-\hatl^{\a\b}\bar g^{\chi\e}\ri)_{|\b\a}\;,
\ea
where we have taken into account that the expression enclosed in the brackets, is symmetric with respect to indices $\a$ and $\b$. 
We substitute (\ref{ss2}) to (\ref{ss3}) and recollect definition of $\hatl^{\m\n}=\sqrt{-\bar g}l^{\m\n}$ along with the constancy of the background metric tensor $\bar g_{\m\n}$ with respect to the covariant derivative. It results in the differential operator
\be\la{xpq1}
F^{\rm\st G}_{\mu\nu}=\frac{1}{2}\lef(l_{\mu\nu}{}^{|\a}{}_{|\a} +\bar{g}_{\mu\nu}l^{\a\b}{}_{|\a\b}-l^\a{}_{\m|\n\a}
-l^\a{}_{\n|\m\a}\ri)\;,
\ee
where each vertical bar denotes a covariant derivative with respect to the background metric $\bar g_{\mu\nu}$, and $l_{\a\b}\eq\hatl_{\a\b}/\sqrt{-\bar g}$ (the indices are raised and lowered with the background metric $\bar g_{\a\b}$). We emphasize that expression (\ref{xpq1}) is {\it exact}.

Operator $F^{\rm\st M}_{\mu\nu}$ describes perturbation of the stress-energy tensor $\bar T_{\m\n}$ of the background matter governing the on-shell evolution of the background manifold $\bar\M$. Hence, it vanishes on any Ricci-flat spacetime manifold ($\bar R_{\m\n}=0$) in general relativity as a consequence of the background Einstein's equations (\ref{p9m5}). Cosmological FLRW spacetime is not Ricci flat. Therefore, $F^{\rm\st M}_{\mu\nu}$ makes a non-trivial contribution to the field equations (\ref{(2.17)}) for gravitational perturbations. Variational derivative in definition (\ref{(2.19)}) of $F^{\rm\st M}_{\mu\nu}$ is taken from the Lagrangian, $\lag^{\rm\st M}$, characterizing the background matter fields $\bar\Phi^A$, and depends crucially on its particular form which must be specified in each individual case of physical fields under consideration. We can bring (\ref{(2.19)}) to a more explicit form by accounting for the definition of the metrical stress-energy tensor of the background matter \citep{kopeikin_2011book}
\be
\la{k9w}
\bar T^{\rm\st M}_{\mu\nu}\eq\frac{2}{\sqrt{-\bar g}}\frac{\de\bar \lag^{\rm\st M}}{\de\bar g^{\mu\nu}}\;,
\ee 
and introducing a new function
\be\la{zom1}
\bar I^{\rm\st M}_A\eq\frac{2}{\sqrt{-\bar g}}\frac{\delta\bar\lag^{\rm\st M}}{\delta\bar\Phi^A}\;.
\ee
We notice that $\bar I^{\rm\st M}_A$ vanishes on-shell because of the field equation (\ref{aq3}). However, this equation should not be applied immediately in the definition (\ref{(2.19)}) of $F^{\rm\st M}_{\mu\nu}$ as we, first, have to take the variational derivative with respect to the metric tensor which is off-shell operation. With these remarks equation (\ref{(2.19)}) takes on the following form
\be
\la{zom2}
F^{\rm\st M}_{\mu\nu}= -\frac{8\pi}{\sqrt{-\bar g}} \frac{\delta}{\delta\bar{g}^{\mu\nu}}
\lef(\hatl^{\r\s}\bar T^{\rm\st M}_{\r\s}-\frac12\hatl\bar T^{\rm\st M} + \sqrt{-\bar g}\phi^A \bar I^{\rm\st M}_A\ri)\;.
\ee
We shall calculate (\ref{zom2}) later on for an ideal fluid (dark matter) and a scalar field (dark energy) in the case of the FLRW universe governed by dark matter and dark energy.

The right side of equation (\ref{(2.17)}) contains the effective stress-energy tensor consisting of two contributions
\be\la{pmy6}
\Lambda_{\m\n}=\T_{\mu\nu}+{\cal T}_{\mu\nu}\;,
\ee
where 
\be
\la{ia8}
\T_{\mu\nu}\equiv\frac{2}{\sqrt{-\bar g}}\frac{\delta\lag^{\rm\st P}}{\delta\bar{g}^{\mu\nu}}\;,
\ee
is the stress-energy tensor of the bare perturbation,
and 
\be
\la{(2.20)}
{\cal T}_{\mu\nu}\equiv\frac{2}{\sqrt{-\bar g}}\frac{\delta\lag^{\rm dyn}}{\delta\bar{g}^{\mu\nu}}\;,
\ee
is the stress-energy tensor associated with the dynamic field perturbations $\hatl^{\m\n}$ and $\phi^A$.

It is important to emphasize that stress-energy tensor $\T_{\mu\nu}$ of the bare perturbations is defined as a variational derivative with respect to the background metric, $\bar g_{\m\n}$. Hence, it differs from the similar tensor 
\be\la{q19z}
T_{\a\b}=\frac{2}{\sqrt{-g}}\frac{\d\lag^{\rm\st P}}{\d g^{\a\b}}\;,
\ee
which was introduced earlier (see (\ref{n7x4})) and is defined in terms of the variational derivative with respect to the full metric $g_{\m\n}=\bar g_{\m\n}+\varkappa_{\m\n}+...$. These two tensors, $\T_{\m\n}$ and $T_{\m\n}$, are closely related. The relation between them can be found by making use of equations
\ba\la{gh61}
\frac{\pd\gag^{\m\n}}{\pd\bar g^{\a\b}}&=&\sqrt{-\bar g}\lef[\d^{(\m}_\a\d^{\n)}_\b-\frac12\bar g^{\m\n}\bar g_{\a\b}\ri]\;,\\\la{hy6s3}
\frac{\pd g^{\r\s}}{\pd\gag^{\m\n}}&=&\frac1{\sqrt{-g}}\lef[\d^{(\r}_\m\d^{\s)}_\n-\frac12 g^{\r\s} g_{\m\n}\ri]\;,
\ea
and
\be\la{gh62}
\frac{\d}{\d\bar g^{\a\b}}=\frac{\pd\gag^{\m\n}}{\pd\bar g^{\a\b}}
\frac{\pd g^{\r\s}}{\pd\gag^{\m\n}}\frac{\d}{\d g^{\r\s}}\;.
\ee
It yields an exact relation  
\be\la{rtei1}
\T_{\m\n}=T_{\m\n}-\frac12 g_{\m\n}T-\frac12\bar g_{\m\n}\bar g^{\a\b}\lef(T_{\a\b}-\frac12 g_{\a\b}T\ri)\;,
\ee
where the trace of the stress energy tensor is defined as $T\eq g^{\a\b}T_{\a\b}$. Relation (\ref{rtei1}) can be inverted leading to another exact formula 
\be\la{rtei2}
T_{\m\n}=\T_{\m\n}-\frac12 \bar g_{\m\n}\T-\frac12 g_{\m\n} g^{\a\b}\lef(\T_{\a\b}-\frac12 \bar g_{\a\b}\T\ri)\;,
\ee
where $\T=\bar g^{\a\b}\T_{\a\b}$.

Tensor ${\cal T}_{\mu\nu}$ can be split in two algebraically-independent parts
\be
\la{qq5a}
{\cal T}_{\mu\nu}=\mt_{\mu\nu}+\tau_{\mu\nu}\;,
\ee
where $\mt_{\mu\nu}$ is the stress-energy tensor of pure gravitational perturbations $\hatl^{\mu\nu}$ while  $\t_{\mu\nu}$ is the stress-energy tensor characterizing gravitational coupling of the matter field $\phi^A$ with the gravitational perturbations $\hatl^{\mu\nu}$.
For example, in the second-order approximation, when $\lag^{\rm dyn}=\lag_2$, the corresponding stress-energy tensors are given by equations
\ba\la{qq5z}
\mt_{\mu\nu}&=&-\frac1{16\pi\sqrt{-\bar g}}\frac{\d}{\d\bar g^{\m\n}}\lef(\hatl^{\r\s}F^{\rm\st G}_{\r\s}-\frac12\hatl F^{\rm\st G}\ri)\;,\\
\la{qq5x}
\t_{\mu\nu}&=&-\frac1{16\pi\sqrt{-\bar g}}\frac{\d}{\d\bar g^{\m\n}}\lef(\hatl^{\r\s}F^{\rm\st M}_{\r\s}-\frac12\hatl F^{\rm\st M}+\sqrt{-\bar g}\phi^A F^{\rm\st M}_A\ri)\;,
\ea
where $F^{\rm\st M}_A$ is defined below in (\ref{yu2}).

As soon as the differential operators and the source terms in the field equation (\ref{(2.17)}) are specified, it can be solved by successive iterations. It requires decomposition of the perturbations $\hatl^{\m\n}$ and $\phi^a$ in the post-Friedmanian series
\ba\la{pfser1}
\hatl^{\m\n}&=&G\hatl^{\m\n}_1+G^2\hatl^{\m\n}_2+G^3\hatl^{\m\n}_3+\ldots\;,\\
\la{pfser2}
\phi^A&=&G\phi^A_1+G^2\phi^A_2+G^3\phi^A_3+\ldots\;,
\ea
where the terms with indices $n=1,2,3,\ldots$ represent the successive approximations of the corresponding order of magnitude with respect to the universal gravitational constant $G$ (which we showed in these equations explicitly). These series generalize analogous series in the post-Minkowskian approximation scheme applied to solve Einstein's equations in asymptotically-flat spacetime \citep{damour_1987,bd1,bd2,beld}. We conjecture that the series (\ref{pfser1}), (\ref{pfser2}) are analytic and convergent for a sufficiently small magnitude of the perturbations. However, the proof of this conjecture requires dedicated mathematical efforts and a special study which we do not pursue in the present paper because of its enormous mathematical difficulty.

The post-Friedmannian iteration procedure starts off the substitution of the unperturbed values of $\hatl^{\m\n}=\phi^a=0$ to the right side of (\ref{(2.17)}) and finding the linear perturbation $\hatl^{\m\n}_1$. The solution is substituted back to the right side of (\ref{(2.17)}), which is solved again to find $\hatl^{\m\n}_2$, and so on. In addition to the field equations for the metric tensor perturbations, we need additional set of differential equations to find out the perturbations of the matter fields $\phi^A$.

\subsection{Field equations for matter perturbations}\label{zzzz8}

Equations for the background matter field perturbation $\phi^A$ are derived from the effective Lagrangian (\ref{efla3}) by taking the variational derivative with respect to the dynamic variable $\phi^A$. The Lagrangian $\lag^{\rm\st P}$ does not depend on $\phi^A$ and drops out from further calculations. Moreover, we assume that the background field equations (\ref{aq3}) and the stationary conditions (\ref{lar5}) are satisfied. Thus, the stationarity of the Lagrangian (\ref{lar6}) with respect to the perturbations $\phi^A$ yields
\be
\la{xsk}
\frac{\de\lag^{\rm dyn}}{\de\phi^A}=0\;,
\ee
which is equivalent to
\be
\la{xk3a}
-\frac{\de\lag_1}{\de\bar\Phi^A}=\frac{\de\lag^{\rm dyn}}{\de\bar\Phi^A}\;.
\ee

After taking the variational derivatives, equation (\ref{xk3a}) assumes the following form
\be
\la{bbv2}
F^{\rm\st M}_A = 8\pi\Sigma^{\rm\st M}_A\;,
\ee
where the linear (Laplace-Beltrami) differential operator
\ba
\la{yu2}
F^{\rm\st M}_A&\equiv& -\frac{16\pi}{\sqrt{-\bar g}}\frac{\delta}{\delta\bar\Phi^A}\lef(
\hatl^{\mu\nu}  \frac{\delta\bar\lag^{\rm\st M}}{\delta\bar\gag^{\mu\nu}} +
\phi^A \frac{\delta\bar\lag^{\rm\st M}}{\delta\bar\Phi^A}\ri)\;,\ea
and the source density
\ba
\la{yu3}
\Sigma^{\rm\st M}_A&\equiv&\frac2{\sqrt{-\bar g}}\frac{\delta\lag^{\rm dyn}}{\delta\bar\Phi^A}\;.
\ea
All linear with respect to $\hatl^{\a \beta}$ and $\phi^A $ terms are included in the left side of equation (\ref{bbv2}) while the non-linear terms have been put in $\Sigma^{\rm\st M}_A$. More explicit form of the operator $F^{\rm\st M}_A$ can be obtained with the help of (\ref{k9w}), (\ref{zom1}) that results in
\be
\la{koj2}
F^{\rm\st M}_A= -\frac{8\pi}{\sqrt{-\bar g}} \frac{\delta}{\delta\bar\Phi^A}
\lef(\hatl^{\r\s}\bar T^{\rm\st M}_{\r\s}-\frac12\hatl\bar T^{\rm\st M} + \sqrt{-\bar g}\phi^A \bar I^{\rm\st M}_A\ri)\;.
\ee
Further specification of the operator $F^{\rm\st M}_A$ requires a particular model of the background matter Lagrangian $\bar\lag^{\rm\st M}$ which will be discussed in section \ref{sec4}.

Field equations for bare perturbations, $\Theta^B$, are obtained after taking the variational derivative from the Lagrangian (\ref{efla3}) with respect to the variable $\Theta^B$. Because the only part of the Lagrangian which depends on this field, is $\lag^{\rm\st P}$, the equations are reduced to
\be
\la{ko9a}
\frac{\delta\lag^{\rm\st P}}{\delta\Theta^B}=0\;.
\ee
Particular form of this equation depends on a specific choice of the Lagrangian $\lag^{\rm\st P}$ of the bare perturbation.
In the lowest order of approximation the field equations (\ref{bbv2}) and (\ref{ko9a}) describe evolution of the dynamic perturbations $\phi^A$ and $\Theta^B$ on the unperturbed cosmological background. The next-order approximations take into account the back reaction of the background perturbations on these fields. 

\subsection{Gauge invariance of the field equations}\la{gife}

Gauge invariance of the dynamic perturbations is an important geometric property that allows us to distinguish physical degrees of freedom of gravitational and matter fields from the spurious modes generated by transformations of the local coordinates on manifold. Any self-consistent perturbation theory must clearly separate the coordinate-dependent effects from physical perturbations which do not depend on the choice of coordinates. The gauge transformation is generated by the exponential mapping of spacetime manifold to itself, $\M\rightarrow\M$, that is induced by a non-singular vector flow having a tangent vector $\xi^\a\equiv\xi^\beta(x^\a )$ associated with a finite translation of each point of the manifold
\be
\la{qq2a}
x'^\alpha=\exp\lef(\xi^\beta\pd_\beta\ri)x^\a =x^\a +\xi^\a +\frac{1}{2!}\xi^\beta\pd_\b\xi^\a+\ldots\;.
\ee 
It drags the coordinate grid on manifold along the vector field $\xi^\a$, and makes a point-wise change of any geometric object $\F(x^\alpha)$ to $\F'(x'^\alpha)$. The transformed object $F'(x'^\alpha)$ is, then, pulled back to its value $F'(x^\alpha)$ taken at the point on the manifold having the same coordinates.  It defines the gauge transformation of $\F$ that is found to be an exponential Lie transform \citep{1984CMaPh..94..379G,1988IJMPA...3.2651P},
\be
\la{qq3w}
\F'(x^\alpha)=\lef(\exp\pounds_{\bm\xi}\ri)\F(x^\alpha)=\F(x^\alpha)+\pounds_{\bm\xi}\F(x^\alpha)+\frac{1}{2!}\pounds^2_{\bm\xi}\F(x^\alpha)+\ldots\;,
\ee
where the Lie derivative $\pounds_{\bm\xi}\F$ has been defined in equation (\ref{gde1}), $\pounds^2_{\bm\xi}\equiv\pounds_{\bm\xi}\pounds_{\bm\xi}$, $\pounds^3_{\bm\xi}\equiv\pounds_{\bm\xi}\pounds_{\bm\xi}\pounds_{\bm\xi}$, and so on.

For the gauge transformation of a geometric object is generated by the change of coordinates it has no real physical meaning and should be considered as spurious. The gauge freedom should be carefully studied in order to eliminate the non-physical degrees of freedom.  
The gauge transformation of the metric tensor, $g_{\m\n}$, and the matter fields, $\Phi^A$, $\Theta^B$, leads to appearance of the gauge-dependent perturbations which imply that the background values of $g_{\m\n}$, $\Phi^A$, $\Theta^B$ do not change under the gauge transformation -- only the dynamic perturbations, $\hatl^{\mu\nu}$, $\phi^A$, $\theta^B$ change. Hence, the gauge transformation (\ref{qq3w}) applied to these fields induces the following gauge transformations of the perturbations of these fields,
\ba
\la{qq5ab}
\hatl'^{\mu\nu}&=&\hatl^{\mu\nu}+\lef(\exp\pounds_{\bm\xi}-1\ri)\lef(\hatg^{\mu\nu}+\hatl^{\mu\nu}\ri)\;,\\
\la{qq6x}
\phi'^A&=&\phi^A+\lef(\exp\pounds_{\bm\xi}-1\ri)\lef(\bar\Phi^A+\phi^A\ri)\;,\\
\la{qq6a}
\theta'^B&=&\theta^B+\lef(\exp\pounds_{\bm\xi}-1\ri)\lef(\bar\Theta^B+\theta^B\ri)\;,
\ea
that depend on the gauge vector field $\xi^\a$.

Let us consider the gauge transformation of the total Lagrangian (\ref{(2.1)}) induced by the gauge transformations of its arguments. The transformed  Lagrangian $\lag'$ has the same functional form as $\lag$ but depends now on the transformed (denoted with a prime) values of the dynamic variables, $\lag'\equiv\lag\lef(\hatg^{\mu\nu} +
{\hatl}'^{\mu\nu},\,\bar\Phi^A +\phi'^A, \bar\Theta^B+\theta'^B\ri)$. We replace the transformed variables with their original values by making use of equations (\ref{qq5ab})--(\ref{qq6a}). It yields
\be
\la{sd5a}
\lag'=\lag\lef[\exp\pounds_{\bm\xi}\Bigl(\hatg^{\mu\nu} +{\hatl}^{\mu\nu}\Bigr)\;,\exp\pounds_{\bm\xi}\lef(\bar\Phi^A+\phi^A\ri)\;,\exp\pounds_{\bm\xi}\lef(\bar\Theta^B+\theta^B\ri)\ri]\;.
\ee
This equation can be further transformed by making use of the following relation \citep{1988IJMPA...3.2651P}
\ba
\la{qq8w}
\lag\lef[\exp\pounds_{\bm\xi}\Bigl(\hatg^{\mu\nu} +{\hatl}^{\mu\nu}\Bigr)\;,\exp\pounds_{\bm\xi}\Bigl(\bar\Phi^A+\phi^A\Bigr)\;,\exp\pounds_{\bm\xi}\lef(\bar\Theta^B+\theta^B\ri)\ri]
=\;\exp\pounds_{\bm\xi}\left[\lag\lef(\hatg^{\mu\nu} +{\hatl}^{\mu\nu}\;,\bar\Phi^A+\phi^A\;,\bar\Theta^B+\theta^B\ri)\ri]\;,
\ea
that is valid modulo total divergence which is inessential in the variational calculus. We expand the right side of (\ref{qq8w}) in a Taylor series, like in (\ref{qq3w}), and take into account that the Lagrangian is a scalar density of weight $+1$ for which the Lie derivative, $\pounds_{\bm\xi}\lag=\pd_\a\lef(\xi^\a\lag\ri)$. It eventually yields the gauge transformation of the Lagrangian in the following form
\ba
\la{qq4a}
\lag'=\lag+\pd_\a \Bigl(\xi^\a \lag\Bigr)
+\frac{1}{2!}\pd_\a\lef(\xi^\a\pd_\b \lef(\xi^\b \lag\ri)\ri)+\frac{1}{3!}\pd_\a\lef(\xi^\a\pd_\beta\lef(\xi^\beta\pd_\g \Bigl(\xi^\g \lag\Bigr)\ri)\ri)+\ldots\;,
\ea
where the second, third, and all other terms in the right side of this infinite series represent a divergence. The divergence vanishes when one takes the variational derivative from it and, hence, it can be omitted from the action functional $S$ given in (\ref{(2.0)}). The conclusion is that the action $S$ and the Lagrangian (\ref{(2.9)}) are gauge-invariant with respect to the gauge transformation of their arguments. This assertion does not involve any background equations of motion and/or field equations and, thus, is valid both on-shell and off-shell.

On the other hand, the effective Lagrangian (\ref{efla3}) is gauge-invariant only on-shell that is only when the background field equations (\ref{aq3}), (\ref{(2.7)}) are satisfied. Indeed, the effective Lagrangian can be represented as a difference $\lag^{\rm eff}=\lag-\lag_1-\bar\lag$. After making the gauge transformations (\ref{qq5ab})--(\ref{qq6a}) of the dynamic variables, we get a new effective Lagrangian $\lag'^{\rm eff}=\lag'-\lag'_1-\bar\lag$ where the background value of the Lagrangian stays the same. The difference $\d\lag^{\rm eff}=\lag'^{\rm eff}-\lag^{\rm eff}$ is
\be\la{di4v}
\d\lag^{\rm eff}=\d\lag+\lef(\exp\pounds_{\bm\xi}-1\ri)\lef[\lef(\hatg^{\mu\nu}+\hatl^{\mu\nu}\ri)
\frac{\d\bar\lag}{\d\hatg^{\mu\nu}}+\lef(\bar\Phi^A+\phi^A\ri)\frac{\d\bar\lag}{\d\bar\Phi^A}\ri]\;,
\ee
where the terms being enclosed in the square brackets, vanish on-shell due to the background field equations (\ref{aq3}), (\ref{(2.7)}). Therefore, $\d\lag^{\rm eff}=\d\lag$ which is a total divergence as follows from (\ref{qq4a}). Hence, $\lag^{\rm eff}$ it is gauge-invariant on-shell. 

The gauge invariance of the Lagrangian suggests that the Einstein equations
(\ref{(2.17)}) for metric perturbations are gauge invariant as well. It is straightforward to prove it by direct but otherwise tedious calculation which technical details are given in \cite{1988IJMPA...3.2651P}. Gauge transformations (\ref{qq5ab})--(\ref{qq6a}) applied to the Einstein equations (\ref{(2.17)}) transform them as follows
\ba
\la{azw5}
F'^{\rm\st G}_{\mu\nu} +F'^{\rm\st M}_{\mu\nu} -8\pi \Lambda'_{\mu\nu}&=&F^{\rm\st G}_{\mu\nu} +F^{\rm\st M}_{\mu\nu} -8\pi \Lambda_{\m\n}+\exp\pounds_{\bm\xi}{\cal F}\;,
 \ea
where function ${\cal F}$ vanishes on-shell due to the background field equations (\ref{aq3}), (\ref{(2.7)}). Therefore, if the field equations for gravitational perturbations are valid at least in one gauge, they are valid in any other gauge as well. We have checked that the field equations (\ref{bbv2}) for the matter perturbations are also gauge-invariant.

\section{The Dynamic Field Theory in Cosmology}\la{sec4}
We shall implement the formalism of the dynamic field theory in cosmology to derive the field equations for cosmological perturbations of gravitational field and matter. We shall rely in our analysis upon the cosmological model that is in the most close agreement with modern observational data. In this model the background manifold represents the spatially homogeneous and isotropic FLRW universe which temporal evolution is governed by an ideal fluid  with an arbitrary equations of state and a scalar field with an arbitrary potential. The ideal fluid models a self-interacting dark matter \citep{2013pdmg.conf10001B} while the scalar field describes dark energy in the form of quintessence \citep{Steinhardt_1999PhRvL}. The dark matter without self-interaction is included in our theoretical scheme as a pressureless ideal fluid. The dynamic field variables of the dark matter and dark energy are two scalar fields, $\Phi^1$ and $\Phi^2$ which form a doublet, $\Phi^A=\{\Phi^1,\Phi^2\}$. We identify the scalar field $\Phi^1$ with the (scalar) Clebsch potential $\Phi$ of the ideal fluid, and $\Phi^2$ with a scalar field $\Psi$ having the potential $W=W(\Psi)$ depending only on the scalar field $\Psi$. The third matter component in our model is the baryonic matter making stars, galaxies, etc. as well as neutrino. The baryonic matter and neutrino make up a small fraction ($\simeq 4$\%) of the overall mass of the observed universe. Thus, it can be associated with the bare field perturbation. We limit ourselves with a single scalar field $\Theta^B\equiv\Theta$ describing this perturbation. This model may be still too simple to describe the real universe but it nicely demonstrates the richness and flexibility of the formalism of the dynamic field theory in doing cosmological applications without involving too many secondary details.  

The overall Lagrangian of the cosmological model under consideration is given by equation (\ref{(2.1)}) with the Lagrangian of the background matter consisting of two non-directly interacting pieces
\be\la{xrx1}
\lag^{\rm\st M}= \lag^{\rm m}+\lag^{\rm q}\;,
\ee
where $\lag^{\rm m}$ is the Lagrangian of dark matter, and $\lag^{\rm q}$ is the Lagrangian of dark energy. The Lagrangian of the baryonic matter perturbation is $\lag^{\rm\st P}$. We describe the specific structure of the Lagrangians in next sections.

\subsection{The Lagrangian of dark matter}\la{ldmt}

Dark matter is modelled as an ideal fluid that is characterized by four thermodynamic variables \citep{mitowh}: the rest-mass density ${\r_{\rm m}}$, the specific internal energy per unit mass $\Pi_{\rm m}$, pressure $p_{\rm m}$, and entropy per unit mass $s_{\rm m}$, where the sub-index 'm' stands for the dark matter. We shall assume that the entropy of the ideal fluid remains constant, $s_{\rm m}=0$, and dissipative processes are neglected (isentropic motion). This assumption can be relaxed by adding some other thermodynamic variables \citep{1970PhRvD...2.2762S} but we do not discuss this extension in the present paper.

The total energy density of the ideal fluid is
\be\la{gug1}
\e_{\rm m}={\r_{\rm m}}(1+\Pi_{\rm m})\;.
\ee
A physically meaningful thermodynamic variable is formed from the energy density, pressure and the rest-mass density. It is called the specific enthalpy of fluid, $\mu_{\rm m}$, and defined as \citep{mitowh}
\be\la{pf2}
\mu_{\rm m}\eq\frac{\e_{\rm m}+p_{\rm m}}{{\r_{\rm m}}}=1+\Pi_{\rm m}+\frac{p_{\rm m}}{{\r_{\rm m}}}\;.
\ee

We shall consider a barotropic fluid which thermodynamic equation of state is given by equation $p_{\rm m}=p_{\rm m}({\r_{\rm m}},\Pi_{\rm m})$, where the specific internal energy $\Pi_{\rm m}$ is related to pressure and the rest-mass density by the first law of thermodynamics
\be\la{pf1}
d\Pi_{\rm m}+p_{\rm m}d\lef(\frac{1}{{\r_{\rm m}}}\ri)=0\;.
\ee
This equation along with the definition of the specific enthalpy and the energy density given above, allow us to derive the following differential relations
\ba\la{pf3} dp_{\rm m}&=&{\r_{\rm m}} d\m_{\rm m}\;,\\\la{pf3a} d\e_{\rm m}&=&\mu_{\rm m} d{\r_{\rm m}}\;.
\ea
which immediately tells us that the partial derivatives 
\begin{subequations}\la{jk3e}
\ba\la{jk3ea}
\frac{\pd p_{\rm m}}{\pd\m_{\rm m}}&=&\r_{\rm m}\;,\\ \la{jk3eb}
\frac{\pd \e_{\rm m}}{\pd\r_{\rm m}}&=&\m_{\rm m}\;.
\ea
\end{subequations}
Equations (\ref{jk3e}) elucidate that all thermodynamic quantities are functions of only one thermodynamic variable. For the reasons which are explained below, we accept that this variable is the specific enthalpy $\m_{\rm m}$. The equation of state, relating pressure and the energy density, becomes $p_{\rm m}=p_{\rm m}(\e_{\rm m})$, and it is also an implicit, single-valued function of the thermodynamic variable $\m_{\rm m}$ because $\e_{\rm m}=\e_{\rm m}(\m_{\rm m})$.

Partial derivatives of the thermodynamic quantities with respect to $\m_{\rm m}$ can be calculated by making use of (\ref{pf3}), (\ref{pf3a}), the equation of state $p_{\rm m}=p_{\rm m}(\e_{\rm m})$, and definition of the (adiabatic) speed of sound $c_{\rm s}$ propagating in the fluid
\be\la{pf4a}
\frac{\pd p_{\rm m}}{\pd\e_{\rm m}}=\frac{c^2_{\rm s}}{c^2}\;,
\ee
where the partial derivative is taken under a condition that the entropy, $s_{\rm m}$, does not change. Notice that the speed of sound in dark matter is {\it not} constant in the most general case of a non-linear equation of state. In this case, the speed of sound depends on the thermodynamic potential $\m_{\rm m}$ through the equation of state, that is $c_{\rm s}=c_{\rm s}(\m_{\rm s})$. It is also worth emphasizing that the speed-of-sound-defining equation (\ref{pf4a}) is valid for any wavelength of sound waves in the ideal fluid, not only for short wavelengths. In cosmology, the equation of state of dark matter is postulated as having the following form 
\be\la{i9n4}
p_{\rm m}=w_{\rm m}\e_{\rm m}\;,
\ee
where $w_{\rm m}$ is an implicit function of the specific enthalpy, $w_{\rm m}=w_{\rm m}(\mu_{\rm m})$. Taking a partial derivative from both sides of equation (\ref{i9n4}) with respect to $\mu_{\rm m}$ and making use of (\ref{pf4a}) yield 
\be\la{io3b6}
\frac{\pd w_{\rm m}}{\pd\mu_{\rm m}}=-\frac{c^2}{c_{\rm s}^2}\frac{w_{\rm m}-c_{\rm s}^2/c^2}{1+\Pi_{\rm m}}\;,
\ee
which is naturally reduced to $w_{\rm m}=c_{\rm s}^2/c^2$ in case of the constant parameter $w_{\rm m}$ of the cosmological equation of state. However, in more general cosmological studies $w_{\rm m}$ is not constant and changes as the universe evolves.

Other partial derivatives of the thermodynamic quantities can be calculated with the help of the equation of state and (\ref{jk3e}), (\ref{pf4a}) which can be inverted, if necessary, because all thermodynamic relations in the ideal fluid are single-valued. We have, for example,
\be\la{pf5}
\frac{\pd\e_{\rm m}}{\pd\m_{\rm m}}=\frac{c^2}{c^2_{\rm s}}{\r_{\rm m}}\;,\qquad \frac{\pd{\r_{\rm m}}}{\pd\m_{\rm m}}=\frac{c^2}{c^2_{\rm s}}\frac{{\r_{\rm m}}}{\m_{\rm m}}\;,
\ee
where all partial derivatives are performed under the same condition of the constant entropy. 

Theoretical description of the ideal fluid as a dynamic field system evolving on space-time manifold is given the most conveniently in terms of the Clebsch scalar potential \citep{LL_1959flme,Carter_1988JFM}, $\Phi$ which is also known either as a velocity potential or the Taub potential \citep{1970PhRvD...2.2762S}. The Clebsch potential is a scalar function on spacetime manifold which can be taken as an independent dynamic variable characterizing dynamic evolution of the ideal fluid. This description is complimentary (dual) to the Lagrangian formulation of the ideal fluid based on the coordinates and four-velocity of the fluid particles \citep{fock_book,comer_lrr2007,rezzolla_2013} which is more familiar in the field of cosmology. Nonetheless, the description of the ideal fluid (dark matter) in terms of the Clebsch vector field $\Phi$ and its derivative $\Phi_{|\a}$ considered as independent dynamic variables, makes it very similar to the description of dark energy also given in terms of (another) scalar field. It allows us to consider physical effects of dark matter and dark energy on the same fundamental level of the Lagrangian formalism. It seems the first researcher who realized the advantages of using the Clebsch potential description of the ideal fluid in cosmology, was V.~N. Lukash \citep{1980JETPL..31..596L}. 

In the case of a single-component fluid the Clebsch potential $\Phi$ is introduced by the following relationship
\be\la{pf6}
\mu_{\rm m} w_\a=-\Phi_\a\;,
\ee
where $w^\a=dx^\a/d\t$ is the four-velocity of the fluid, $w_\a=g_{\a\b}w^\b$, $\t$ is the proper time of the fluid element taken along its world line, and we denote $\Phi_\a\eq\Phi_{,\a}=\Phi_{|\a}$ from now on. This type of representation of fluid's velocity has been introduced by A. Clebsch \citep{clebsch_1859}. 
Equation (\ref{pf6}) solves the relativistic Euler equation of motion of the ideal fluid which justifies the connection between the specific enthalpy, four-velocity and the Clebsch potential $\Phi_\a$ \citep{LL_1959flme,rezzolla_2013}.
The four-velocity is normalized, $g_{\a\b}w^\a w^\b=-1$, so that the specific enthalpy can be expressed in terms of the the metric tensor and the derivative from the Clebsch potential,
\be 
\la{pf7}
\mu_{\rm m}=\sqrt{-g^{\a\b}\Phi_\a\Phi_\b}\;.
\ee
One may also notice that the normalization condition for the four-velocity allows us to re-write (\ref{pf6}) in the following form,
\be\la{pf7a}
\m_{\rm m}=w^\a\Phi_\a\;.
\ee
The Clebsch potential $\Phi$ has no direct physical meaning as it can be changed to another value: $\Phi\rightarrow\Phi'=\Phi+\tilde\Phi$ such that the gauge function, $\tilde\Phi$, is constant along the worldlines of the fluid in the sense that $w^\a\tilde\Phi_\a=0$. This gauge transformation of the Clebsch potential does not change the value of the specific enthalpy $\mu_{\rm m}$.

The Lagrangian of the ideal fluid is usually taken in the form of the total energy density, $\lag^{\rm m}=\sqrt{-g}\e_{\rm m}$ \citep{LL_1959flme,comer_lrr2007,rezzolla_2013}. However, this form of the Lagrangian implicitly assumes that the equation of continuity is valid and has been used as a constraint in the form of the Lagrange multiplier \citep{Poplawski2009} so that the rest mass density $\r_{\rm m}$ of the fluid is solely an explicit function of the metric tensor $g_{\a\b}$. The equation of continuity is used, then, to derive the variational derivative of the rest mass density of the fluid \citep{fock_book,comer_lrr2007,rezzolla_2013}. This way, however, becomes coordinate dependent as it relies upon using coordinates and velocities of the fluid particles for doing variational analysis. It prevents us from making use of the full power of the dynamic field theory on the manifolds because coordinates and velocities are not field variables. An attempt to use the fluid density $\r_{\rm m}$ as a dynamic variable is not satisfactory because $\r_{\rm m}$ has no a corresponding conjugated counterpart as contrasted to the Clebsch potential, $\Phi$, and its derivative, $\Phi_{|\a}$, which are truly independent pair of canonically-conjugated dynamic variables on manifold. We avoid the approach based on the Lagrangian $\lag^{\rm m}=\sqrt{-g}\e_{\rm m}$ by taking the Lagrangian of the ideal fluid in the form of pressure, $L^{\rm m}=-\sqrt{-g}p_{\rm m}$, and demand that thermodynamic equations like (\ref{jk3ea}), (\ref{jk3eb}) are valid. This allows us to treat all thermodynamic quantities entering the Lagrangian as single-valued explicit functions of the specific enthalpy $\m_{\rm m}$. Any dependence on the metric tensor in this treatment of the ideal fluid is only through the specific enthalpy as given in (\ref{pf7}). The equation of continuity is not a priory imposed on the dynamic system but is a consequence of the Euler-Lagrange equation for the Clebsch potential $\Phi$ considered as a dynamic variable (more details in \citep[pp. 334-336]{kopeikin_2011book}).

The Lagrangian in the form of pressure differs from the Lagrangian in the form of energy by a total divergence \citep[pp. 334-335 ]{kopeikin_2011book}. The Lagrangian of the ideal fluid in the form of pressure is
\be\la{pfl1}
\lag^{\rm m}=\sqrt{-g}\lef(\e_{\rm m}-{\r_{\rm m}}\m_{\rm m}\ri)\;,
\ee
where $\e_{\rm m}=\e_{\rm m}(\m_{\rm m})$ and $\r_{\rm m}=\r_{\rm m}(\m_{\rm m})$ are functions of the specific enthalpy $\m_{\rm m}=\sqrt{-g^{\a\b}\Phi_\a\Phi_\b}$.
It is important to notice that the Lagrangian of dark matter is a singled-valued function of $\mu_{\rm m}$ and depends only on the derivative $\Phi_\a$ of the Clebsch potential. There is no explicit dependence on the field $\Phi$ whatsoever. It could appear if the ideal fluid had some special kind of potential interaction between the fluid's particles like in plasma which is not electrically neutral. However, we exclude such type of fluids from further consideration. 

The metrical stress-energy tensor of the ideal fluid is obtained by taking a variational derivative of the Lagrangian (\ref{pfl1}) with respect to the metric tensor,
\be\la{pf2a}
T^{\rm m}_{\a\b}=\frac{2}{\sqrt{-g}}\frac{\d\lag^{\rm m}}{\d g^{\a\b}}\;.
\ee
In our field-theoretical description of the ideal fluid the metric tensor enters all thermodynamic quantities only through the specific enthalpy in the form of equation (\ref{pf7}). Therefore, taking the variational derivative in (\ref{pf2a}) with respect to the metric tensor can be done with the help of the chain rule 
\be\la{pm4b7c}
\frac{\d\lag^{\rm m}}{\d g^{\a\b}}=\frac{\pd\lag^{\rm m}}{\pd\mu_{\rm m}}\frac{\d\m_{\rm m}}{\d g^{\a\b}}\;,
\ee
where the variational derivative from $\mu_{\rm m}$ is given in (\ref{pr5n}) of Appendix \ref{appvdm}. Calculation shows that the stress-energy tensor (\ref{pf2a}) is as follows,
\be\la{pf2b}
T^{\rm m}_{\a\b}=\lef(\e_{\rm m}+p_{\rm m}\ri)w_\a w_\b+p_{\rm m}g_{\a\b}\;,
\ee
which is just the standard form of the stress-energy tensor of the ideal fluid \citep{LanLif,fock_book}. Many studies in cosmology and general relativity take the stress-energy tensor (\ref{pf2b}) as a starting point. However, the dynamic field theory discloses that there is more deep underlying structure - the Clebsch potential which drastically simplifies theoretical analysis of hydrodynamic behaviour of the ideal fluid. 

\subsection{The Lagrangian of dark energy}\la{lsfi}

The Lagrangian of dark energy is taken in the form of a quintessence of a scalar field $\Psi$ \citep{mukh_book,Amendola_2010}
\be\la{hl4}
\lag^{\rm q}=\sqrt{-g}\lef(\frac12 g^{\a\b}\Psi_\a\Psi_\b+W\ri)\;,
\ee
where $W\equiv W\lef(\Psi\ri)$ is the scalar field potential, and we denote the partial derivative of the field, $\Psi_\a\eq\Psi_{,\a}=\Psi_{|\a}$ from now on. We assume that there is no direct coupling between the Lagrangian of dark energy and that of dark matter. They interact only indirectly through the gravitational field.
Many various forms of the potential $W$ are used in cosmology \citep{Amendola_2010,rubak_2011} but at the present paper we do not need to specify it further on, and keep it arbitrary. The scalar field $\Psi$ does not admit the gauge transformation like that of the Clebsch potential $\Phi$ for the ideal fluid. The reason is that the quintessence scalar field has a potential $W(\Psi)$ which is not gauge invariant. This makes the true scalar field different from the Clebsch potential.

The metrical stress-energy tensor of the scalar field is obtained by taking a variational derivative
\be\la{h15a}
T^{\rm q}_{\a\b}=\frac{2}{\sqrt{-g}}\frac{\d\lag^{\rm q}}{\d g^{\a\b}}\;,
\ee
that yields
\be\la{h15b}
T^{\rm q}_{\a\b}=\Psi_\a\Psi_\b-g_{\a\b}\biggl[\frac12g^{\m\n}\Psi_\m\Psi_\n+W(\Psi)\biggr]\;.
\ee
We can {\it formally} reduce tensor (\ref{h15b}) to the form being similar to that of the ideal fluid by making use of the following procedure.
First, we define the analogue of the specific enthalpy of the quintessence "fluid"
\be\la{h15c}
\mu_{\rm q}=\sqrt{-g^{\a\b}\Psi_\a\Psi_\b}\;,
\ee
and the effective four-velocity, $v^\a$, of the "fluid"
\be\la{h15ca}
\mu_{\rm q}v_\a= -\Psi_\a\;.
\ee
The four-velocity $v^\a$ is normalized to $g_{\a\b}v^\a v^\b=-1$. Therefore, the scalar field enthalpy $\mu_{\rm q}$ can be expressed in terms of the partial derivative from the scalar field
\be\la{h15cb}
\mu_{\rm q}=v^\a\Psi_\a\;.
\ee
We introduce the analogue of the rest mass density $\r_{\rm q}$ of the quintessence "fluid" by identification,
\be\la{h15d}
\r_{\rm q}=\mu_{\rm q}\;.
\ee
As a consequence of the above definitions, the energy density, $\e_{\rm q}$ and pressure $p_{\rm q}$ of the quintessence "fluid" can be introduced as follows
\ba\la{h16a}
\e_{\rm q}&\equiv&-\frac12 g^{\a\b}\Psi_\a\Psi_\b+W(\Psi)=\frac12\r_{\rm q}\m_{\rm q}+W(\Psi)\;,\\\la{h16b}
p_{\rm q}&\equiv&-\frac12g^{\a\b}\Psi_\a\Psi_\b-W(\Psi)=\frac12\r_{\rm q}\m_{\rm q}-W(\Psi)\;.
\ea
We notice that relation
\be\la{h16d}
\m_{\rm q}=\frac{\e_{\rm q}+p_{\rm q}}{\r_{\rm q}}\;,
\ee
between the specific enthalpy $\m_{\rm q}$, density $\r_{\rm q}$, pressure $p_{\rm q}$ and the energy density, $\e_{\rm q}$, of the scalar field "fluid" formally holds on the same form (\ref{pf2}) as in the case of the barotropic ideal fluid.

After substituting the above-given definitions of various ``thermodynamic'' quantities into equation (\ref{h15b}), it formally reduces to the stress-energy tensor of an ideal ``fluid''
\be\la{h16c}
T^{\rm q}_{\a\b}=\lef(\e_{\rm q}+p_{\rm q}\ri)v_\a v_\b+p_{\rm q}g_{\a\b}\;.
\ee
It is worth emphasizing that the analogy between the stress-energy tensor (\ref{h16c}) of the scalar field "fluid" with that of the barotropic ideal fluid (\ref{pf2b}) is rather formal since the scalar field, in the most general case, does not satisfy {\it all} required thermodynamic equations because of the presence of the potential $W=W(\Psi)$ in the energy density $\e_{\rm q}$, and pressure $p_{\rm q}$ of the scalar field. The dark energy in the form of quintessence is physically different from dark matter in the form of the ideal fluid! In particular, the ``speed of sound'' of the quintessence ``fluid'' is always equal to the speed of light $c$ independently of the equation of state of the quintessence, $p_{\rm q}=w_{\rm q}\e_{\rm q}$, where
parameter 
\be\la{b3c7s}
w_{\rm q}=\frac{\disp\frac12\r_{\rm q}\m_{\rm q}-W(\Psi)}{\disp\frac12\r_{\rm q}\m_{\rm q}+W(\Psi)}\;,
\ee
and it can take the values in the range from $-1$ to $+1$ depending on how large is the kinetic energy of the scalar field as compared to its potential energy $W$ \citep{Amendola_2010}. 

\subsection{The Lagrangian of baryonic matter}\la{laloas}

The Lagrangian $\lag^{\rm\st P}$ of the baryonic matter represents a bare perturbation of the cosmological manifold. It enters the total Lagrangian (\ref{(2.1)}) and can be chosen in accordance with the specific problem we want to solve. We keep it unspecified as long as the theory permit. We assume that the baryonic matter of the bare perturbation is described by dynamic fields $\Theta^B$ which geometric nature depends on the type of the baryonic matter. In what follows, we shall omit index $B$ from the baryonic fields to simplify notations as it does not lead to confusion. Metrical stress-energy tensor of the baryonic matter, $T_{\a\b}$, has been defined in terms of the variational derivative in (\ref{q19z}).
Tensor $T_{\a\b}$ is a source of the bare gravitational perturbation of the background manifold which generates the small-scale structures in the universe. A particularly familiar form of the stress-energy tensor of the baryonic matter is given by that of the ideal fluid 
\be\la{qze1k}
T_{\a\b}=\lef(\e+p\ri)\mathfrak{u}_\a \mathfrak{u}_\b+pg_{\a\b}\;,
\ee
where $\e$, $p$ are the energy density and pressure of the fluid comprising the bare perturbation, and $\mathfrak{u}^\a$ is its four-velocity normalized to $g_{\a\b}\mathfrak{u}^\a \mathfrak{u}^\b=-1$. It is worth emphasizing that the four-velocity $\mathfrak{u}^\a$ of the baryonic matter has a peculiar component and differs from the velocity of the Hubble flow (see below). Notice that the stress-energy tensor $\T_{\m\n}$ of the baryonic matter defined in (\ref{pmy6}), is not fully identical with $T_{\m\n}$. We have derived relation between the two tensors in (\ref{rtei1}) and (\ref{rtei2}).

\subsection{Background manifold}\la{bmnf}

All geometric objects on background cosmological manifold $\bar\M$ will be denoted with a bar over the object. The FLRW metric on the background manifold is given in (\ref{frm1}). It is convenient to introduce global isotropic coordinates $X^\a=(X^0,X^i)$ by changing the cosmic time $T$ to the conformal time $\eta\eq X^0$ via differential equation $dT=a(\eta)d\eta$, where a cosmological scale factor $a(\eta)\eq R[T(\eta)]$. The FLRW metric tensor in the isotropic coordinates reads \citep{weinberg_2008}
\be\la{as12}
\bar g_{\m\n}=a^2(\eta)\mm_{\m\n}\;
\ee
where $\mm_{\m\n}=(-1,\mm_{ij})$, and
\be\la{nxe3}
\mm_{ij}=\lef(1+\disp\frac{1}4 kr^2\ri)^{-2}\d_{ij}\;,
\ee
depends on the curvature of the spatial hypersurfaces, $k=\{-1,0,+1\}$. In case, $k=0$, the metric $\mm_{\m\n}$ is reduced to the Minkowski metric, $\eta_{\m\n}$ so that the physical metric $g_{\m\n}$ is confromally-flat. In fact, FLRW metric $g_{\m\n}$ is conformally-flat in any case but the conformal factor is not reduced to $a(\eta)$ but is given a more complicated function of time and space. This question is discussed in more detail in an excellent article by M. Ibison \citep{Ibison_2007JMP}. Congruence of world lines of freely-falling particles which have constant spatial coordinates makes up the Hubble flow. Four-velocity of each such a particle in the isotropic coordinates is $\bar U^\a=dX^\a/dT=(a^{-1},0,0,0)$.

Due to the maximal symmetry of FLRW spacetime, all background geometric objects (like the metric, the affine connection, etc.) when expressed in the isotropic coordinates, depend only on time $X^0=\eta$ but do not depend on spatial coordinates $X^i$. Nonetheless,
we can, and will, use arbitrary coordinates $x^\a=(x^0,x^i)$ on the manifold which are connected to the isotropic coordinates $X^\a$ by diffeomorphism $x^\a=x^\a(X^\b)$. Partial coordinate derivative of a background geometric object, $\bar \F=\bar \F(\eta)$, in the arbitrary coordinates is given by
\be\la{nxe4}
\bar \F_{,\a}=-\frac{\bar \F'}{a}\bar u_\a=-\dot{\bar \F}\bar u_\a\;,
\ee
where $\bar u^\a$ is four-velocity of the Hubble flow in the arbitrary coordinates, $\bar\F'=d\bar\F/d\eta$, and $\dot{\bar\F}\eq d\bar\F/dT$.
Equation (\ref{nxe4}) applied to the scale factor, yields $a_{,\a}=-\dot{a}\bar u_\a=-{\cal H}\bar u_\a$, and the partial derivative from the conformal Hubble parameter ${\cal H}_{,\a}=-\dot{\cal H}\bar u_\a$. These expressions for the partial derivatives are very useful in calculations.

Einstein's field equations on the background cosmological manifold with FLRW metric are given by (\ref{(2.7)})-(\ref{aq2}). After substitution FLRW metric to these equations they yield two Friedmann equations describing the temporal evolution of the scale factor $a$,
\ba\la{a12} H^2&=&~~\frac{8\pi}{3}\bar\e-\frac{k}{a^2}\;,
\\\la{a13}
2{\dot H}+3H^2&=&-8\pi\bar p-\frac{k}{a^2}\;
\ea
where $\bar\e$ and $\bar p$ are the effective energy density, $\bar\e=\bar\e_{\rm m}+\bar\e_{\rm q}$, and pressure, $\bar p=\bar p_{\rm m}+\bar p_{\rm q}$, of the background dark matter and dark energy.
A consequence of the Friedmann equations (\ref{a12}), (\ref{a13}) is equation
\be\la{13a}
\dot{H}=-4\pi\lef(\bar\e+\bar p\ri)+\frac{k}{a^2}\;,
\ee
that relates the time derivative of the Hubble parameter to the sum of the overall energy density and pressure of dark matter and dark energy
\be\la{13b}
\bar\e+\bar p=\bar\r_{\rm m}\bar\m_{\rm m}+\bar\r_{\rm q}\bar\m_{\rm q}\;.
\ee

Equation of continuity for the rest mass density $\bar{\r}_{\rm m}$ of the background dark matter is given by (\ref{aq3}) where we have to make a replacement, $\lag^{\rm\st M}\rightarrow\lag^{\rm m}$ and $\bar\Phi^A\rightarrow\bar\Phi$ for the background value of the Clebsch potential. The equation reads
\be\la{cr30}
\lef(\bar\r_{\rm m}\bar u^\a\ri)_{|\a}=0\;,
\ee
that is equivalent to
\be\la{cr3}
{\bar\r}_{\rm m|\a}-3H{\bar\r}_{\rm m}\bar u_\a=0\;.
\ee
The background equation of the conservation of the energy density $\e_{\rm m}$ of dark matter is derived from its definition (\ref{gug1}), the law of conservation of thermal energy (\ref{pf1}), and the continuity equation (\ref{cr3}). It yields,
\be\la{cr2}
{\bar\e}_{\rm m|\a}-3H\left({\bar\e}_{\rm m}+{\bar p}_{\rm m}\right)\bar u_\a=0\;.
\ee

Background equation for the evolution of the dark energy is also given by the Euler-Lagrange equation (\ref{aq3}) after replacements $\lag^{\rm\st M}\rightarrow\lag^{\rm q}$ and $\bar\Phi^A\rightarrow\bar\Psi$. It reads
\be\la{cr0}
\bar g^{\a\b}\bar\Psi_{|\a\b}-\frac{\pd \bar W}{\pd\bar\Psi}=0\;.
\ee
After making use of definition of the background specific enthalpy of the scalar field $\bar\m_{\rm q}\equiv\bar u^\a\bar\Psi_{|\a}$, an equality $\bar\m_{\rm q}=\bar\r_{\rm q}$, and 
definition (\ref{h16a}) of the specific energy $\bar\e_{\rm q}$ of the scalar field, equation (\ref{cr0}) can be recast to
\be\la{cr1a}
{\bar\e}_{\rm q|\a}-3H\left({\bar\e}_{\rm q}+{\bar p}_{\rm q}\right)\bar u_\a=0\;,
\ee
that is completely similar to the hydrodynamic equation (\ref{cr2}) of conservation of the energy density of dark matter. Because of this similarity, the second Friedmann equation (\ref{a13}) is not really independent, and can be derived directly from the first Friedmann equation (\ref{a12}) by taking a time derivative and applying the energy conservation equations (\ref{cr2}) and (\ref{cr1a}) to simplify the result.

Equation of continuity for the density of dark energy, $\bar{\r}_{\rm q}$, is obtained by differentiating definition (\ref{h15d}) of $\bar{\r}_{\rm q}$, and making use of (\ref{cr0}). It yields
\be\la{crq}
\lef(\bar\r_{\rm q}\bar u^\a\ri)_{|\a}=-\frac{\pd\bar W}{\pd\bar\Psi}\;,
\ee
or, equivalently,
\be\la{cr3a}
{\bar\r}_{\rm q|\a}-3H{\bar\r}_{\rm q}\bar u_\a=\frac{\pd\bar W}{\pd\bar\Psi}\bar u_\a\;,
\ee
which shows that the density $\bar\r_{\rm q}$ is not conserved. This fact again points out that the similarity of the scalar field and an ideal fluid is not complete. Dark energy is not thermodynamically equivalent to dark matter. Only if the scalar field $\bar\Psi$ is potential-free, the quintessence can be treated as an ideal fluid. We should emphasize that non-conservation of the density $\bar\r_{\rm q}$ does not violate any physical law since (\ref{cr3a}) is simply another way of writing the evolution equation (\ref{cr0}) for dark energy.

\subsection{Perturbations of the dynamic variables}\la{pfri}

In the present paper, FLRW background manifold is defined by the metric $\bar g_{\a\b}$ which dynamics is governed by the two scalar fields -
the Clebsch potential $\bar\Phi$ of dark matter and the scalar field $\bar\Psi$ of dark energy. We assume that the background metric and the fields are perturbed intrinsically (the primordial perturbations)  and extrinsically (the bare perturbations) by the presence of baryonic matter described by the field $\Theta$.
The perturbed metric and the matter fields can be split in their background values and the corresponding perturbations,
\be\la{pf8}
\gag^{\a\b}=\bar\gag^{\a\b}+\hatl^{\a\b}\;,\qquad\quad \Phi=\bar\Phi+\p\;,\qquad\quad \Psi=\bar\Psi+\psi\;.
\ee
These equations are exact. Equation for the perturbation of the metric tensor
\be\la{m6z2a}
g_{\a\b}=\bar g_{\a\b}+\varkappa_{\a\b}\;
\ee
will be also treated as exact. Corresponding perturbation of the contravariant component of the metric is not independent and is determined from the isomorphism $g_{\alpha\gamma}g^{\gamma\beta}=\bar g_{\alpha\gamma}\bar g^{\gamma\beta}=\delta_\alpha^\beta$, yielding
\begin{equation}\label{pf8q}
g^{\a\b}=\bar g^{\a\b}-\varkappa^{\a\b}+\varkappa^\alpha{}_\gamma \varkappa^{\gamma\beta}+\ldots\;,
\end{equation}
where the ellipses denote terms of the higher order.

We consider perturbation of the metric - $\varkappa_{\a\b}$, that of the potential of dark matter - $\p$, and that of the potential of dark energy - $\psi$ as weak with respect to their corresponding background values $\bar g_{\a\b}$, $\bar\Phi$, and $\bar\Psi$, which dynamics is governed by equations that have been explained in section \ref{bmnf}. Perturbations $\varkappa_{\a\b}$, $\p$, and $\psi$ have the same order of magnitude as $\Theta$. 
Calculations also prompt us to single out $\sqrt{-\bar g}$ from $\mathfrak{h}^{\a\b}$, and operate with a variable
\be\la{mcv1}
l^{\a\b}\equiv\frac{\mathfrak{h}^{\a\b}}{\sqrt{-\bar g}}\;.
\ee
Tensor indices of the metric tensor perturbations, $l^{\a\b}$, $\hatl^{\a\b}$, etc., are raised and lowered with the help of the background metric, for example, $l_{\a\b}\equiv \bar g_{\a\m}\bar g_{\b\n} l^{\m\n}$.
The field variable $l^{\a\b}$ relates to the perturbation $\varkappa_{\a\b}$ of the metric tensor as follows
\be\la{ok9}
l^{\a\b}=-\varkappa^{\a\b}+\frac12\bar g^{\a\b}\varkappa+\varkappa^{\g\alpha}\varkappa^{\b}{}_\g-\frac12 \varkappa^{\a\b}\varkappa-\frac14\bar g^{\a\b}\left(\varkappa^{\mu\nu}\varkappa_{\mu\nu}-\frac12 \varkappa^2\right)+\ldots\;,
\ee
where $\varkappa\equiv \varkappa^\sigma{}_\sigma=\bar g^{\rho\sigma}\varkappa_{\rho\sigma}$, and ellipses denote terms of the higher orders in $\varkappa_{\a\b}$.

Perturbations of four-velocities, $w^\alpha$ and $v^\alpha$, entering definitions of the stress-energy tensors (\ref{pf2b}), (\ref{h16c}), are fully determined by the perturbations of the metric and the potentials of dark matter and dark energy. Indeed, according to definitions (\ref{pf6}) and (\ref{h15ca}) the four-velocities are defined by the following equations
\be\la{pf8aa}
w_\alpha=-\frac{\Phi_\a}{\m_{\rm m}}\;,\qquad\qquad v_\alpha=-\frac{\Psi_\a}{\m_{\rm q}}\;.
\ee
where $\m_{\rm m}$ and $\m_{\rm q}$ are given by (\ref{pf7}) and (\ref{h15c}) respectively.
We define perturbations $\delta w_\a$ and $\delta v_\a$ of the covariant components of the four-velocities as follows
\be\la{pf8s}
w_\a=\bar u_\a+\delta w_\a\;,\qquad\qquad v_\a=\bar u_\a+\delta v_\a\;,
\ee
where the unperturbed values of the four-velocities coincide and are equal to the four-velocity of the Hubble flow due to the requirement of the homogeneity and isotropy of the background FLRW spacetime, that is $\bar w^\a=\bar v^\a=\bar u^\a$. Hence, we have
\be\la{pf8tn}
\bar u_\alpha=-\frac{\bar\Phi_\a}{\bar\m_{\rm m}}\;,\qquad\qquad \bar u_\alpha=-\frac{\bar\Psi_\a}{\bar\m_{\rm q}}\;.
\ee
Making use of (\ref{pf8aa}) and (\ref{pf8tn}) in the left side of definitions (\ref{pf8aa}), and expanding its right side by making use of expansions (\ref{m6z2a}) and (\ref{pf8q}), yield relation between the four-velocity perturbations and the perturbations of the dynamic field variables
\be\la{kio}
\delta w_\a=-\frac{1}{\bar\m_{\rm m}}{\bar P}^\b{}_\a\phi_{\b}-\frac12\mathfrak{q}\bar u_\a\;,\qquad\qquad
\delta v_\a=-\frac{1}{\bar\m_{\rm q}}{\bar P}^\b{}_\a\psi_{\b}-\frac12\mathfrak{q}\bar u_\a\;,
\ee
where $\phi_{\b}\eq\phi_{|\b}$, $\psi_{\b}\eq\psi_{|\b}$,  $\bar P_{\a\b}=\eq\bar g_{\a\b}+\bar u_\a\bar u_\b$ is a projector tensor onto the hypersurface orthogonal to the Hubble flow, and
\be\la{mt6k}
\mathfrak{q}\equiv-\bar u^\a \bar u^\b \varkappa_{\a\b}=\bar u^\a \bar u^\b l_{\a\b}+\frac{l}{2}\;,
\ee
is a scalar-type projection of the metric tensor perturbation on the Hubble flow ($l\eq\bar g^{\a\b}l_{\a\b}$). Equations (\ref{kio}) are valid in linear approximations. Higher-order corrections can be obtained by the same procedure by keeping more terms in the expansions.

\subsection{Field equations}
\subsubsection{Gravitational field}
Field equations for metric tensor perturbation are given by the Euler-Lagrange equations (\ref{(2.17)}) where $F_{\m\n}^{\rm\st  G}$ is given by {\it exact} expression (\ref{xpq1}), and operator $F_{\m\n}^{\rm\st M}$ is a linear superposition of two pieces 
\be\la{zxb6}
F_{\m\n}^{\rm\st  M}=F_{\m\n}^{\rm m}+F_{\m\n}^{\rm q}\;,
\ee 
corresponding to dark matter (index 'm') and dark energy (index 'q'). These pieces are defined in accordance with (\ref{(2.19)}) that is
\ba\la{xrc1}
F^{\rm m}_{\mu\nu}&\equiv& -\frac{16\pi}{\sqrt{-\bar g}} \frac{\delta}{\delta\bar{g}^{\mu\nu}}
\lef(\hatl^{\rho\s}  \frac{\delta\bar\lag^{\rm m}}{\delta\bar
\gag^{\rho\s}} + \phi \frac{\delta\bar\lag^{\rm m}}{\delta\bar\Phi}\ri)\;,\\
\la{xrc2}
F^{\rm q}_{\mu\nu}&\equiv& -\frac{16\pi}{\sqrt{-\bar g}} \frac{\delta}{\delta\bar{g}^{\mu\nu}}
\lef(\hatl^{\rho\s}  \frac{\delta\bar\lag^{\rm q}}{\delta\bar
\gag^{\rho\s}} + \psi \frac{\delta\bar\lag^{\rm q}}{\delta\bar\Psi}\ri)\;.
\ea
Calculation of variational derivatives from various functions entering $\lag^{\rm m}$ and $\lag^{\rm q}$ is straightforward and follows from their definitions, the chain rule (\ref{xcr1}), and a set of variational derivatives from thermodynamic quantities given in Appendix \ref{vdwrmt}. 

Making use of the Lagrangian's definitions (\ref{pfl1}) and (\ref{hl4}) taken on the background manifold and calculating variational derivatives in (\ref{xrc1}), (\ref{xrc2}), we obtain 
\ba\la{mt6a}
F^{\rm m}_{\m\n}&=&-4\pi(\bar p_{\rm m}-\bar\e_{\rm m})l_{\mu\nu}+8\pi\bar\r_{\rm m}\lef(\bar u_\mu\p_{\nu}+\bar u_\nu\p_{\mu}-\bar g_{\mu\nu}\bar u^\a\p_\a\ri)\\\nonumber
&&+8\pi\bar\r_{\rm m}\left(1-\frac{c^2}{c^2_{\rm s}}\right)\lef(\bar u^\a\p_\a-\frac12\bar\m_{\rm m}\mathfrak{q}\ri)\bar u_\mu \bar u_\nu
\;, \\\nonumber\\
\la{mt6c}
F^{\rm q}_{\m\n}&=&-4\pi\lef(p_{\rm q}-\e_{\rm q}\ri) l_{\mu\nu}+8\pi\bar\r_{\rm q}\left(\bar u_\m\psi_\n+\bar u_\n\psi_\m-\bar g_{\mu\nu}\bar u^\a\psi_\a\ri)+8\pi\bar g_{\m\n}\frac{\pd\bar W}{\pd\bar\Psi}\,\psi
\;, \ea
where $\bar\r_{\rm q}=\m_{\rm q}\equiv\dot{\bar\Psi}/a$ in accordance with definition (\ref{h15d}) projected on the background manifold. The dark energy potential function, $\bar W=\bar W(\bar\Psi)$, is arbitrary. We emphasize that expressions (\ref{mt6a}), (\ref{mt6c}) are {\it exact}.

Substituting (\ref{mt6a}), (\ref{mt6c}) along with (\ref{xpq1}) to the left side of (\ref{(2.17)}) yields the field equations for gravitational perturbations $l^{\a\b}$ in a covariant form \citep[Eq. 161]{kopetr}
\be\la{axz4}
l_{\m\n}{}^{|\a}{}_{|\a}+\bar g_{\m\n}A^\a{}_{|\a}-2A_{(\m|\n)}-2\bar R^\a{}_{(\m}l_{\n)\a}-2\bar R_{\m\a\b\n}l^{\a\b}+2\left(F_{\m\n}^{\rm m}+F_{\m\n}^{\rm q}\ri)=16\pi\Lambda_{\m\n}\;,
\ee
where $A^\a\eq l^{\a\b}{}_{|\b}$ is the gauge vector function. This form of the field equation for gravitational perturbation $l_{\a\b}$ of the background FLRW manifold is {\it exact}, gauge-invariant and covariant. The left side of (\ref{axz4}) contains only linear terms while all quadratic, cubic, etc. perturbations are included in its right side to $\Lambda_{\m\n}$ which also contains the stress-energy tensor $\T^{\m\n}$ of the baryonic matter (the bare perturbation). The linear operator in the left side of (\ref{axz4}) is rather complicated but it can be significantly simplified by choosing a gauge condition imposed on the variable $A^\a\eq l^{\a\b}{}_{|\b}$ in the following form \citep{kopetr}
\be
\la{qe6}
A^\a=-2H l^{\a\b}\bar u_\b+16\pi\lef(\bar\r_{\rm m}\p+\bar\r_{\rm q}\psi\ri)\bar u^\a\;.
\ee
This gauge condition is analogous to the de Donder (harmonic) gauge condition that is frequently used in the approximation schemes of solving Einstein's equations in asymptotically flat spacetime \citep{2006LRR.....9....4B,kopeikin_2011book}. Equation (\ref{qe6}) extrapolates the harmonic gauge condition to the realm of cosmological FLRW spacetime. 

This gauge condition (\ref{qe6}) cancels a large number of terms in the field equations (\ref{axz4}) and allows us to decouple the field equations for different components of the metric tensor perturbation, $l^{\a\b}$. Picking up the isotropic coordinates of the Hubble observers we bring the gravity field equations to the following form \citep{kopetr}
\bsu\la{c6q2}
\ba\la{qe7a}
\Box q+2H q_{,0}+4k q-4\pi\lef(1-\frac{c^2}{c^2_{\rm s}}\ri)\bar\r_{\rm m}\bar\m_{\rm m}q
&=&8\pi\lef(\Lambda_{00}+\Lambda_{kk}\ri)-
8\pi a\bar\r_{\rm m}\lef(1-\frac{c^2}{c^2_{\rm s}}\ri)\phi_{0}-\\\nonumber
&&16\pi a^2\frac{\pd\bar W}{\pd\bar\Psi}\psi+
32\pi aH\lef(\bar\r_{\rm m}\phi+\bar\r_{\rm q}\psi\ri)\;,\\\nonumber\\\la{qe7b}
\Box l_{0i}+2 H l_{0i,0}+2k l_{0i}&=&16\pi \Lambda_{0i}\;,\\\la{qe7c}
\Box l_{<ij>}+2 H l_{<ij>,0}+2\lef(\dot H-k\ri) l_{<ij>} &=&16\pi \Lambda_{<ij>}\;,\\\la{qe7d}
\Box l+2H l_{,0}+2\lef(\dot H+2k\ri) l &=&16\pi \Lambda_{kk}\;.
\ea\esu
where we denoted the wave operators $\Box q\eq\mm^{\m\n}g_{;\m\n}$ and $\Box l_{\m\n}\eq\bar\mm^{\a\b}l_{\m\n;\a\b}$. Other notations in (\ref{c6q2}) are $\phi_0\equiv \phi_{,0}$, $q\equiv\lef(l_{00}+l_{kk}\ri)/2$, $l\eq l_{kk}=l_{11}+l_{22}+l_{33}$, $l_{<ij>}=l_{ij}-(1/3)\d_{ij}l$, and the same index notations are applied to the effective stress-energy tensor $\Lambda_{kk}=\Lambda_{11}+\Lambda_{22}+\Lambda_{33}$, $\Lambda_{<ij>}=\Lambda_{ij}-(1/3)\d_{ij}\Lambda_{kk}$.

Solution of the linearised gravitational field equation (\ref{axz4}) (and (\ref{c6q2})) consists of two parts - a general solution, $l_{\m\n}^{\rm\st H}$, of homogeneous equation (\ref{axz4}) with the source $\Lambda_{\m\n}=0$, and a particular solution, $l_{\m\n}^{\rm\st P}$ of inhomogeneous equation (\ref{axz4}) with the source $\Lambda_{\m\n}\not=0$. They form a linear superposition
\be\la{b72c5}
l_{\m\n}=l_{\m\n}^{\rm\st H}+l_{\m\n}^{\rm\st P}\;,
\ee
which is crucial for understanding the physical effects of cosmological perturbations. The homogeneous solution, $l_{\m\n}^{\rm\st H}$, is not trivial but associated with the primordial cosmological perturbations originating at the Big Bang \citep{weinberg_2008,rubak_2011}. This perturbation dominates on the horizon and super-horizon scales, and its gauge-invariant scalar part (which exact definition and equations are given in \citep[section 7]{kopetr}) evolves over time to form the large-scale structure of the universe governed by dark matter. The tensor part of the homogeneous solution represents relic gravitational waves.  The particular solution, $l_{\m\n}^{\rm\st P}$, represents gravitational perturbations produced by the small-scale structures in the universe consisting of baryonic matter. We notice that $l_{\m\n}^{\rm\st H}$ and $l_{\m\n}^{\rm\st P}$ correspond to the long wavelength and short wavelength perturbations introduced by Green and Wald \citep{wald_2011}, and denoted in their paper as $\g_{\m\n}^{(L)}$ and $h_{\m\n}^{(S)}$ respectively (see \citep[eq. 70]{wald_2011}). In what follows, we operate with a single value of the perturbation $l_{\m\n}$ without substituting the explicit decomposition (\ref{b72c5}) into subsequent formulas as it was not a primary goal of the present paper. Decomposition \eqref{b72c5} is required for discussion the problem of averaging, back-reaction and precise definition of the Newtonian limit in cosmology \citep{wald_2011,wald_2012}.

\subsubsection{Dark matter}\la{b6e4m}

Evolution of dark matter perturbation is described by the perturbation $\p$ of the Clebsch potential. Equation for $\p$ is derived from a general equation (\ref{bbv2}) is, in case of dark matter, reads
\be
\la{gd4}
F^{\rm m}_\Phi = 8\pi\Sigma^{\rm m}\;.
\ee
where all terms can now be explicitly written down because the Lagrangian of dark matter is fully determined by (\ref{pfl1}). 
The linear differential operator $F^{\rm m}_\Phi$ is derived from (\ref{koj2}) which is split in two independent parts for dark matter and dark energy. The dark matter part reads
\be
\la{gd5}
F^{\rm m}_\Phi\eq -\frac{8\pi}{\sqrt{-\bar g}} \frac{\delta}{\delta\bar\Phi}
\lef(\hatl^{\r\s}\bar T^{\rm m}_{\r\s}-\frac12\hatl\bar T^{\rm m} + \sqrt{-\bar g}\phi \bar I^{\rm m}\ri)\;,
\ee
where 
\be\la{gd6}
\bar I^{\rm m}\eq 2\lef(\bar\r_{\rm m}\bar u^\a\ri)_{|\a}\;.
\ee
The source density in the right side of (\ref{gd4}) represent contribution of non-linear perturbations
\ba
\la{gd6a}
\Sigma^{\rm m}&\equiv&\frac2{\sqrt{-\bar g}}\frac{\delta\lag^{\rm dyn}}{\delta\bar\Phi}\;,
\ea
and we shall calculate it explicitly in section \ref{dam476}. Notice that because of the non-linearity the source term, $\Sigma^{\rm m}$, depends not only on the dark matter variables but on the dynamic variables describing the gravitational and dark energy perturbations as well. 

Calculation of the variational derivative in (\ref{gd5}) requires taking the variational derivative from various thermodynamic quantities like the energy density $\e_{\rm m}$, pressure $p_{\rm m}$, etc. with respect to the background value of the Clebsch potential $\bar\Phi$. All of them are functions of the specific enthalpy $\m_{\rm m}$ which, in its own turn, depends only on the derivatives $\Phi_\a$ of the potential. As an example, let us consider the density of the ideal fluid $\bar\r_{\rm m}=\bar\r_{\rm}\lef(\bar\m_{\rm m}\ri)$. We have
\be\la{djt3}
\frac{\d\r_{\rm m}}{\d\bar\Phi}= -\frac{\pd}{\pd x^\a}\frac{\pd\r_{\rm m}}{\pd\bar\Phi_\a}=-\frac{\pd}{\pd x^\a}\lef(\frac{\pd\r_{\rm m}}{\pd\bar\m_{\rm m}}\frac{\pd\m_{\rm m}}{\pd\bar\Phi_\a}\ri)\;,
\ee
where the partial derivative of the density with respect to the specific enthalpy is calculated with the help of (\ref{pf5}) by making use of equation of state of the ideal fluid, and the partial derivative 
\be\la{djt4}
\frac{\pd\m_{\rm m}}{\pd\bar\Phi_\a}=\bar u^\a\;,
\ee
as follows from the definition of $\m_{\rm m}$. The same procedure is applied for calculation of the variational derivative from other thermodynamic quantities. The variational derivative from the Hubble four-velocity $\bar u^\a$ is calculated from the relation, $\bar\m_{\rm m}\bar u_\a=-\bar\Phi_\a$, between the specific enthalpy, four-velocity and the gradient of the Clebsch potential. All variational derivatives that enter the calculation are given in Appendix \ref{fff5v6} of the present paper. Finally, we come to the following result,  
\be \la{mtj10}
F^{\rm m}_\Phi\eq 8\pi Y^\a{}_{|\a}\;,
\ee
where the vector field
\be\la{mt10a}
Y^\a\equiv\frac{\bar\r_{\rm m}}{\bar\mu_{\rm m}}~\p^\a-\bar\r_{\rm m} l^{\a\b}\bar u_\b+\left(1-\frac{c^2}{c^2_{\rm s}}\right)\lef(\frac{\bar\r_{\rm m}}{\bar\mu_{\rm m}} \bar u^\b\p_\b -\frac12\bar\r_{\rm m}  \mathfrak{q}\ri)\bar u^\a\;.
\ee
It shows that in the linear approximation of the dynamic perturbation theory, where $\Sigma^{\rm m}=0$, the current $\sqrt{-\bar g}Y^\a$ is conserved.

Taking covariant derivative in (\ref{mtj10}) brings about the field equations for $\p$ 
\ba\la{qe8}
\p^\a{}_{\a}-\m_{\rm m}A^\a\bar u_\a-4\bar\m_{\rm m}H \lef(4\mathfrak{q}-l\ri)
+\lef(1-\frac{c^2}{c^2_{\rm s}}\ri)\lef(\bar u^\a\bar u^\b\p_{\a\b}-\frac12\bar\m_{\rm m}\bar u^\a \mathfrak{q}_\a\ri)&&\\\nonumber
-3H\bar\m_{\rm m}\frac{\pd\ln c^2_{\rm s}}{\pd\bar\m_{\rm m}}\lef(\bar u^\a\p_\a-\frac12\bar\m_{\rm m}\mathfrak{q}\ri)&=&\Sigma^{\rm m}\;,
\ea
where $\p^\a{}_{\a}\eq \p^{|\a}{}_{|\a}$, $\mathfrak{q}_\a\eq\mathfrak{q}_{|\a}$, and the very last term accounts for the fact that the speed of sound is not constant in inhomogeneous medium - the effect which is important for a more adequate treatment of precise cosmological observations.
Indeed, the speed of sound, $c_{\rm s}$, relates to other thermodynamic quantities by equation of state making the speed of sound a function of the specific enthalpy, $c_{\rm s}=c_{\rm s}(\bar\m_{\rm m})$. Covariant derivative from the speed of sound is $c_{{\rm s}|\a}=\lef(\pd c_{\rm s}/\pd\bar\m_{\rm m}\ri)\bar\m_{{\rm m}|\a}$, where the covariant derivative $\bar\m_{{\rm m}|\a}=\lef(\pd\bar\m_{\rm m}/\pd\bar\r_{\rm m}\ri)\bar\r_{{\rm m}|\a}$ and, according to equation of continuity, $\bar\r_{{\rm m}|\a}=3H\bar\r_{\rm m}\bar u_\a$. It yields
\be\la{ir5s}
\lef(1-\frac{c^2}{c^2_{\rm s}}\ri)_{|\a}=3H\bar\m_{\rm m}\frac{\pd\ln c^2_{\rm s}}{\pd\bar\m_{\rm m}}\bar u_\a\;,
\ee
that explains how the last term in (\ref{qe8}) originates from (\ref{mtj10}), (\ref{mt10a}).

After imposing the gauge condition (\ref{qe6}), the covariant equation (\ref{qe8}) is reduced to
\ba\la{qe8aa}
\p^\a{}_{\a}+16\pi\bar\m_{\rm m}\lef(\bar\r_{\rm m}\p+\bar\r_{\rm q}\psi\ri)-2\bar\m_{\rm m}H\mathfrak{q}
+\lef(1-\frac{c^2}{c^2_{\rm s}}\ri)\lef(\bar u^\a\bar u^\b\p_{\a\b}-\frac12\bar\m_{\rm m}\bar u^\a \mathfrak{q}_\a\ri)&&\\\nonumber
-3H\bar\m_{\rm m}\frac{\pd\ln c^2_{\rm s}}{\pd\bar\m_{\rm m}}\lef(\bar u^\b\p_\b-\frac12\bar\m_{\rm m}\mathfrak{q}\ri)&=&\Sigma^{\rm m}\;,
\ea
which is linearly coupled to the dynamic perturbation, $\psi$, of dark energy besides of coupling with gravitational field perturbation $l^{\a\b}$. Equation (\ref{qe8aa}) is to be solved by iterations starting from equating the right side of it, $\Sigma^{\rm m}=0$, and solving for $\phi$, which is used then for calculation of $\Sigma^{\rm m}$, and solving (\ref{qe8aa}) again, and so on. Since equation (\ref{qe8aa}) is linearly coupled with $\psi$, we will need equation for the dark energy perturbation.

We should underline that the field equation (\ref{qe8aa}) for the perturbations of the dark matter is nothing else but the covariant form of equation for the sound waves propagating through the substance of the background dark matter, $\bar\M$, with the speed of sound $c_{\rm s}$. Indeed, in the isotropic coordinates $x^\a=(\eta,x^i)$, the operator
\be\la{mn6v3}
\p^\a{}_{\a}+\lef(1-\frac{c^2}{c^2_{\rm s}}\ri)\bar u^\a\bar u^\b\p_{\a\b}=\frac{1}{a^2}\left(-\frac{1}{c^2_{\rm s}}\frac{\pd^2\p}{\pd\eta^2}+\mm^{ij}\frac{\pd^2\p}{\pd x^i\pd x^j}\right)\;,
\ee
which is a wave operator describing propagation of the perturbations of dark matter with the speed of sound. We also notice that the wave equation (\ref{qe8aa}) is homogeneous in linearised order of approximation in which the source $\Sigma^{\rm m}$ can be neglected because it is quadratic with respect to perturbations. It means that solution of the linearised equation (\ref{qe8aa}) corresponds only to the primordial excitations of the sound waves in dark matter. Baryonic matter (stress-energy tensor of the bare perturbation) cannot produce any direct perturbation of the background distribution of dark matter in the linearised approximation. 

\subsubsection{Dark energy}\la{b6e3q}

Calculation of the field equation for dark energy perturbation, $\psi$, follows the similar path like we did in the previous subsection \ref{b6e4m}. The field equations follow from (\ref{bbv2}), and they are 
\be
\la{gd7}
F^{\rm q}_\Psi = 8\pi\Sigma^{\rm q}\;,
\ee
where $F^{\rm q}_\Psi$ and $\Sigma^{\rm q}$ are determined by the variational derivatives from the Lagrangian of dark energy (\ref{hl4}) and the dynamic Lagrangian (\ref{ia4}) respectively. The linear operator $F^{\rm q}_\Psi$ is calculated by substituting the Lagrangian (\ref{hl4}) into (\ref{koj2}) which yields
\be\la{gd8}
F^{\rm q}_\Psi\eq -\frac{8\pi}{\sqrt{-\bar g}} \frac{\delta}{\delta\bar\Psi}
\lef(\hatl^{\r\s}\bar T^{\rm q}_{\r\s}-\frac12\hatl\bar T^{\rm q} + \sqrt{-\bar g}\psi \bar I^{\rm q}\ri)\;,
\ee
where 
\be\la{gd9}
\bar I^{\rm q}\eq 2\lef[\lef(\bar\r_{\rm q}\bar u^\a\ri)_{|\a}+\frac{\pd\bar W}{\pd\bar\Psi}\ri]\;.
\ee
The source density
\ba
\la{gd6b}
\Sigma^{\rm q}&\equiv&\frac2{\sqrt{-\bar g}}\frac{\delta\lag^{\rm dyn}}{\delta\bar\Psi}\;,
\ea
and we shall calculate it explicitly in section \ref{aa23w}. 

According to equation (\ref{hl4}), the Lagrangian density of the scalar field $\lag^{\rm q}$ depends on both the field $\Psi$ and its first derivative, $\Psi_\a$. For this reason, unlike the operator $F^{\rm m}$, the differential operator $F^{\rm q}$ is not reduced to the covariant divergence from a vector field as the partial derivative of the Lagrangian $\lag^{\rm q}$ with respect to $\Psi$ does not vanish. We have
\be\la{mt10c}
F^{\rm q}_\Psi\equiv 8\pi\lef(Z^\a{}_{|\a}-\frac{l}{2}\frac{\pd\bar W}{\pd\bar\Psi}-\psi\frac{\pd^2\bar W}{\pd\bar\Psi^2}\ri)
\ee
where $l\equiv \bar g^{\a\b}l_{\a\b}$, and vector field
\be\la{mt12}
Z^\a\equiv\psi^{\a}-\bar\r_{\rm q}l^{\a\b}\bar u_\b\;,
\ee
where we have used equation $\bar\Psi_{\a}=-\bar u^\b\bar\Psi_{\b}\bar u_\a=-\bar\r_{\rm q}\bar u_\a$. The current $Z^\a$ is not conserved unlike $Y^\a$ for the dark matter.

Taking covariant derivative in (\ref{mt10c}) and making use of the gauge condition (\ref{qe6}) yield the field equations for $\psi$ 
\be\la{qe9}
\psi^\a{}_\a+16\pi\bar\m_{\rm m}\lef(\bar\r_{\rm m}\p+\bar\r_{\rm q}\psi\ri)-\lef(2\bar\m_{\rm q}H+\frac{\pd\bar W}{\pd\bar\Psi}\ri)\mathfrak{q}-\frac{\pd^2\bar W}{\pd\bar\Psi^2}\,\psi=\Sigma^{\rm q}\;,
\ee
where  $\psi^\a{}_{\a}\eq \psi^{|\a}{}_{|\a}$, and we have use the equality $\bar\r_{\rm q}=\bar\m_{\rm q}$. Equation (\ref{qe9}) is to be solved by iterations starting from equating the right side of it, $\Sigma^{\rm q}=0$, and solving for $\psi$, which is used then for calculation of $\Sigma^{\rm q}$, and solving (\ref{qe9}) again, and so on. This procedure is going on along by simultaneously solving (\ref{qe8aa}) for $\phi$. Of course, before solving equation (\ref{qe9}) we have to specify the structure of the scalar potential $\bar W$. 

Like in the case of dark matter, the wave equation (\ref{qe9}) is homogeneous in linearised order of approximation in which the source $\Sigma^{\rm q}$ can be neglected because it is quadratic with respect to perturbations. Hence, solution of the linearised equation (\ref{qe9}) corresponds only to the primordial excitations of the scalar field in dark energy. Baryonic matter (stress-energy tensor of the bare perturbation) cannot produce any direct perturbation of dark energy in the linearised approximation.

\section{Stress-Energy Tensor of the Dynamic Field Perturbations}\la{sec5}

In order to solve the field equations (\ref{axz4}) for the gravitational perturbations in the quadratic and higher order approximations, we have to know the effective stress-energy tensor $\Lambda_{\m\n}$ entering the right side of these equations.
The effective stress-energy tensor $\Lambda_{\m\n}$ is defined as a variational derivative (\ref{kwn8}) taken from the effective Lagrangian (\ref{efla3}). According to (\ref{pmy6}), it consists of two parts -- the stress-energy tensor of matter of the baryonic (bare) perturbation $\T_{\m\n}$, and the stress-energy tensor of the dynamic field perturbations, ${\cal T}_{\m\n}$. We shall keep tensor $\T_{\m\n}$ unspecified as long as theory permits and focus on calculation of ${\cal T}_{\m\n}$ which also consists of two parts, $\mt_{\m\n}$ and $\tau_{\m\n}$, according to (\ref{qq5a}). Tensor $\mt_{\m\n}$ is the stress-energy tensor of gravitational field perturbations (\ref{qq5z}). Tensor $\tau_{\m\n}$ is the stress-energy tensor originating from the coupling of the background dark matter and dark energy perturbations to the gravitational field perturbations. General formula for calculating $\tau_{\m\n}$ is given in (\ref{qq5x}). In case of FLRW universe gowerved by dark matter and dark energy, tensor $\tau_{\m\n}$ is linearly split in two counterparts 
\be\la{jkw5}
\tau_{\a\b}=\tau^{\rm m}_{\a\b}+\tau^{\rm q}_{\a\b}\;,
\ee
where $\tau^{\rm m}_{\a\b}$ and $\tau^{\rm q}_{\a\b}$ describe contributions of dark matter and dark energy respectively. In this section we calculate all the components of the effective stress-energy tensor in the quadratic approximation. Higher-order terms will be published somewhere else.

\subsection{Stress-energy tensor of gravitational field perturbations}\la{setgf}

Stress-energy tensor of gravitational perturbations, $\mt_{\m\n}$, has a universal and unique presentation on any pseudo-Riemannian manifold because it originates from a pure geometric part of the perturbed Hilbert Lagrangian. We begin calculation of $\mt_{\m\n}$ from its definition which is given by (\ref{qq5z}) in the form of variational derivative from the following scalar density  
\be\la{n7v2s}
{\cal F}^{\rm\st G}\eq\hatl^{\r\s}F^{\rm\st G}_{\r\s}-(1/2)\hatl F^{\rm\st G}\;,
\ee
where the tensor $F^{\rm\st G}_{\r\s}$ is given by (\ref{xpq1}), $F^{\rm\st G}=\bar g^{\r\s}F^{\rm\st G}_{\r\s}$, $\hatl^{\r\s}=\sqrt{-\bar g}l^{\r\s}$, $\hatl\eq\bar g^{\r\s}\hatl_{\r\s}$. We put together all terms entering ${\cal F}^{\rm\st G}$, and, then, employ the Leibniz rule to single out the total divergence from the products of two functions - the metric tensor perturbation and its second derivative. It results in
\be\la{rr1}
{\cal F}^{\rm\st G}=\hatl^{\r\s|\l}l_{\l\r|\s}-\frac12\hatl^{\r\s|\l} l_{\r\s|\l}+\frac14 \hatl^{|\l}l_{|\l}+{\rm div}\;,
\ee
where $l\eq \bar g^{\m\n}l_{\m\n}$, and ${\rm div}$ denote the terms which form a total divergence that vanishes upon taking a variational derivative and, hence, can be discarded. For this reason, we drop it off from further calculation. 
Next step is to apply the covariant definition (\ref{er8}) of variational derivative to (\ref{rr1}) in definition (\ref{qq5z}) of $\mt_{\a\b}$ which can be written as follows
\be\la{rtb5}
16\pi\mt_{\a\b}=\frac{1}{\sqrt{-\bar g}}\bar g_{\a\m}\bar g_{\b\n}\frac{\d{\cal F}^{\rm\st G}}{\d\bar g_{\m\n}}\;,
\ee
that conforms with the lower (subscript) position of indices of the metric tensor entering in the denominator of definition (\ref{er8}) of variational derivative. 

It is worthwhile to remind the reader that perturbation $\hatl^{\r\s}$ is an independent variable which has been used in derivation of (\ref{qq5z}). It means that the partial derivative
\be\la{rr1a}
\frac{\pd\hatl^{\a\b}}{\pd\bar g_{\m\n}}=0\;.
\ee 
On the other hand, the covariant components of the gravitational perturbation, $\hatl_{\a\b}=\bar g_{\a\k}\bar g_{\b\l}\hatl^{\k\l}$, contain explicitly the background metric tensor and, hence, cannot be considered as independent from it. Therefore, we have for the partial derivatives
\be\la{rr1b}
\frac{\pd\hatl_{\a\b}}{\pd\bar g_{\m\n}}=\frac{\pd \bar g_{\a\k}}{\pd\bar g_{\m\n}}\bar g_{\b\l}\hatl^{\k\l}+\frac{\pd\bar g_{\b\l}}{\pd\bar g_{\m\n}}\bar g_{\a\k}\hatl^{\k\l}=\d^{(\m}_{\a}\hatl^{\n)}{}_{\b}+\d^{(\m}_\b\hatl^{\n)}{}_{\a}\;.
\ee 

Let us consider now a functional dependence of the covariant derivative $\hatl^{\a\b}{}_{|\l}$ on the metric tensor. We notice that the perturbation $\hatl^{\a\b}$ is a tensor density of weight $-1$. Therefore, its covariant derivative has one more term as compared with that of a tensor of a second rank. More specifically,
\be\la{rr1d}
\hatl^{\a\b}{}_{|\k}=\hatl^{\a\b}{}_{,\k}+\bar\G^\a{}_{\s\k}\hatl^{\s\b}+\bar\G^\b{}_{\s\k}\hatl^{\s\a}-\bar\G^\s{}_{\s\k}\hatl^{\a\b}\;.
\ee
It reveals that the derivative $\hatl^{\a\b}{}_{|\k}$ depends merely on the Christoffel symbols and is independent of the metric tensor $\bar g_{\a\b}$. Hence, the partial derivative
\be\la{m3d5}
\frac{\pd\hatl^{\a\b}{}_{|\l}}{\pd\bar g_{\m\n}}=0\;.
\ee
It agrees with our postulate that the metric tensor and the Christoffel symbols are true independent variables along with the tensor density $\hatl^{\a\b}$ and its covariant derivative $\hatl^{\a\b}{}_{|\l}$.
Equation (\ref{rr1}) given in terms of the independent variables, reads (with the divergence term discarded)
\be\la{rr1c}
{\cal F}^{\rm\st G}=\frac{\bar g^{\k\l} \bar g_{\b\r}}{\sqrt{-\bar g}}\lef(\bar g_{\a\l}\hatl^{\r\s}{}_{|\k}\hatl^{\a\b}{}_{|\s}-\frac12 \bar g_{\a\s}\hatl^{\r\s}{}_{|\k}\hatl^{\a\b}{}_{|\l}+\frac14 \bar g_{\a\s}\hatl^{\a\s}{}_{|\k}\hatl^{\b\r}{}_{|\l}\ri)\;,
\ee
where we have discarded the total divergence. 

Variational derivative in the form of (\ref{er8}) taken from (\ref{rr1c}) engages partial derivatives with respect to the background metric tensor, $\bar g_{\m\n}$, and those with respect to the background Christoffel symbols, $\bar\Gamma^\a{}_{\b\g}$. 
The partial derivative with respect to the metric tensor yields
\ba\la{rr1e}
\frac{1}{\sqrt{-\bar g}}\frac{\pd{\cal F}^{\rm\st G}}{\pd\bar g_{\m\n}}&=&-\frac12\bar g^{\m\n}\lef(l^{\r\s|\l}l_{\l\r|\s}-\frac12 l^{\r\s|\l} l_{\r\s|\l}+\frac14 l^{|\l}l_{|\l}\ri)\\\nonumber
&&-l^{\m\s|\r}l^\n{}_{\s|\r}+l^{\m\s|\r}l^\n{}_{\r|\s}+\frac12 l^{\r\s|\m}l_{\r\s}{}^{|\n}+\frac12 l^{\m\n|\s}l_{|\s}-\frac 14 l^{|\m}l^{|\n}\;,
\ea
where we have used (\ref{m3d5}).
The partial derivative with respect to the Christoffel symbols taken from $\hatl^{\r\s}{}_{|\k}$ is calculated from its presentation in the form of (\ref{rr1d}) with the help of (\ref{er10}). It gives
\ba
\la{rr2} 
\frac{\pd \hatl^{\r\s}{}_{|\k}}{\pd\bar\G^\a{}_{\l\g}}&=&\d^{\r}_\a\d^{(\l}_\k \hatl^{\g)\s}+\d^{\s}_\a\d^{(\l}_\k \hatl^{\g)\r}-\d^{(\l}_\a\d^{\g)}_\k \hatl^{\r\s}\;.
\ea
After making use of this formula, the partial derivative of ${\cal F}^{\rm\st G}$ with respect to the Christoffel symbols results in
\ba
\la{rr3} 
\frac{1}{\sqrt{-\bar g}}\frac{\pd{\cal F}^{\rm\st G}}{\pd\bar\G^\a{}_{\m\g}}&=&2l^{\r\g}l^\m{}_{(\r|\a)}+2l^{\r\m}l^\g{}_{(\r|\a)}-2l_{\r\a}{}^{|(\m}l^{\g)\r}\\\nonumber
&-&2\d^{(\m}_\a l^{\g)}{}_{\r|\s}l^{\r\s}+l_\a{}^{(\g}l^{|\m)}+l^{\r\s}l_{\r\s}{}^{|(\m}\d^{\g)}_\a-\frac12\d^{(\m}_\a l^{\g)}l\;.
\ea
It allows us to calculate the linear combination of the partial derivatives entering definition of variational derivative (\ref{er8}), namely,
\ba
\la{rr4}
-\frac12\frac{1}{\sqrt{-\bar g}}\lef(\bar g^{\s\n}\frac{\pd{\cal F}^{\rm\st G}}{\pd\bar\G^\s{}_{\m\g}}+\bar g^{\s\m}\frac{\pd{\cal F}^{\rm\st G}}{\pd\bar\G^\s{}_{\n\g}}-\bar g^{\s\g}\frac{\pd{\cal F}^{\rm\st G}}{\pd\bar\G^\s{}_{\m\n}}\ri)
&=&
2l_\r{}^{(\m} l^{\n)\r|\g}-2l^{\g\r|(\m}l^{\n)}{}_\r-l_\r{}^\g l^{\m\n|\r}-\frac12 l^{\m\n}l^{|\g}\\\nonumber
&+&\,\bar g^{\m\n}\lef(l^{\r\s}l^{\g}{}_{\r|\s}-\frac12l^{\r\s} l_{\r\s}{}^{|\g}+\frac14 ll^{|\g}\ri)\;.
\ea
After making use of (\ref{rr1e}) and (\ref{rr4}) in expression (\ref{rtb5}) for variational derivative defined by the rule (\ref{er8}), the stress-energy tensor of the gravitational field perturbations takes on the following form 
\ba\la{qq5f}
16\pi\mt_{\m\n}&=&-\frac12\bar g_{\m\n}\lef(l^{\r\s|\g}l_{\g\r|\s}-\frac12 l^{\r\s|\g} l_{\r\s|\g}+\frac14 l^{|\g}l_{|\g}\ri)\\\nonumber
&&-l_{\m\s|\r}l_\n{}^{\s|\r}+l_{\m\s|\r}l_\n{}^{\r|\s}+\frac12 l_{\r\s|\m}l^{\r\s}{}_{|\n}+\frac12 l_{\m\n|\s}l^{|\s}-\frac 14 l_{|\m}l_{|\n}
\\\nonumber
&&+\bar g_{\m\n}\lef(l^{\r\s}l^{\g}{}_{\r|\s}-\frac12l^{\r\s} l_{\r\s}{}^{|\g}+\frac14 ll^{|\g}\ri)_{|\g}\\\nonumber
&&+
\lef(2l_{\r(\m} l_{\n)}{}^{\r|\g}-l_{\n\r}l^{\g\r}{}_{|\m}-l_{\m\r}l^{\g\r}{}_{|\n}-l^{\g\r} l_{\m\n|\r}-\frac12 l_{\m\n}l^{|\g}\ri)_{|\g}\;.
\ea
It apparently depends on the second derivatives of the gravitational perturbation, $l_{\m\n}$ which is a consequence of our covariant field-theoretical approach for description of perturbations of gravitational field \citep{Deser_1970GReGr,1984CMaPh..94..379G}.  Most of alternative approaches to construct the stress-energy tensor of gravitational field perturbations without second derivatives unavoidably make it non-covariant that is coordinate-dependent. For this reason such ``tensors'' of gravitational field perturbations are commonly-known as pseudo-tensors \citep{Szabados_1992}. Babak and Grishchuk \citep{2000NuPhS..80C1204B,2000PhRvD..61b4038B} proposed an interesting method to constructing a {\it tensor} of gravitational field perturbations which does not include the second derivatives of the field perturbations. The method requires an introduction of an additional (Lagrange multiplier) term to the gravitational field Lagrangian which is proportional to the Riemann tensor of the background manifold. This procedure has been worked out in \citep{2000NuPhS..80C1204B,2000PhRvD..61b4038B} for the case of Minkowski-flat background. Further research should be conducted to extend it to the case of an arbitrary curved background manifold.    

Significant number of the second derivatives in expression (\ref{qq5f}) can be eliminated on-shell by making use of the covariant field equation (\ref{axz4}). To this end we write down the terms with the second covariant derivatives in (\ref{qq5f}) and express the commutator of the second-order derivatives from the metric tensor perturbation in terms of the Riemann tensor, 
\be\la{com8f}
l^\a{}_{\r|\s\b}=l^\a{}_{\r|\b\s}-l^\g{}_{\r}\bar R^\a{}_{\g\s\b}+l^\a{}_{\g}\bar R^\g{}_{\r\s\b}\;.
\ee
A useful consequence of this equation is a contraction with respect to index $\a$ which gives
\be\la{com9f}
l^\a{}_{\r|\s\a}=A_{\r|\s}+l^\a{}_{\r}\bar R_{\s\a}+l^\a{}_{\g}\bar R^\g{}_{\r\s\a}\;,
\ee
where $A^\a\eq l^{\a\b}{}_{|\b}$, and $A_\a=\bar g_{\a\b}A^\b$. Straightforward but tedious rearrangement of the second-order derivatives from the metric tensor perturbations with the help of (\ref{com8f}), (\ref{com9f}) allows us to put (\ref{qq5f}) into the following form  
\ba\la{5f}
16\pi\mt_{\m\n}&=&2l_{\m\r|\s}l_\n{}^{(\r|\s)}-l^{\r\s}{}_{|\m}l_{\n\r|\s}-l^{\r\s}{}_{|\n}l_{\m\r|\s}
+\frac12 l_{\r\s|\m}l^{\r\s}{}_{|\n}-\frac 14 l_{|\m}l_{|\n}-l^{\r\s}l_{\m\n|\r\s}
\\\nonumber
&&+\frac12\bar g_{\m\n}\lef(l^{\r\s|\g}l_{\g\r|\s}-\frac12 l^{\r\s|\g} l_{\r\s|\g}+\frac14 l^{|\g}l_{|\g}\ri)+2l^{\r}{}_{(\m}A_{\n)|\r}-l_{\m\n}A^\r{}_{|\r}-l_{\m\n|\r}A^\r
\\\nonumber
&&+8\pi\lef[4l^{\r}{}_{(\m}\Theta_{\n)\r}-l_{\m\n}\Theta-\bar g_{\m\n}\lef(l^{\r\s}\Theta_{\r\s}-\frac{l}2\Theta\ri)\ri]
+2l^\r{}_\m l^{\s}{}_\n\bar R_{\r\s}+2l^{\a\b}l^{\r}{}_{(\m}\bar R_{\n)\a\b\r}
\ea
where 
\be\la{thet3}
\Theta_{\a\b}=\mathfrak{T}_{\a\b}-\frac1{8\pi}\lef(F^{\rm m}_{\a\b}+F^{\rm q}_{\a\b}\ri)\;,
\ee
and the spur, $\Theta\eq\bar g^{\a\b}\Theta_{\a\b}$. 

As we can see, most of the second-order derivatives from the metric tensor perturbations have vanished. The remaining second-order derivatives remain as the very last term in the first line of (\ref{5f})) and in the terms which depend on the gauge function $A^\a$ in the second line of (\ref{5f}). On-shell expression (\ref{5f}) of the tensor of gravitational field perturbations also depends on the Riemann (curvature) tensor of the background manifold. Had the background manifold $\bar\M$ been flat such terms would not be present. The last but not least notice is that $\mt_{\m\n}$ includes on-shell coupling of the gravitational field perturbations with the perturbations of the background matter as well as with the bare perturbations. These terms are proportional to the terms with $\Theta_{\a\b}$ which come from $F^{\st\rm G}_{\a\b}$ because on shell, $F^{\st\rm G}_{\a\b}=\Theta_{\a\b}$, due to the field equations (\ref{(2.17)}).

We expressed the stress-energy tensor of gravitational field perturbations $\mt_{\m\n}$ in terms of the variable $l_{\a\b}$. It is instructive to reformulate it in terms of the perturbation of the metric tensor, $\varkappa_{\a\b}$, defined in (\ref{m6z2a}) and related to $l_{\a\b}$ according to (\ref{ok9}). Calculations are done in two steps. First, we replace all $l_{\a\b}$ in (\ref{qq5f}) with $\varkappa_{\a\b}$ and retain only linear terms in (\ref{ok9}). Second step is to replace $l_{\a\b}$ in the linear operator (\ref{xpq1}) with $\varkappa_{\a\b}$ by taking into account quadratic terms in expansion (\ref{ok9}). All quadratic terms with respect to $\varkappa_{\a\b}$ are combined together to produce the stress-energy tensor of gravitational field perturbations expressed in terms of the dynamic variable $\varkappa_{\a\b}$. This tensor coincides (up to the sign convention) with that given in the textbook by S. Weinberg \citep[equation 7.6.15]{weinberg_1972} which also depends on the second-order derivatives from the metric tensor perturbations. The advantage of our perturbation scheme as compared with S. Weinberg's book \citep{weinberg_1972} is that we have worked out an iterative procedure of calculation of the field perturbations at any order of approximation. In particular, we can derive an exact analytic form of
the gravitational stress-energy tensor $\mt_{\mu\nu}$ which reads \citep{1984CMaPh..94..379G} 
\ba
\label{GPP-tei}
 \mt_{\mu\nu}&=& \frac1{8\pi}\lef(\delta^\rho_\mu \delta^\s_\nu - \frac12
\bar g_{\mu\nu}\bar
g^{\rho\s}\ri)\lef(
\mathfrak{G}^\a {}_{\rho\beta}\mathfrak{G}^\beta{}_{\s\a }-\mathfrak{G}^\a {}_{\rho\s}\mathfrak{G}^\beta{}_{\a \beta}\ri)\\
\nonumber
& +&  \frac1{8\pi}\lef[\frac12 \hatl_{\mu\nu}\bar g^{\rho\beta}\mathfrak{G}^\a {}_{\a \beta}
-\frac12\bar
g_{\mu\nu}\hatl^{\a \beta}\mathfrak{G}^\rho{}_{\a \beta}-
\hatl^{\rho}{}_{(\mu}\mathfrak{G}^\a {}_{\nu)\a } + \hatl^{\rho}{}_{\beta}\bar
g_{\a (\mu}\mathfrak{G}^\a {}_{\nu)\beta} +
\hatl^\beta{}_{(\mu}\mathfrak{G}^\rho{}_{\nu)\beta} - \hatl^{\beta}{}_{(\mu}\,\bar
g_{\nu)\a }\bar g^{\rho\s} \mathfrak{G}^\a {}_{\beta\s}\ri]_{|\rho}\;,
\ea
where
$\mathfrak{G}^\a {}_{\beta\gamma} \equiv \Gamma^\a {}_{\beta\gamma} - \Bar
\Gamma^\a {}_{\beta\gamma}$ is the difference between the Christoffel
symbols of the perturbed, $\M$, and the background, $\bar\M$, manifolds
\ba
\la{GBarG}
\mathfrak{G}^\a {}_{\beta\gamma}&=&\frac{1}{2} g^{\a \rho}\lef(\varkappa_{\rho\beta|\gamma}
+\varkappa_{\rho\gamma|\beta} - \varkappa_{\beta\gamma|\rho}\ri)\;,
\ea
where $\varkappa_{\mu\nu}\equiv g_{\mu\nu}-\bar g_{\mu\nu}$. We emphasize that the geometric object
$\mathfrak{G}^\a {}_{\beta\gamma}$ is a tensor with respect to coordinate transformation on the background manifold since it
represents the difference between the two Christoffel symbols \citep{waldorf}. It does
not mean, of course, that we employ a bi-metric theory of gravity being different from
general theory of relativity. The background metric $\bar g_{\mu\nu}$ is simply the lowest (unperturbed)
state of the gravitational field which dynamical properties are described by the full
metric $g_{\mu\nu}$. Since both the background metric $\bar g_{\mu\nu}$, its perturbation
$\varkappa_{\mu\nu}$, and the object $\mathfrak{G}^\a {}_{\beta\gamma}$ are tensors,
$\mt_{\mu\nu}$ is a stress-energy tensor of the gravitational field perturbations. It defines energy, a linear momentum, and other physical characteristics of the perturbations at each point of the background spacetime \citep{petkatz}. Expansion of (\ref{GPP-tei}) in Taylor series with respect to perturbations and leaving only quadratic terms yields (\ref{qq5f})

\subsection{Stress-energy tensor of dark matter perturbations}\la{setdmp}

The part of the stress-energy tensor describing the dark matter perturbation is given in (\ref{jkw5}) by $\tau^{\rm m}_{\m\n}$ that, according to (\ref{qq5x}), is calculated as a variational derivative 
\be\la{axrt5}
16\pi\tau^{\rm m}_{\a\b}=\frac{1}{\sqrt{-\bar g}}\bar g_{\a\m}\bar g_{\b\n}\frac{\d{\cal F}^{\rm m}}{\d\bar g_{\m\n}}\;,
\ee
from the Lagrangian density given by
\be\la{ax5}
{\cal F}^{\rm m}\eq\hatl^{\r\s}F^{\rm m}_{\r\s}-\frac12\hatl F^{\rm m}+\sqrt{-\bar g}\phi F^{\rm m}_\Phi\;,
\ee
where the individual terms entering the right side of (\ref{ax5}) are taken from (\ref{mt6a}) and (\ref{mtj10}) respectively, and $F^{\rm m}\eq\bar g^{\a\b}F^{\rm m}_{\a\b}$. We single out the total divergence in (\ref{ax5}), discard it, and brings (\ref{ax5}) to the following form
\be\la{ax345}
{\cal F}^{\rm m}\eq\hatl^{\r\s}F^{\rm m}_{\r\s}-\frac12\hatl F^{\rm m}-8\pi\sqrt{-\bar g}\phi_\a Y^\a\;,
\ee
where the total divergence has been dropped off, $\p_\a\eq\p_{|\a}$, and the current $Y^\a$ is given in (\ref{mt10a}). After reducing similar terms equation (\ref{ax345}) takes on the following form
\ba\la{ax6}  
{\cal F}^{\rm m}&\eq&-8\pi\sqrt{-\bar g}\lef(\frac{\bar\r_{\rm m}}{\bar\m_{\rm m}}\p^\a\p_\a-3\bar\r_{\rm m}l^{\a\b}{\bar u}_\a\p_\b\ri)-4\pi\sqrt{-\bar g}\lef(\bar p_{\rm m}-{\bar\e}_{\rm m}\ri)\lef(l^{\a\b}l_{\a\b}-\frac12l^2\ri) \\\nonumber
&&-8\pi\sqrt{-\bar g}\frac{\bar\r_{\rm m}}{\bar\m_{\rm m}}\lef(1-\frac{c^2}{c^2_{\rm s}}\ri)\lef[\lef(\bar u^\a\p_\a\ri)^2-\frac32\bar\m_{\rm m}\mathfrak{q}\bar u^\a\p_\a+\frac12\m_{\rm m}^2\mathfrak{q}^2\ri]\;.
\ea
where again we have used notation $\p_\a\eq\p_{|\a}$ and $l\eq\bar g^{\a\b}l_{\a\b}$.

Taking variational derivative in (\ref{axrt5}) is rather straightforward but tedious procedure. Because the Lagrangian ${\cal F}^{\rm m}$ depends neither on the Christoffel symbols nor on the curvature tensor, the variational derivative (\ref{axrt5}) is reduced to a partial derivative with respect to the metric tensor $\d{\cal F}^{\rm m}/\d g_{\m\n}=\pd {\cal F}^{\rm m}/\pd g_{\m\n}$. Calculation of the partial derivative is done with the help of the chain rule and equations in Appendix \ref{cvdhl}. It yields the stress-energy tensor of dark matter
\ba\la{ax7}
\tau^{\rm m}_{\m\n}&=&\frac{\bar\r_{\rm m}}{2\bar\m_{\rm m}}\p_\m\p_\n-\frac{\bar\r_{\rm m}}{4\bar\m_{\rm m}}\lef[\p^\a\p_\a+\lef(1-\frac{c^2}{c^2_{\rm s}}\ri)\lef(\bar u^\a\p_\a\ri)^2\ri]\bar g_{\m\n}\\\nonumber
&-&\frac{\bar\r_{\rm m}}{2\bar\m_{\rm m}}\lef(1-\frac{c^2}{c^2_{\rm s}}\ri)\lef[\lef(\frac32\bar\m_{\rm m}\mathfrak{q}-2\bar u^\a\p_\a\ri)\bar u_{(\m}\p_{\n)}-\frac34\bar\m_{\rm m}\bar u^\a\p_\a l_{\m\n}\ri]\\\nonumber
&+&\frac{\bar\r_{\rm m}}{4\bar\m_{\rm m}}\lef(1-\frac{c^2}{c^2_{\rm s}}\ri)\lef(\p^\a\p_\a-3\bar\m_{\rm m}l^{\a\b}\bar u_\a\p_\b+\frac32\bar\m_{\rm m}l\bar u^\a\p_\a\ri)\bar u_\m\bar u_\n\\\nonumber
&+&\frac{\bar\r_{\rm m}}{4\bar\m_{\rm m}}\lef[\lef(1-\frac{c^2}{c^2_{\rm s}}\ri)\lef(3-\frac{c^2}{c^2_{\rm s}}\ri)-\bar\r_{\rm m}\bar\m_{\rm m}\frac{c^2}{c^2_{\rm s}}\frac{\pd\ln c^2_{\rm s}}{\pd\bar p_{\rm m}}\ri]
\lef[\lef(\bar u^\a\p_\a\ri)^2-\frac32\bar\m_{\rm m}\mathfrak{q}\bar u^\a\p_\a+\frac12\m_{\rm m}^2\mathfrak{q}^2\ri]\bar u_\m\bar u_\n
\\\nonumber
&-&\frac18\bar\r_{\rm m}\bar\m_{\rm m}\lef(1-\frac{c^2}{c^2_{\rm s}}\ri)\lef[2\mathfrak{q}l_{\m\n}-\mathfrak{q}^2\bar g_{\m\n}+\lef(l^{\a\b}l_{\a\b}-\frac12l^2+2\mathfrak{q}l\ri)\bar u_\m\bar u_\n\ri]\\\nonumber
&-&\frac12\lef(\bar p_{\rm m}-\bar\e_{\rm m}\ri)\lef[l_{\a\m}l_{\n}{}^{\a}-\frac12ll_{\m\n}-\frac14\lef(l^{\a\b}l_{\a\b}-\frac12l^2\ri)\bar g_{\m\n}\ri]
\;,
\ea
where we have used thermodynamic relation (\ref{pf2}) to make a replacement $\bar\e_{\rm m}+{\bar p}_{\rm m}=\bar\r_{\rm m}\bar\m_{\rm m}$.

As we have modelled dark matter by the ideal fluid, equation (\ref{ax7}) represents the stress-energy tensor of sound waves propagating on the background cosmological manifold. This tensor depends on the speed of sound, $c_{\rm s}$, which enters denominators in some terms of (\ref{ax7}). It may cause an impression that in case of dust, when $c_{\rm s}\rightarrow 0$, the tensor $\tau^{\rm m}_{\m\n}$ is divergent. This impression is not true as the numerators of the corresponding terms also approach to zero with the same rate as the denominator. It leaves $\tau^{\rm m}_{\m\n}$ well-defined even in case of a model of dark matter consisting of non-interacting dust particles.

\subsection{Stress-energy tensor of dark energy perturbations}\la{setdep}

The part of the stress-energy tensor describing the dark energy perturbation is given in (\ref{jkw5}) by $\t^{\rm q}_{\m\n}$ that, according to (\ref{qq5x}), is calculated as a variational derivative 
\be\la{axrh5}
16\pi\tau^{\rm q}_{\a\b}=\frac{1}{\sqrt{-\bar g}}\bar g_{\a\m}\bar g_{\b\n}\frac{\d{\cal F}^{\rm q}}{\d\bar g_{\m\n}}\;,
\ee
from the Lagrangian density given by
\be\la{ax5q}
{\cal F}^{\rm q}\eq\hatl^{\r\s}F^{\rm q}_{\r\s}-\frac12\hatl F^{\rm q}+\sqrt{-\bar g}\psi F^{\rm q}_\Psi\;,
\ee
where the individual terms entering the right side of (\ref{ax5q}) are taken from (\ref{mt6c}) and (\ref{mt10c}) respectively, and $F^{\rm q}\eq\bar g^{\a\b}F^{\rm q}_{\a\b}$. We single out the total divergence and bring (\ref{ax5q}) to the following form
\be\la{ax99}
{\cal F}^{\rm q}\eq\hatl^{\r\s}F^{\rm q}_{\r\s}-\frac12\hatl F^{\rm q}-8\pi\sqrt{-\bar g}\psi_\a Z^\a
-4\pi\sqrt{-\bar g}\psi\lef(l\frac{\pd\bar W}{\pd\bar\Psi}+2\psi\frac{\pd^2\bar W}{\pd\bar\Psi^2}\ri)\;,
\ee
where the total divergence has been dropped off. More explicitly,
\ba\la{ax6q}  
{\cal F}^{\rm q}&\eq&-8\pi\sqrt{-\bar g}\lef[\psi^{\a}\psi_{\a}+\frac32l\psi\frac{\pd\bar W}{\pd\bar\Psi}+\psi^2\frac{\pd^2\bar W}{\pd\bar\Psi^2}-3\bar\m_{\rm q}l^{\a\b}{\bar u}_\a\psi_{\b}\ri]+8\pi\sqrt{-\bar g}\bar W(\bar\Psi)\lef(l^{\a\b}l_{\a\b}-\frac12l^2\ri)\;,
\ea
where $\psi_\a\eq\psi_{|\a}$.
Taking variational derivative from the left side of (\ref{ax6q}) with respect to $\bar g_{\m\n}$, we obtain the stress-energy tensor of dark energy perturbation
\ba\la{ax7q}
\tau^{\rm q}_{\m\n}&=&\frac12\psi_\m\psi_\n-\frac14\lef(\psi^{\a}\psi_{\a}+\psi^2\frac{\pd^2\bar W}{\pd\bar\Psi^2}\ri)\bar g_{\m\n}-\frac34l_{\m\n}\psi\frac{\pd\bar W}{\pd\bar\Psi}
+\bar W(\bar\Psi)\lef[l_{\a\m}l_{\n}{}^\a-\frac12ll_{\m\n}-\frac14\lef(l^{\a\b}l_{\a\b}-\frac12l^2\ri)\bar g_{\m\n}\ri]
\;.
\ea
This tensor depends on the potential $\bar W(\bar\Psi)$ of the scalar field and on its first and second derivatives. The potential has been kept arbitrary which makes expression (\ref{ax7q}) rather general and applicable to discussion of a wide spectrum of physical situations.

\section{Post-Friedmanian Equations of Motion in Cosmology}\la{sec6}

In this section we shall derive equations of motion of the baryonic matter in the universe governed by dark matter and dark energy. Baryonic matter falls freely in the gravitational field produced by dark matter and dark energy primordial perturbations which are responsible for the formation of the large scale structure in the universe \citep{weinberg_2008,rubak_2011}. Since luminous matter is made of baryons, its astronomical observations traces the gravitational potential of dark matter and helps us to identify where it confines and clumps to clusters. We shall also take into account the self-gravitational interaction of the baryonic matter, thus, extending the post-Newtonian treatment of equations of motion in asymptotically-flat spacetime \citep{kopeikin_2011book,futamase_2007LRR} to the realm of cosmology where FLRW background metric is not flat.

\subsection{General Formulation}

Let us consider a background spacetime manifold, $\bar\M$, with the effective Lagrangian 
\be\la{r5v7}
\lag^{\rm eff}=\lag^{\rm eff}\lef(\bar g^{\mu\nu},\bar \G^\a_{\b\g};\bar\Phi^A,\bar\Phi_\a;\Theta^B,\Theta^B_\a;\hatl^{\mu\nu},\hatl^{\m\n}{}_{|a};\phi^A,\phi^A_\a\ri)\;,
\ee
depending on a set of the independent dynamic variables and their conjugated counterparts which are covariant derivatives on the background manifold. We have proved in section \ref{gife} that the effective Lagrangian $\lag^{\rm eff}$ is gauge-invariant {\it on shell} modulo a total divergence. The gauge invariance of $\lag^{\rm eff}$ suggests that its Lie derivative along an arbitrary vector field, $\xi^\a$, must be also nil modulo a total divergence: $\pounds_{\bm\xi}\lag^{\rm eff}=\pd_\a U^\alpha$, where $U^\alpha$ is a vector field. Because a total divergence added to the Lagrangian do not affect the field equations we drop it out of the subsequent equations. 

We compute the Lie derivative of the effective Lagrangian by making use of (\ref{li4}) that reduce calculation of the Lie derivative to that of variational derivatives modulo a total divergence. After dropping off the divergence, we have
\ba
\la{zim1}
\pounds_{\bm\xi}\lag^{\rm eff}&=&\frac{\de\lag^{\rm eff}}{\de\bar g^{\a\b}}\pounds_{\bm\xi}\bar g^{\a\b}+\frac{\de\lag^{\rm eff}}{\de\bar\Phi^A}\pounds_{\bm\xi}\bar\Phi^A+\frac{\de\lag^{\rm eff}}{\de\hatl^{\mu\nu}}\pounds_{\bm\xi}\hatl^{\mu\nu}+\frac{\de\lag^{\rm eff}}{\de\phi^A}\pounds_{\bm\xi}\phi^A+\frac{\de\lag^{\rm eff}}{\de\Theta^B}\pounds_{\bm\xi}\Theta^B\;.
\ea
Field equations (\ref{tyu7}), (\ref{xsk}), (\ref{ko9a}) describing evolution of the dynamic field perturbations $\hatl^{\mu\nu}$, $\phi^A$, $\theta^B$ on the background manifold exterminate the last three terms in the right side of (\ref{zim1}). The first term in the right side of (\ref{zim1}) can be written down as follows
\be\la{nkz2}
\frac{\de\lag^{\rm eff}}{\de\bar g^{\a\b}}\pounds_{\bm\xi}\bar g^{\a\b}=-\sqrt{-\bar g}\Lambda_{\a\b}\xi^{\a|\b}\;,
\ee
where we have used definitions (\ref{kwn8}) and equation for the Lie derivative of the background metric
\be\la{in483}
\pounds_{\bm\xi}\bar g^{\a\b}=-\xi^{\a|\b}-\xi^{\b|\a}\;,
\ee 

In order to develop a second term in the right side of (\ref{zim1}), we have to know the Lie derivative of the field, $\bar\Phi^A$, which depends on its geometric properties. In a particular case of a tensor density $\bar\Phi^A\eq\lef(\bar\Phi^A\ri)^{\mu_1\ldots\mu_p}_{\nu_1\ldots\nu_q}$ of weight $m$, the Lie derivative is given by (\ref{gde1}) that can be written symbolically as follows
\be
\la{zim2}
\pounds_{\bm\xi}\bar\Phi^A=\xi^\a\bar\Phi^A_{\a}+\bar{K}^A_{\a\b}\xi^{\a|\b}\;,
\ee
where $\bar\Phi^A_{\a}\equiv\lef(\bar\Phi^A\ri)^{\mu_1\ldots\mu_p}_{\nu_1\ldots\nu_q|\a}$, $\bar{K}^A_{\a\b}=\bar{K}^{A\s}_\a\bar g_{\s\b}$, and
\ba
\la{zim2a}
\bar{K}^{A\s}_\a&\equiv& m\d^\s_\a\lef(\bar\Phi^A\ri)^{\mu_1\ldots\mu_p}_{\nu_1\ldots\nu_q}\\\nonumber&&
-\delta^{\mu_1}_\a\lef(\bar\Phi^A\ri)^{\s\mu_2\ldots\mu_p}_{\nu_1\ldots\nu_q}-\ldots-\delta^{\mu_p}_\a\lef(\bar\Phi^A\ri)^{\mu_1\ldots\mu_{p-1}\s}_{\nu_1\ldots\nu_q}
+\delta^{\s}_{\nu_1}\lef(\bar\Phi^A\ri)^{\mu_1\ldots\mu_p}_{\a\nu_2\ldots\nu_q}+\ldots+\delta^{\s}_{\nu_q}\lef(\bar\Phi^A\ri)^{\mu_1\ldots\mu_p}_{\nu_1\ldots\nu_{q-1}\a}\;.
\ea
Making use of definition (\ref{yu3}) and (\ref{zim2}) we can present the second term in the right side of (\ref{zim1}) in the following form
\be\la{zzh0}
\frac{\de\lag^{\rm eff}}{\de\bar\Phi^A}\pounds_{\bm\xi}\bar\Phi^A=\frac12\sqrt{-\bar g}\Sigma^{\rm\st M}_A\lef(\xi^\a\bar\Phi^A_{\a}+\bar{K}^A_{\a\b}\xi^{\a|\b}\ri)\;.
\ee

Substituting (\ref{nkz2}), (\ref{zzh0}) to the right side of (\ref{zim1}) results in
\be\la{klop9}
\pounds_{\bm\xi}\lag^{\rm eff}=\frac12\sqrt{-\bar g}\Sigma^{\rm\st M}_A\bar\Phi^A_{\a}\xi^\a+\sqrt{-\bar g}\lef(-\Lambda_{\a\b}+\frac12\Sigma^{\rm\st M}_A\bar{K}^A_{\a\b}\ri)\xi^{\a|\b}\;.
\ee
Applying the Leibniz rule to change the order of differentiation in the terms depending on $\xi^{\a|\b}$, we can recast (\ref{klop9}) to the following form
\be\la{mnw7}
\pounds_{\bm\xi}\lag^{\rm eff}=\sqrt{-\bar g}\lef[\frac12\Sigma^{\rm\st M}_A\bar\Phi^A_{\a}+\Lambda_{\a\b}{}^{|\b}-\frac12\lef(\Sigma^{\rm\st M}_A\bar{K}^{A}_{\a\b}\ri)^{|\b}\ri]\xi^\a+\sqrt{-\bar g}W_\b{}^{|\b}\;,
\ee
where the vector field
\be\la{lon2}
W_\b\equiv\lef(-\Lambda_{\a\b}+\frac12\Sigma^{\rm\st M}_A\bar{K}^{A}_{\a\b}\ri)\xi^\a\;.
\ee
The last term in (\ref{mnw7}) is reduced to the total divergence of a vector density
\be\la{bys3}
\sqrt{-\bar g}W_\b{}^{|\b}=\partial_\b\lef(\sqrt{-\bar g}W^\b\ri)\;,
\ee
where $W^\b=\bar g^{\a\b}W_\a$. The Lie derivative (\ref{mnw7}) of the effective Lagrangian vanishes modulo the divergence of the vector field $U^\beta\equiv \sqrt{-\bar g}W^\b$ if, and only if, the combination of terms enclosed to the square brackets in (\ref{mnw7}) is nil. It yields the equations of motion of matter 
\be
\la{zim4}
\Lambda_{\a\b}{}^{|\b}=-\frac12\Sigma^{\rm\st M}_A\bar\Phi^A_{\a}+\frac12\lef(\Sigma^{\rm\st M}_A\bar{K}^{A}_{\a\b}\ri)^{|\b}\;.
\ee
It should be compared with the law of conservation of matter in flat background spacetime, $\Lambda_{\a\b}{}^{,\b}=0$, with the right side equal to zero \citep{LanLif}. The presence of the background matter fields $\bar\Phi^A$ on the curved background manifold makes the right side of (\ref{zim4}) different from zero. This result was established in \cite{1984CMaPh..94..379G}. 

Equation (\ref{zim4}) can be interpreted as the integrability condition of the gravitational field equation (\ref{(2.17)}). Taking a covariant derivative from both sides of the field equation (\ref{(2.17)}) and applying the equations of motion (\ref{zim4}) yields
\be
\la{zim5}
\lef(F^{\rm\st G}_{\a\b}+F^{\rm\st M}_{\a\b}\ri)^{|\b}=-4\pi\lef[\Sigma^{\rm\st M}_A\bar\Phi^A_{\a}-\lef(\Sigma^{\rm\st M}_A\bar{K}^{A}_{\a\b}\ri)^{|\b}\ri]\;.
\ee
In the linear approximation, when all quadratic and higher-order terms with respect to the perturbations are discarded ($\Sigma^{\rm\st M}_A\rightarrow 0$), the covariant divergence $(F^{\rm\st G}_{\a\b}+F^{\rm\st M}_{\a\b})^{|\b}=0$. It agrees with the assumption that the stress-energy tensor of the bare perturbation is conserved in the linearised perturbative order, $\T_{\a\b}{}^{|\b}=0$.
Now we are set to start calculating equations of motion of matter of the baryonic matter in FLRW universe goiverned by the dark matter and dark energy which we consider in the next few subsections. 

\subsection{Equations of motion in the universe governed by dark matter and dark energy}

The dark matter and dark energy components of matter that governs the temporal evolution of the universe are modelled by two scalar fields $\Phi^1\eq\Phi$ and $\Phi^2\eq\Psi$. For scalar fields the tensor $\bar{K}^A_{\a\b}\eq 0$ and, consequently, the second term in the right side of (\ref{zim4}) is identically nil. Therefore, equations of motion of matter (\ref{zim4}) can be written more explicitly in the following form
\be\la{qxea1}
\mathfrak{T}_{\m\n}{}^{|\n}+\mathfrak{t}_{\m\n}{}^{|\n}+\t^{\rm m}_{\m\n}{}^{|\n}+\t^{\rm q}_{\m\n}{}^{|\n}=\frac12\lef(\bar\m_{\rm m}\Sigma^{\rm m}+\bar\m_{\rm q}\Sigma^{\rm q}\ri)\bar u_\m\;,
\ee
where we have used equation (\ref{pf8tn}) for expressing the gradients of the scalar fields $\bar\Phi$ and $\bar\Psi$ in terms of the background four-velocity $\bar u^\a$ as well as equations (\ref{pmy6}), (\ref{qq5a}), (\ref{jkw5}) defining the effective stress-energy tensor $\Lambda_{\a\b}$.
Equation (\ref{qxea1}) is a covariant equation of motion of the baryonic matter described by the stress-energy tensor $\mathfrak{T}_{\m\n}$, in the presence of dynamic perturbations of gravitational field, dark matter and dark energy. In case of asymptotically flat spacetime the right side of (\ref{qxea1}) would vanish while in the left side of (\ref{qxea1}) only the first two terms would remain among which the stress-energy tensor of gravitational field, $\mathfrak{t}_{\m\n}$, would be made of the perturbations of gravitational field caused by the baryonic matter itself. 

In FLRW universe with dark matter and dark energy, more terms appear in equations of motion (\ref{qxea1}) which should be properly treated. Our goal is to calculate the explicit form of $\Sigma^{\rm m}$ and $\Sigma^{\rm q}$ as well as the covariant divergences of stress-energy tensors entering (\ref{qxea1}). We split the process of calculation in three parts. First, we calculate the divergence, $\mathfrak{t}_{\m\n}{}^{|\n}$, of the stress-energy tensor of gravitational field, then, we proceed to calculation of the divergence, $\t^{\rm m}_{\m\n}{}^{|\n}$, of the stress-energy tensor of dark matter, and that $\t^{\rm q}_{\m\n}{}^{|\n}$ of dark energy. It becomes clear in the course of the calculations, that a large group of terms making up $\Sigma^{\rm m}$ and $\Sigma^{\rm q}$ can be represented in the form of a covariant divergence. Such terms are combined with $\t^{\rm m}_{\m\n}{}^{|\n}$ and $\t^{\rm q}_{\m\n}{}^{|\n}$ respectively to reduce the number of similar terms. We give more detailed description in the text which follows.

\subsubsection{Divergence of the stress-energy tensor of gravitational field}

Covariant divergence from the stress-energy tensor of gravitational field, $\mathfrak{t}_{\m\n}$, is derived by means of direct calculation from its definition (\ref{5f}). In the process of calculation we can simplify a significant number of terms by employing the commutation relations (\ref{com8f}), (\ref{com9f}) for second-order covariant derivatives along with a rule for the third order derivative 
\be\la{com10f}
l^\l{}_{\m|\r\s\n}=l^\l{}_{\m|\r\n\s}-l^\g{}_{\m|\r}\bar R^\l{}_{\g\s\n}+l^\l{}_{\g|\r}\bar R^\g{}_{\m\s\n}+l^\l{}_{\m|\g}\bar R^\g{}_{\r\s\n}\;,
\ee
which allows us (after one more commutation of the covariant derivative in $l^\l{}_{\m|\r\n\s}$) to derive
\ba\la{com11f}
l^\n{}_{\m|\r\s\n}&=&A_{\m|\r\s}+l^\a{}_{\m|\s}\bar R_{\a\r}+l^\a{}_{\m|\r}\bar R_{\a\s}+l^\a{}_{\b|\s}\bar R^\b{}_{\m\r\a}+
\\\nonumber&&l^\a{}_{\b|\r}\bar R^\b{}_{\m\s\a}+l^\a{}_{\m|\b}\bar R^\b{}_{\r\s\a}+l^\a{}_{\m}\bar R_{\a\r|\s}+l^\a{}_{\b}\bar R^\b{}_{\m\r\a|\s}\;.
\ea
A significant number of similar terms is cancelled out, and after a multi-page analytic calculation we obtain a fairly simple result,
\ba\la{thet2} 
\mathfrak{t}_{\m\n}{}^{|\n}&=&\lef(l^\r{}_{\n}\Theta_{\r\m}-\frac12l_{\m\n}\Theta\ri)^{|\n}-\frac12\lef(l^{\r\s}\Theta_{\r\s|\m}-\frac{1}2l\Theta_{|\m}\ri)\;,
\ea
where a tensor $\Theta_{\a\b}$ was defined in (\ref{thet3}), and $\Theta=\bar g^{\a\b}\Theta_{\a\b}$. After taking the covariant divergence, it is convenient to split the right side of (\ref{thet2})algebraically in three parts
\ba\la{hg1a}
\mathfrak{t}_{\m\n}{}^{|\n}&=&\lef(l^\r{}_\n\mathfrak{T}_{\r\m}-\frac12l_{\m\n}\mathfrak{T}\ri)^{|\n}-\frac12\lef(l^{\r\s}\mathfrak{T}_{\r\s|\m}-\frac{1}2l\mathfrak{T}_{|\m}\ri)\\\nonumber&-&
\frac1{8\pi}\lef(l^\r{}_\n F^{\rm m}_{\r\m}-\frac12l_{\m\n}F^{\rm m}\ri)^{|\n}+\frac1{16\pi}\lef(l^{\r\s}F^{\rm m}_{\r\s|\m}-\frac{1}2lF^{\rm m}_{|\m}\ri)\\\nonumber&-&
\frac1{8\pi}\lef(l^\r{}_\n F^{\rm q}_{\r\m}-\frac12l_{\m\n}F^{\rm q}\ri)^{|\n}+\frac1{16\pi}\lef(l^{\r\s}F^{\rm q}_{\r\s|\m}-\frac{1}2lF^{\rm q}_{|\m}\ri)\;,
\ea
where ${F}^{\rm m}_{\a\b}$ and ${F}^{\rm q}_{\a\b}$ have been given in (\ref{mt6a}) and (\ref{mt6c}) respectively, and $F^{\rm m}\eq\bar g^{\a\b}F^{\rm m}_{\a\b}$, $F^{\rm q}\eq\bar g^{\a\b}F^{\rm q}_{\a\b}$. The first line in the right side of this equation describes the coupling of gravitational perturbation with the stress-energy tensor $\mathfrak{T}_{\m\n}$ of the baryonic matter, and the second and the third lines outline the contribution of dark matter (index `m') and dark energy (index `q').

\subsubsection{The dark matter force density}\la{dam476}
The source density $\Sigma^{\rm m}$ for dark matter dynamic perturbations in quadratic and higher orders, is defined by equation (\ref{gd6a}) where we shall take into account in the dynamic Lagrangian only terms of the second order, 
\be\la{b5x89}
\lag^{\rm dyn}=\lag_2\;,
\ee
and keep in $\lag_2$ only dark matter variables. By a simple inspection, we find out that $\Sigma^{\rm m}$ depends only on the derivatives, $\bar\Phi_\a$, of the scalar field $\bar\Phi$ and, thus, can be written in the form of a covariant divergence
\be\la{gh79}
\Sigma^{\rm m}=J^{\rm m}_\n{}^{|\n}\;,
\ee
where
\ba\la{gh80}
J^{\rm m}_\n&=&\frac1{16\pi\sqrt{-\bar g}}\frac{\d{\cal F^{\rm m}}}{\d\bar\Phi^\n}\;,
\ea
is a second order (quadratic) correction to the conserved dark matter current $Y_\m$ given in (\ref{mt10a}).

The current $J^{\rm m}_\n$ can be algebraically split in two components - one being parallel to the Hubble velocity, $\bar u^\a$, and another one being orthogonal to it, 
\ba\la{gh80a} 
J^{\rm m}_\a&=&\bar\r_{\rm m}\bar u_\a j+ \bar\r_{\rm m}\bar P_\a{}^\b j_\b\;,
\ea
where $P_\a{}^\b\eq\d_\a^\b+\bar u_\a\bar u^\b$. The corresponding projections, which appear in (\ref{gh80a}), are given by the following expressions,
\ba\la{gh81f}
j&=&\frac1{2\bar\m^2_{\rm m}} \lef(1-\frac{c^2}{c^2_{\rm s}}\ri)\lef[\p^\a\p_\a+\lef(1-\frac{c^2}{c^2_{\rm s}}\ri)\lef(\bar u^\a\p_\a\ri)^2\ri]
+\frac3{2\bar\m_{\rm m}}\frac{c^2}{c^2_{\rm s}}\lef[l^{\a\b}\bar u_\a\p_\b+\frac12\lef(1-\frac{c^2}{c^2_{\rm s}}\ri)\lef(\bar u^\a\p_\a\ri)\mathfrak{q}\ri]\\\nonumber\\\nonumber&&-\frac1{4}\lef(1-\frac{c^2}{c^2_{\rm s}}\ri)\lef[\lef(1+\frac{c^2}{c^2_{\rm s}}\ri)\mathfrak{q}^2+l^{\a\b}l_{\a\b}-\frac{l^2}2\ri]-\frac1{2\bar\m_{\rm m}}\frac{c^2}{c^2_{\rm s}}\frac{\pd\ln c^2_{\rm s}}{\pd\bar\m_{\rm m}}\lef[\lef(\bar u^\a\p_\a\ri)^2-\frac32\bar\m_{\rm m}\mathfrak{q}\lef(\bar u^\a\p_\a\ri)+\frac12\bar\m_{\rm m}^2\mathfrak{q}^2\ri]\;,
\\\nonumber\\\nonumber\\
\la{gh81h}
j_\b&=&\frac{1}{\bar\m^2_{\rm m}} \lef(1-\frac{c^2}{c^2_{\rm s}}\ri)\lef(\bar u^\a\p_\a\ri)\p_\b-
\frac{3}{2\bar\m_{\rm m}}l_\b{}^\a\p_\a-\frac{3}{2\bar\m_{\rm m}}\lef(1-\frac{c^2}{c^2_{\rm s}}\ri)\lef[\lef(\bar u^\a\p_\a\ri)l_\b{}^\g\bar u_\g+\frac12\mathfrak{q}\p_\b\ri]+\lef(1-\frac{c^2}{c^2_{\rm s}}\ri)\mathfrak{q}l_\b{}^\a\bar u_\a\;.
\ea

Direct calculation of the covariant divergence from $J^{\rm m}_\a$ in (\ref{gh79}) entangles a lot of algebraic operations which number can be significantly reduced by making use of the following procedure. First of all,
we notice that the term, $\bar\m_{\rm m}\Sigma^{\rm m}\bar u_\m$, in the right side of (\ref{qxea1}) can be replaced on shell with $\bar\m_{\rm m}J^{{\rm m}|\n}_\n\bar u_\m$ due to (\ref{gh79}). Then, we use the chain rule and derivatives 
\ba\la{g6c10}
\bar\m_{\rm m}{}^{|\n}&=&\frac{\pd\bar\m_{\rm m}}{\pd\bar\r_{\rm m}}\bar\r_{\rm m}{}^{|\n}=3\frac{c^2_{\rm s}}{c^2}H\bar\m_{\rm m}\bar u^\n\;,\\
\la{n5d7l}
\bar u_\m{}^{|\n}&=&H\bar P_\m{^\n}\;.
\ea 
in order to transform
\be\la{xsy50}
\bar\m_{\rm m}\Sigma^{\rm m}\bar u_\m=\lef(\bar\m_{\rm m}J^{\rm m}_\n\bar u_\m\ri)^{|\n}+3\frac{c^2_{\rm s}}{c^2}H\bar\r_{\rm m}\bar\m_{\rm m}j\bar u_\m-H\bar\r_{\rm m}\bar\m_{\rm m}j_\n\bar P_\m{}^\n\;,
\ee
where we have used (\ref{gh80a}).

Now, we combine the total divergence in the right side of (\ref{xsy50}) with the divergence of the stress-energy tensor of dark matter in the left side of (\ref{qxea1}). In doing so, we notice the following equations
\ba\la{jklw50}
16\pi\sqrt{-\bar g}\t^{\rm m}_{\m\n}&=&\bar g_{\m\r}\bar g_{\n\s}\lef(\frac{\pd\cal F^{\rm m}}{\pd\bar g_{\r\s}}+\frac{\pd\cal F^{\rm m}}{\pd\bar\m_{\rm m}}\frac{\pd\bar\m_{\rm m}}{\pd\bar g_{\r\s}}\ri)\;,\\\la{jklw51}
16\pi\sqrt{-\bar g}J^{\rm m}_\n&=&\frac{\pd\cal F^{\rm m}}{\pd\bar\Phi^\n}+
\frac{\pd\cal F^{\rm m}}{\pd\bar\m_{\rm m}}\frac{\pd\bar\m_{\rm m}}{\pd\bar\Phi^\n}\;,
\ea
which are just more explicit form of the definitions (\ref{axrt5}) and (\ref{gh80}) of the corresponding quantities expressed as the variational derivatives with respect to the metric tensor and the derivative of the scalar field respectively. Accounting for the variational derivatives (\ref{pr5n}), (\ref{subeq2a}) we obtain
\be\la{nmx41}
\t^{\rm m}_{\m\n}-\frac12\bar\m_{\rm m}J^{\rm m}_\n\bar u_\m=\frac1{16\pi\sqrt{-\bar g}}\lef(\bar g_{\m\r}\bar g_{\n\s}\frac{\pd\cal F^{\rm m}}{\pd\bar g_{\r\s}}-\frac12\frac{\pd\cal F^{\rm m}}{\pd\bar\Phi^\n}\bar u_\m\ri)\;.
\ee
This equation elucidates that we do not need to directly calculate a large number of terms depending on the partial derivatives with respect to the specific enthalpy $\bar\m_{\rm m}$ when calculating the covariant divergence in the left side of equations of motion (\ref{qxea1}). It saves us from doing a lot of redundant algebraic operations. 

It is also reasonable to combine (\ref{nmx41}) with the dark matter term representing a total divergence in the second line of (\ref{hg1a}) and denote
\be\la{nmx42}
X_{\m\n}\eq 
\t^{\rm m}_{\m\n}-\frac12\bar\m_{\rm m}J^{\rm m}_\n\bar u_\m-\frac1{8\pi}\lef(l^\r{}_\n F^{\rm m}_{\r\m}-\frac12l_{\m\n}F^{\rm m}\ri)\;.
\ee
Notice that tensor $X_{\m\n}$ is not symmetric with respect to its indices. Making use of (\ref{xrc1}), (\ref{ax7}), (\ref{gh80a}) in the right side of (\ref{nmx42}), and reducing similar terms, we obtain a rather short expression
\ba\la{nmx43}
X_{\m\n}&\eq&
\frac{\bar\r_{\rm m}}{2\bar\m_{\rm m}}\lef(\p_\m\p_\n-\frac12\p^\a\p_\a\bar g_{\m\n}\ri)
+
\frac{\bar\r_{\rm m}}{2\bar\m_{\rm m}}\lef(1-\frac{c^2}{c^2_{\rm s}}\ri)\lef[\lef(\bar u^\a\p_\a\ri)\p_\m\bar u_\n-\frac12\lef(\bar u^\a\p_\a\ri)^2\bar g_{\m\n}\ri]
\\\nonumber
&-&\bar\r_{\rm m}\lef(\p_\m l_\n{}^\r\bar u_\r+\frac14\bar u_\m l_\n{}^\r\p_\r\ri)
- 
\bar\r_{\rm m}\lef(1-\frac{c^2}{c^2_{\rm s}}\ri)\lef[\frac38\mathfrak{q}\p_\m\bar u_\n+\lef(\bar u^\a\p_\a\ri)\lef(\frac18 l_{\m\n}+\frac14\bar u_\m l_\n{}^\r\bar u_\r\ri)\ri]\\\nonumber
&+&\frac18\lef[\bar\r_{\rm m}\bar\m_{\rm m}\lef(1-\frac{c^2}{c^2_{\rm s}}\ri)\mathfrak{q}^2+\lef(\bar p_{\rm m}-\bar\e_{\rm m}\ri)\lef(l^{\r\s}l_{\r\g}-\frac12l^2\ri)\ri]\bar g_{\m\n}\;.
\ea

Let us denote the density of the force caused by dark matter on the motion of the baryonic matter by $f^{\rm m}_\m$. After grouping together all terms in (\ref{qxea1}), belonging to the dark matter sector, the force density is defined by the following expression
\ba\la{nmx44}
f^{\rm m}_\m&\eq& -X_{\m\n}{}^{|\n}-\frac1{16\pi}\lef(l^{\r\s}F^{\rm m}_{\r\s|\m}-\frac{1}2lF^{\rm m}{}_{|\m}\ri)+\frac32\frac{c^2_{\rm s}}{c^2}H\bar\r_{\rm m}\bar\m_{\rm m}j\bar u_\m-\frac12H\bar\r_{\rm m}\bar\m_{\rm m}j_\n\bar P_\m{}^\n\;,
\ea
where the second term in the right side was taken from (\ref{hg1a}), and the last two terms -- from (\ref{xsy50}). We can split the force density, $f^{\rm m}_\m$, in two orthogonal components
\be\la{nmx45}
f^{\rm m}_\m=a^{\rm m}\bar u_\m+a^{\rm m}_\n\bar P^\n{}_\m\;,
\ee
where $a^{\rm m}\eq -\bar u^\n f^{\rm m}_\n$ and $a^{\rm m}_\m\eq \bar P_\m{}^\n f^{\rm m}_\n$.
We have, more explicitly,
\ba\la{nmx46}
a^{\rm m}&=&\frac14\bar\r_{\rm m}\lef[l^{\a\b}\p_{\a\b}+A^\a\p_\a-2\lef(\bar u^\a\p_\a\ri)\lef(\bar u^\b A_\b\ri)\ri]\\\nonumber
&+&\frac18\bar\r_{\rm m}\lef(1-\frac{c^2}{c^2_{\rm s}}\ri)\lef[\bar u^\a\bar u^\b l_\b{}^\g\p_{\a\g}+\mathfrak{q}\bar u^\a\bar u^\b\p_{\a\b}-
\lef(\bar u^\a\p_\a\ri)\lef(\bar u^\b\mathfrak{q}_\b\ri)+\lef(\bar u^\a\p_\a\ri)\lef(\bar u^\b A_\b\ri)\ri]\\\nonumber
&+&
2\bar\r_{\rm m}H\lef(\bar u^\a\p_\a\ri)\lef(2\mathfrak{q}-\frac12l\ri)\\\nonumber
&+&\frac18\bar\r_{\rm m}H\lef(1-\frac{c^2}{c^2_{\rm s}}\ri)\lef[l^{\a\b}\bar u_\a\p_\b+\lef(\bar u^\a\p_\a\ri)\lef(3\mathfrak{q}-l\ri)\ri]\\\nonumber
&+&\frac38\bar\r_{\rm m}\bar\m_{\rm m}H\frac{\pd\ln c^2_{\rm s}}{\pd\bar\m_{\rm m}}\lef(\mathfrak{q}-\frac{l}2\ri)\lef(\bar u^\a\p_\a\ri)\;,\\\nonumber\\\nonumber\\
\la{nmx47}
a^{\rm m}_\m&=&\frac12\bar\r_{\rm m}\lef(\bar u^\a A_\a\ri)\p_\m
\\\nonumber&+&\frac18\bar\r_{\rm m}\lef(1-\frac{c^2}{c^2_{\rm s}}\ri)\lef[ l_\m{}^\a\bar u^\b\p_{\a\b}-\mathfrak{q}\bar u^\a\p_{\m\a}+\lef(\bar u^\a\mathfrak{q}_\a\ri)\p_\m+\lef(\bar u^\a\p_\a\ri) A_\m\ri]\\\nonumber
&-&
2\bar\r_{\rm m}H\lef(2\mathfrak{q}-\frac12l\ri) \p_\m\\\nonumber
&+&\frac12\bar\r_{\rm m}H\lef(1-\frac{c^2}{c^2_{\rm s}}\ri)\lef[\lef(\bar u^\a\p_\a\ri) l_{\m\b}\bar u^\b-\frac14\mathfrak{q}\p_\m+\frac14l_\m{}^\a\p_\a\ri]\\\nonumber
&+&\frac38\bar\r_{\rm m}\bar\m_{\rm m}H\frac{\pd\ln c^2_{\rm s}}{\pd\bar\m_{\rm m}}\lef[\lef(\bar u^\a\p_\a\ri)l_\m{}^\a\bar u_\a-\mathfrak{q}\p_\m\ri]\;,
\ea
where $\phi_{\a\b}\eq\phi_{|\a\b}$.
Our next goal is to calculate the force density exerted by dark energy on the motion of the baryonic matter in the universe.

\subsubsection{The dark energy force density}\la{aa23w}

The procedure of calculation of the dark energy force density is similar to that described in the previous subsection (\ref{dam476}).
The dark energy source , $\Sigma^{\rm q}$, which is defined in (\ref{gd6b}) as a variational derivative from the dynamic Lagrangian $\lag^{\rm dyn}$, depends not only on the derivatives of the scalar field $\bar\Psi$ but on the field itself through the field potential $W=W(\Phi)$. We take into account in $\lag^{\rm dyn}$ only the quadratic terms with respect to the dynamic perturbations which yield 
\be\la{go2}
\Sigma^{\rm q}=\frac1{16\pi\sqrt{-\bar g}}\lef[-\frac{\pd{\cal F^{\rm q}}}{\pd\bar\Psi}+\lef(\frac{\pd{\cal F^{\rm q}}}{\pd\bar\Psi_\n}\ri)_{|\n}\ri]\;,
\ee
where the Lagrangian density $\cal F^{\rm q}$ is given in (\ref{ax99}).
After taking the variational derivatives in (\ref{go2}), we obtain
\ba\la{gh80b}
\Sigma^{\rm q}&=&\frac12\psi^2\frac{\pd^3\bar W}{\pd\bar\Psi^3}+\frac34l\psi\frac{\pd^2\bar W}{\pd\bar\Psi^2}-\frac12\frac{\pd\bar W}{\pd\bar\Psi}\lef(l^{\a\b}l_{\a\b}-\frac12l^2\ri)+\frac32\lef(l^{\a\b}{}_{|\g}\psi_a\bar u_\b\bar u^\g+l^{\a\b}\psi_{\a\g}\bar u_\b\bar u^\g+3Hl^{\a\b}\bar u_\a\psi_\b\ri)\;.
\ea
The force density exerted by dark energy on the motion of the baryonic matter is combined from all terms in (\ref{qxea1}) which depend on dark energy components, 
\ba\la{gh81}
f^{\rm q}_\m&\eq &-\t^{\rm q}_{\m\n}{}^{|\n}+\frac1{8\pi}\lef(l^\r{}_\n F^{\rm q}_{\r\m}-\frac12l_{\m\n}F^{\rm q}\ri)^{|\n}
-\frac1{16\pi}\lef(l^{\r\s}F^{\rm q}_{\r\s|\m}-\frac{1}2lF^{\rm q}{}_{|\m}\ri)+\frac12\bar\m_{\rm q}\Sigma^{\rm q}\bar u_\m\;,
\ea
where the second and third terms standing in the irght side of this definition come from the third line of (\ref{hg1a}).
In order to calculate the right side of (\ref{gh81}) we use equation (\ref{ax7q}) for $\t^{\rm q}_{\m\n}$, equation (\ref{mt6c}) for $F^{\rm q}_{\m\n}$, and equation (\ref{gh80b}) for $\Sigma^{\rm q}$. 
After long but straightforward calculation and reduction of many similar terms, we get
\ba\la{gh82}
f^{\rm q}_\m&=&\bar\r_{\rm q}\bar u_\m
\lef(l^{\a\b}\psi_{\a\b}+\frac34 l^{\a\b}{}_{|\g}\psi_\a\bar u_\b\bar u^\g+\frac34l^{\a\b}\psi_{\a\g}\bar u_\b\bar u^\g+\frac94Hl^{\a\b}\bar u_\a\psi_\b\ri)
\\\nonumber
&&+\lef(A^\a\psi_\a\ri)\bar u_\m+\frac12\lef(A^\a\bar u_\a\ri)\psi_\m
+\bar\r_{\rm q}H\lef(2\mathfrak{q}-\frac12 l\ri)\psi_\m\\\nonumber
&&-\frac14\frac{\pd\bar W}{\pd\bar\Psi}\lef(\psi A_\m+l_\m{}^\n\psi_\n-2\mathfrak{q}\psi_\m\ri)+\frac14\bar\r_{\rm q}\psi\frac{\pd^2\bar W}{\pd\bar\Psi^2}\lef(l_{\m\n}\bar u^\n-\frac{1}2 l\bar u_\m\ri)
\;,
\ea
where we denoted $\psi_{\a\b}\eq\psi_{|\a\b}$, and $\bar\r_{\rm q}=\m_{\rm q}$.

\subsection{Final form of the equations of motion}\la{bb34f7}

After making use of the results of the presiding section, equations of motion (\ref{qxea1}) of the baryonic matter take on the following form
\ba\la{eqmos1}
\mathfrak{T}_\m{}^\n{}_{|\n}+l^{\r\n}\lef(\mathfrak{T}_{\r\m|\n}-\frac12\mathfrak{T}_{\r\n|\m}\ri)-
\frac12\lef(l_\m{}^\n-\frac{l}2\d_\m{}^\n\ri)\mathfrak{T}_{|\n}+A^\r\mathfrak{T}_{\r\m}-\frac12A_\m\mathfrak{T}&=&f^{\rm m}_\m+f^{\rm q}_\m\;.
\ea
The left side of this equation can be brought to a more conventional form of a covarinat derivative with respect to the full metric, if we use relation (\ref{rtei2}) between the stress-energy tensor of the bare perturbation $T_{\m\n}$ given in (\ref{q19z}) and $\T_{\m\n}$ defined in (\ref{ia8}). 

Let us take a covariant divergence of $T_{\m\n}$ with respect to the full metric $g_{\m\n}$ that is $\nabla_\n T_\m{}^\n\eq g^{\n\r}\nabla_\n T_{\r\m}$ where the $\nabla_\n$ denotes a covariant derivative with respect to the full metric, and we rise and lowered indices with the help of the full metric. Covariant derivatives from the stress-energy tensor $\T_{\m\n}$ are calculated with the help of
\be\la{asud1}
\nabla_\a\T_{\m\n}=\T_{\m\n|\a}-\mathfrak{G}^\b_{\a\m}\T_{\n\b}-\mathfrak{G}^\b_{\a\n}\T_{\m\b}\;,
\ee
where $\mathfrak{G}^\b_{\a\m}$ is the Christoffel symbol being associated with the full metric. It is rather straightforward to prove that they have the following {\it exact} form,
\be\la{asud2}
\mathfrak{G}^\b_{\a\m}=\frac12\bar g^{\b\g}\lef(\varkappa_{\g\a|\m}+\varkappa_{\g\m|\a}-\varkappa_{\a\m|\g}\ri)\;.
\ee
In the linear approximation with respect to the $l_{\m\n}$ equation (\ref{asud2}) reads
\be\la{asud3}
\mathfrak{G}^\b_{\a\m}=-\frac12\bar g^{\b\g}\lef(l_{\g\a|\m}+l_{\g\m|\a}-l_{\a\m|\g}\ri)+\frac14\lef(\d^\b_\a l_{|\m}+\d^\b_\m l_{|\a}-\bar g_{\a\m}l^{|\b}\ri)\;.
\ee
Two contracted values of the Christoffel symbols are
\be\la{asud4}
\mathfrak{G}_\a\eq\mathfrak{G}^\b_{\a\b}=\frac12 l_{|\a}\qquad,\qquad \bar g^{\a\b}\mathfrak{G}^\g_{\a\b}=-l^{\g\b}{}_{|\b}=-A^\g\;,
\ee
Making use of these notations and definitions, and doing a direct calculation results in
\ba\la{asud5}
\nabla_\n T_\m{}^\n&=&\mathfrak{T}_\m{}^\n{}_{|\n}+l^{\r\n}\lef(\mathfrak{T}_{\r\m|\n}
-\frac12\mathfrak{T}_{\r\n|\m}\ri)-
\frac12\lef(l_\m{}^\n-\frac{l}2\d_\m{}^\n\ri)\mathfrak{T}_{|\n}+A^\r\mathfrak{T}_{\r\m}-\frac12A_\m\mathfrak{T}\;.
\ea
It elucidates that equation of motion (\ref{eqmos1}) has the following form
\be\la{asud6}
\nabla_\n T_\m{}^\n=f^{\rm m}_\m+f^{\rm q}_\m\;.\ee
Had the background spacetime been flat, the right side of (\ref{asud6}) would vanish yielding the conventional law of conservation, $\nabla_\n T_\m{}^\n=0$. However, in cosmology the spacetime manifold is given by the perturbed FLRW metric. The perturbations interact with themselves causing an effective force $f_\m=f^{\rm m}_\m+f^{\rm q}_\m$ which disturbs microscopic motion of the baryonic matter and ``violates'' the law of conservation of its stress-energy tensor.

The important case of the baryonic matter is a perfect fluid with the stress-energy tensor (\ref{qze1k}),
\be\la{v5x0k}
T^{\m\n}=(\e+p)\mathfrak{u}^\m \mathfrak{u}^\n+p g^{\m\n}\;,
\ee
where $\e$ and $p$ are the energy density and pressure of the fluid, $\mathfrak{u}^\m$ is the four-velocity of the fluid element, and $g^{\m\n}$ is a full (contravariant) metric. The pressure, $p=p(\m)$, and the energy density, $\e=\e(\m)$, are functions of the specific enthalpy, $\m$, of the fluid
defined by 
\be\la{0m4a}
\mu=\sqrt{-g^{\m\n}\Theta_\m\Theta_\n}\;,
\ee
where $\Theta_\m\eq\pd_\m\Theta$, and $\Theta$ is the Clebsch potential of the baryonic fluid. The baryonic mass density $\rho$ is defined by thermodynamic equation
\be\la{u73c5}
\rho\mu=\e+p\;.
\ee

Substituting (\ref{v5x0k}) to the left side of (\ref{asud6})and projecting this equation on the four-velocity $\mathfrak{u}^\a$ of the baryonic fluid, we get the post-Friedmannian law of conservation of energy
\be\la{g6c2h}
\mathfrak{u}^\m\nabla_\m\e+(\e+p)\nabla_\m\mathfrak{u}^\m=-\mathfrak{u}^\m\left(f^{\rm m}_\m+f^{\rm q}_\m\right)\;,
\ee
and the post-Friedmannian  Euler equation
\be\la{g5d3v}
(\e+p)\mathfrak{u}^\n\nabla_\n\mathfrak{u}^\m=\left(g^{\m\n}+\mathfrak{u}^\m\mathfrak{u}^\n\right)\left(-\nabla_\n p+f^{\rm m}_\n+f^{\rm q}_\n\right)\;.
\ee
One more equation is obtained by direct variation of the Lagrangian of the baryonic matter with respect to the Clebsch potential, leading to the law of conservation of baryonic mass density $\rho$
\be\la{n6x3aa}
\nabla_\m\left(\rho\mathfrak{u}^\m\right)=0\;,
\ee
which is an exact relation.

\section{Discussion}\label{diss56}

The present paper employs a new gauge-invariant approach to the theory of cosmological perturbations. This approach utilizes the dynamic field theory on curved geometric manifolds introduced by Bruce DeWitt \citep{dewitt_book}, and represents a systematic development of the iterative scheme for deriving a decoupled system of field equations for the perturbations of the metric tensor and material fields considered as dynamic variables on background FLRW manifold. We also demonstrate how to formulate the covariant equations of motion for the perturbations of the material variables like density, pressure, velocity of matter, etc., on the expanding spacetime of FLRW universe.

The original motivation for the development of the dynamic field theory of the gauge-invariant perturbations in cosmology was the task of generalization of the post-Minkowskian (PMA) and post-Newtonian (PNA) approximation schemes used in experimental gravitational physics for testing general relativity in the solar system, binary pulsars, other localized astronomical systems like the Milky Way \citep{damour_1987,will_1993,kopeikin_2011book,kopeikin_editor1,kopeikin_editor2}, and in gravitational wave astronomy  \citep{2006LRR.....9....4B,bd1,bd2,Schaefer_2011mmgr,blkos,beld} for studying the process of generation, propagation, and emission of gravitational waves by the isolated system comprised of massive bodies. Standard PMA and PNA schemes assume that the background spacetime is asymptotically flat which does not correspond to cosmological observations clearly indicating that the background spacetime is described by the curved FLRW metric. Therefore, the standard 
PMA and PNA schemes are totally missing cosmological effects which can become important in discussion of certain experimental situations \citep{2010RvMP...82..169C,Kopeikin_2012ephemerid}.

Earlier existing perturbation frameworks in cosmology developed by Lifshitz \citep{1964SvPhU...6..495L,lif}, Bardeen \citep{1980PhRvD..22.1882B}, Mukhanov {\it et al} \citep{mukh,mukh_book}, Ellis {\it et al} \citep{ellis1,ellis2,1992CQGra...9..921B} made use a principle of separation of the metric tensor perturbations in scalar, vector, and tensor harmonics but it does not comply with the theoretical foundation of experimental gravitational physics in asymptotically-flat spacetime \citep{will_1993,kopeikin_2011book}. For this reason, we do not use the scalar-vector-tensor decomposition of the metric tensor but operate directly with the components of the metric tensor perturbations and scalar fields for which we derive the gauge-invariant equations as described in sections \ref{v5x8n} and \ref{zzzz8}.  Moreover, those perturbative approaches of previous researchers did not clearly separate the gravitational effects of small-scale and large-scale inhomogenities of matter so that it remained fuzzy how to split the matter and gravitational field of an astronomical N-body system, which is an external perturbation of the background geometry, from the matter and gravitational field caused by the primordial perturbations of the background matter including cosmological gravitational waves. Recently, Green and Wald \citep{wald_2011,wald_2012} have developed a new framework for separation of the gravitational effects of the small and large-scale inhomogeneities in cosmology based on generalized Burnett’s shortwave approximation \citep{Burnett_1989JMP}. Green-Wald's framework resembles the similar approach developed by Futamase \citep{Futamase_1996PhRvD,Takada_1999MNRAS}, but it has considerably wider applicability. 

We compared the Lagrangian-based framework of the present paper with that of Green and Wald and noticed that both frameworks are based on somewhat similar assumptions. In particular, both approaches admit that there is a background spacetime metric, $\bar g_{\a\b}$, which is kept arbitrary for development of general formalism (until section \ref{sec4} in this paper). Both approaches assume that the metric tensor perturbations, $h_{\a\b}=g_{\a\b}-\bar g_{\a\b}$, are small so that the perturbative series are conjectured to be convergent, but no restrictions like, $h_{\a\b,\m\n}\ll h_{\a\b,\m}\ll h_{\a\b}$, are imposed on the first and second derivatives of the metric tensor perturbations. Both Green and Wald, and we, allow the matter perturbations to have a high-density contrasts, $\d\r/\bar\r \gg 1$ which are identified with the progenitors of the bare matter perturbations in the present paper. The bare matter perturbations have their own stress-energy tensor which, in general, is not associated with the stress-energy tensor of the background matter, and is allowed to have a different physical origin depending on the situation under discussion. Further comparative analysis of the results of the present paper and those of Green and Wald \citep{wald_2011,wald_2012} revealed the following:
\begin{enumerate}
\item We consistently rely upon the perturbative approach to develop the dynamic field theory of cosmological perturbations and never assume, for example, that quadratic products of the first derivatives $h_{\a\b,\m}$ are of the same order as the curvature of the background metric. Thus, we do not include explicitly to our scheme the case of generation of the background metric $\bar g_{\a\b}$ by the small-scale perturbations of the metric and/or its derivatives, via short-wave averaging of Einstein's equations like Green and Wald \citep{wald_2011,wald_2012} did. We do admit the back-reaction of the metric perturbations (both small and large scale) on the background metric but it can produce in our approach only small pertubative corrections to the expansion rate of the universe. In this sense the short-wave approximation approach in cosmology developed by Green and Wald \citep{wald_2011,wald_2012} seems to have wider application, at least in the geometric sector of the theory.
\item We assume that the background metric, $\bar g_{\a\b}$, obeys Friedmann's equations exactly (see section \ref{bmnf}) while Green and Wald derived the differential equations governing the evolution of the background metric by making use of the short-wave approximation of Burnett \citep{Burnett_1989JMP}. It means that in our approach the dynamic evolution of the background metric is driven exclusively by the background value of the stress-energy tensor of the background dark matter and dark energy while the effective stress-energy tensors of the gravitational and matter perturbations (see section \ref{sec5}) do not contribute to the background value of the metric of FLRW manifold.  
\item Green and Wald \citep{wald_2011,wald_2012} separate the cosmological perturbations of the background matter and gravitational field in short wavelength (index $(S)$) and long wavelength (index $(L)$) perturbations which are treated differently by making use of additional assumptions and/or limitations on the mathematical behaviour of the perturbations depending on the expansion parameter $\lambda$ (see section III in \citep{wald_2011}). The analogue of the short wavelength perturbations in our approach are the bare perturbations. They are described by the particular solutions of the field equations \eqref{axz4} while the long wavelength perturbations of the background metric tensor are given by their homogeneous solution. The bare perturbations correspond to the gravitational field in the Newtonian limit of N-body problem in cosmology and are caused by the bare stress-energy tensor of baryonic matter making up stars, galaxies and their clusters. 
\item The present paper makes use of a systematic dynamic-field approach based on the Noether's variational principle to disentangle the background quantities from the perturbations in the iterative sense, and to derive the field equations for the perturbations at each iteration by taking variational derivatives. This makes our approach fully algorithmic and the process of calculation of the variational derivatives can be written down as a recursive computer program. In principle, we can calculate the field equations for perturbations of any order starting from the background Lagrangian while Green-Wald's approach \citep{wald_2011,wald_2012} is less algorithmically formalized and gets more and more laborious as one goes to higher approximations. 
\item Green and Wald \citep{wald_2011,wald_2012} directly operated with the Einstein tensor, $G_{\a\b}=R_{\a\b}-(1/2)g_{\a\b}R$, to decompose it into the background and perturbative parts and to derive the field equations for the perturbations. However, they did not pay sufficient attention to the structure of the perturbation of the stress-energy tensor which is clearly shown in their equations (see, for example, $T^{(1)}_{ab}$ in \citep[eq. 87]{wald_2011}) but remains unspecified.  The perturbation of the background stress-energy-tensor of matter contains the linear-in-metric-tensor perturbation terms which should be included to the left side of the field equations for the metric tensor perturbations. Our approach carefully treats this problem of extracting the linear-in-metric tensor perturbations terms to formulate the linear operator of the field equations for the perturbations of the dynamic variables (see how the left side of the linearised field equation \eqref{axz4} is defined). 
\item At last, but not least, we notice the difference in the choice of the gauge condition \eqref{qe6} used in the present paper and that in \citep{wald_2011} (see discussion at the end of section III in \citep{wald_2011} and \citep[eq. 91]{wald_2011}). No doubt, that the gauge condition is, in a sense, a matter of taste of a researcher serving to one or another particular task. The advantage of our gauge condition \eqref{qe6} is that it allows to decouple the field equations for the metric perturbations in time domain and put them into the form being very similar to that implemented in the canonical PMA and PNA approaches used for testing general relativity. It allows to compare the results of the cosmological tests of general relativity to those performed in the solar system much more easier. Furthermore, our gauge condition \eqref{qe6} reduces the field equations for the dynamic variables to the Bessel-type wave equations which have well-defined retarded Green functions and can be solved in terms of the retarded integrals \citep{Ramirez_2001PhLA,Ramirez_2002PhLB,Chu_2011PhRvD}.
\end{enumerate}

Our field-theoretical approach to cosmological perturbations can be extended to incorporate more general physical situations. One of the main advantages of our formalism is the method of treatment of the perfect fluid as a dynamic field that makes its effects to be very similar to those produced by a scalar field \citep{Sasaki_2010PhRvD}. It is straightforward to include in our approach more realistic fluids with entropy, viscosity, anisotropic stresses, etc. The Lagrangian for such fluids have been discussed in a number of papers \citep{1970PhRvD...2.2762S,Brown_1993CQGra,Mitskievich_1999GReGr,Poplawski2009}, and is well-established. It is also possible to incorporate to our formalism additional vector and tensor fields which may be important for researchers doing quantum gravity and/or looking for violations of general relativity on cosmological scales \citep{kostel_2007PhRvL,kostel_2010PhRvD,kostel_2014PhRvD}.

Finally, it would be highly desirable to apply our dynamic field theory approach to perform calculations of gravitational radiation emitted by isolated astronomical sources (like binary stars) with taking into account various cosmological effects. This will yield a key to precise measurement of cosmological parameters with gravitational wave detectors. So far, this problem was considered only under assumption that spacetime is asymptotically flat (see review \citep{2006LRR.....9....4B} and references therein), thus, severely limiting the domain of possible fundamental applications of gravitational wave astronomy to cosmology.

\section*{Acknowledgements}
We are grateful to two anonymous referees for critical reading of the manuscript and for their fruitful comments and suggestions which allowed us to strengthen its scientific standing. We thank T. Oliynyk and Yu. Baryshev  for useful conversations.  

\newpage

\bibliographystyle{elsarticle-num}
\bibliography{kopeikin_bibliography}

\newpage
\renewcommand\appendix{\par
  \setcounter{section}{0}
  \setcounter{subsection}{0}
  \setcounter{equation}{0}
  \renewcommand\theequation{A{\arabic{equation}}}
  \renewcommand\thesection{\Alph{section}}
 
}

\appendix\la{sec7}

\section{Variational Derivatives}\la{appB2}
\setcounter{equation}{0}
  \renewcommand\theequation{A{\arabic{equation}}}
\subsection{Variational derivative from the Hilbert Lagrangian}\la{cvdhl}
The goal of this section is to prove relation (\ref{aq1}) being valid on the background manifold $\bar\M$. We shall omit the bar over the background geometric objects as it does not bring about confusion. We notice that the Hilbert Lagrangian density, $\lag^{\rm\st G}=-(16\pi)^{-1}\sqrt{-g}R$, differs from the Einstein Lagrangian density $\lag^{\rm\st E}=-(16\pi)^{-1}\sqrt{-g}L$ by a total derivative that is a consequence of (\ref{ri9}). Due to relation (\ref{gde}) the Lagrangian derivatives from $\lag^{\rm\st G}$ and $\lag^{\rm\st E}$ coincides
\be\la{qqz1}
\frac{\d\lag^{\rm\st G}}{\d\gag^{\m\n}}=\frac{\d\lag^{\rm\st E}}{\d\gag^{\m\n}}\;,
\ee
thus, pointing out that we can safely operate with the Einstein Lagrangian density $\lag^{\rm\st E}$. Because of (\ref{ia6}), we have
\be
\la{qq2}
\frac{\d\lag^{\rm\st E}}{\d\gag^{\m\n}}=\frac{1}{\sqrt{-g}}A^{\r\s}_{\mu\nu}\frac{\d\lag^{\rm\st E}}{\d g^{\r\s}}\;,
\ee
which suggests that calculation of the variational derivative with respect to the metric tensor is sufficient.

Calculation of the variational derivative $\d\lag^{\rm\st E}/\d g^{\r\s}$ demands the partial derivatives of the contravariant metric and Christoffel symbols with respect to $g^{\m\n}$. The partial derivatives of the metric are calculated with the help of (\ref{ay1}), (\ref{ay4}). The Christoffel symbols are given in terms of the partial derivatives from covariant metric tensor, $g_{\a\b,\g}$ which are not conjugated with the dynamic variable $g^{\a\b}$. Thus, calculation of the partial derivative with respect to $g^{\m\n}$ from the Christoffel symbols demands its transformation to the form where the conjugated variables $g^{\a\b}{}_{,\g}$ are used instead. This form of the Christoffel symbols is 
\be
\la{qq3}
\G^\a{}_{\b\g}=\frac12 \lef(g^{\r\k}{}_{,\s}g^{\a\s}g_{\r\b}g_{\k\g}-g^{\a\s}{}_{,\b}g_{\g\s}-g^{\a\s}{}_{,\g}g_{\b\s}\ri)\;.
\ee
Taking the partial derivative of (\ref{qq3}) with respect to the contravariant metric yields
\be
\la{qq4}
\frac{\pd\G^\a{}_{\b\g}}{\pd g^{\m\n}}=-g^{\a\s}\lef\{\G_{[\s\b](\m}g_{\n)\g}+\G_{[\s\g](\m}g_{\n)\b}+\G_{(\b\g)(\m}g_{\n)\s}\ri\}\;,
\ee
and
\be
\la{qq5}
\frac{\pd\Y_\a}{\pd g^{\m\n}}=-\G_{(\m\n)\a}\;,
\ee
where we have used (\ref{aa2a}). Contracting (\ref{qq4}), (\ref{qq5}) with the Christoffel symbols and the metric tensor results in
\ba
\la{qq6}
g^{\s\g}\frac{\pd\G^\a{}_{\b\g}}{\pd g^{\m\n}}\G^\b{}_{\s\a}&=&-2\G^\a{}_{\b\m}\G^\b{}_{\n\a}\;,\\
\la{qq7}
g^{\s\g}\frac{\pd\Y_\b}{\pd g^{\m\n}}\G^\b{}_{\s\g}&=&-\G_{(\m\n)\a}\G^\a\;,\\
\la{qq8}
g^{\s\g}\frac{\pd\G^\b{}_{\s\g}}{\pd g^{\m\n}}\Y_\b&=&\G_{(\m\n)\a}\Y^\a-\G_{\a\m\n}\Y^\a-\Y_\m\Y_\n\;.
\ea

Partial derivatives of the Christoffel symbols with respect to the metric derivatives are calculated from (\ref{qq3}) with the help of (\ref{ay2}). We get
\ba\la{qq9}
\frac{\pd\G^\a{}_{\b\g}}{\pd g^{\m\n}{}_{,\r}}&=&\frac12\lef[g^{\r\a}g_{\b(\m}g_{\n)\g}-\d^\r_\g\d^\a_{(\m}g_{\n)\b}-\d^\r_\b\d^\a_{(\m}g_{\n)\g}\ri]\;,\\
\la{qq10}
\frac{\pd{\cal Y}_\b}{\pd g^{\m\n}{}_{,\r}}&=&-\frac12g_{\m\n}\d^\r_\b\;.
\ea
Contracting (\ref{qq9}), (\ref{qq10}) with the Christoffel symbols and the metric tensor results in
\ba
\la{qq11}
g^{\s\g}\frac{\pd\G^\a{}_{\b\g}}{\pd g^{\m\n}{}_{,\r}}\G^\b{}_{\s\a}&=&-\frac12\G^\r{}_{\m\n}\;,\\
\la{qq12}
g^{\s\g}\frac{\pd\Y_\b}{\pd g^{\m\n}{}_{,\r}}\G^\b{}_{\s\g}&=&-\frac12 g_{\m\n}\G^\r\;,\\
\la{qq13}
g^{\s\g}\frac{\pd\G^\b{}_{\s\g}}{\pd g^{\m\n}{}_{,\r}}\Y_\b&=&\frac12 g_{\m\n}\Y^\r-\d^\r_{(\m}\Y_{\n)}\;.
\ea

Explicit expression for the variational derivative of the Einstein Lagrangian is 
\ba
\la{qq14} 
-16\pi\frac{\d\lag^{\rm\st E}}{\d g^{\m\n}}&=&\lef(\frac{\pd\sqrt{-g}}{\pd g^{\m\n}}g^{\s\g}+\sqrt{-g}\frac{\pd g^{\s\g}}{\pd g^{\m\n}}\ri)\lef(\G^\a{}_{\b\g}\G^\b{}_{\s\a}-\Y_\b\G^\b{}_{\s\g}\ri)\\\nonumber
&+&\sqrt{-g}g^{\s\g}\lef(2\frac{\pd\G^\a{}_{\b\g}}{\pd g^{\m\n}}\G^\b{}_{\s\a}-\frac{\pd\Y_\b}{\pd g^{\m\n}}\G^\b{}_{\s\g}-\frac{\pd\G^\b{}_{\s\g}}{\pd g^{\m\n}}\Y_\b\ri)\\\nonumber
&-&\frac{\pd}{\pd x^\r}\lef[\sqrt{-g}g^{\s\g}\lef(2\frac{\pd\G^\a{}_{\b\g}}{\pd g^{\m\n}{}_{,\r}}\G^\b{}_{\s\a}-\frac{\pd\Y_\b}{\pd g^{\m\n}{}_{,\r}}\G^\b{}_{\s\g}-\frac{\pd\G^\b{}_{\s\g}}{\pd g^{\m\n}{}_{,\r}}\Y_\b\ri)\ri]\;.
\ea
Replacing the partial derivatives in (\ref{qq14}) with the corresponding right sides of equations (\ref{ay1}), (\ref{ay4}), (\ref{qq11})--(\ref{qq13}) and taking the partial derivative with respect to spatial coordinates, yields
\be
\la{qq15}
-16\pi\frac{\d\lag^{\rm\st E}}{\d g^{\m\n}}=\sqrt{-g}\lef(R_{\m\n}-\frac12 g_{\m\n}R\ri)\;,
\ee
where we have used expressions (\ref{ri2}), (\ref{ri4}) for the Ricci tensor and Ricci scalar respectively. Substituting equation (\ref{qq15}) to (\ref{qq2}) yields
\be
\la{qq16}
\frac{\d\lag^{\rm\st E}}{\d\gag^{\m\n}}=-\frac1{16\pi}R_{\m\n}\;.
\ee

\subsection{Variational derivatives of dynamic variables with respect to the metric tensor}\la{vdwrmt}
\subsubsection{Variational derivatives of dark matter variables}\la{appvdm}
The primary thermodynamic variable of dark matter is $\m_{\rm m}$ defined in (\ref{pf7}). Variational derivative from $\m_{\rm m}$ is calculated directly from its definition and yields
\be\la{pr5n}
\frac{\d\bar\m_{\rm m}}{\d\bar g_{\m\n}}=\frac12\bar\m_{\rm m}\bar u^\m\bar u^\n\;.
\ee
Variational derivative of pressure $\bar p_{\rm m}$ is obtained from thermodynamic relation (\ref{jk3ea}) by making use of the chain differentiation rule along with (\ref{pr5n}), that is
\be\la{pr6n}
\frac{\d\bar p_{\rm m}}{\d\bar g_{\m\n}}=\frac12\bar\r_{\rm m}\bar\m_{\rm m}\bar u^\m\bar u^\n\;.
\ee
Variational derivative of the rest mass and energy density are obtained by making use of (\ref{pr5n}) along with equation of state that allows us to express partial derivatives of $\r_{\rm m}$ and $\e_{\rm m}$ in terms of the variational derivative for $\m_{\rm m}$. More specifically, 
\ba\la{pr7n}
\frac{\d\bar\r_{\rm m}}{\d\bar g_{\m\n}}&=&\frac12\frac{c^2}{c^2_{\rm s}}\bar\r_{\rm m}\bar u^\m\bar u^\n\;,\\
\frac{\d\bar\e_{\rm m}}{\d\bar g_{\m\n}}&=&\frac12\frac{c^2}{c^2_{\rm s}}\bar\r_{\rm m}\bar\m_{\rm m}\bar u^\m\bar u^\n\;,
\ea
where the speed of sound appears explicitly.
Variational derivatives from products and/or ratios of the thermodynamic quantities are calculated my applying the chain rule of differentiation and the above equations,
\ba\la{subeq1}
\frac{\d\lef(\bar\r_{\rm m}\bar\m_{\rm m}\ri)}{\d\bar g_{\m\n}}&=&\phantom{+}\frac12\lef(1+\frac{c^2}{c^2_{\rm s}}\ri)\bar\r_{\rm m}\bar\m_{\rm m}\bar u^\m\bar u^\n\;,\\
\frac{\d}{\d\bar g_{\m\n}}\lef(\frac{\bar\r_{\rm m}}{\bar\m_{\rm m}}\ri)&=&-\frac12\lef(1-\frac{c^2}{c^2_{\rm s}}\ri)\frac{\bar\r_{\rm m}}{\bar\m_{\rm m}}\bar u^\m\bar u^\n\;,\\
\frac{\d\lef(\bar p_{\rm m}-\bar\e_{\rm m}\ri)}{\d\bar g_{\m\n}}&=&\phantom{+}\frac12\lef(1-\frac{c^2}{c^2_{\rm s}}\ri)\bar\r_{\rm m}\bar\m_{\rm m}\bar u^\m\bar u^\n\;.
\ea

\subsubsection{Variational derivatives of dark energy variables}
The primary thermodynamic variable of dark energy is $\bar\m_{\rm q}$ defined in (\ref{pf7}). Variational derivative from $\bar\m_{\rm q}$ is calculated directly from its definition,
\be\la{pr51}
\frac{\d\bar\m_{\rm q}}{\d\bar g_{\m\n}}=\frac12\bar\m_{\rm q}\bar u^\m\bar u^\n\;.
\ee
Variational derivative of the mass density $\bar\r_{\rm q}$ of the dark energy ``fluid'' follows directly from $\bar\r_{\rm q}=\bar\m_{\rm q}$, and reads
\be\la{pr71}
\frac{\d\bar\r_{\rm q}}{\d\bar g_{\m\n}}=\frac12\bar\r_{\rm q}\bar u^\m\bar u^\n\;.
\ee
Variational derivative of pressure $\bar p_{\rm q}$ is obtained from definition (\ref{h16b}) along with (\ref{pr51}), which yields
\be\la{pr61}
\frac{\d\bar p_{\rm q}}{\d\bar g_{\m\n}}=\frac12\bar\r_{\rm q}\bar\m_{\rm q}\bar u^\m\bar u^\n\;.
\ee
Variational derivative of energy density $\bar\e_{\rm q}$ is obtained by making use of (\ref{pr51}) along with (\ref{h16a}). More specifically, 
\be\la{pn3b8x}
\frac{\d\bar\e_{\rm q}}{\d\bar g_{\m\n}}=\frac12\bar\r_{\rm q}\bar\m_{\rm q}\bar u^\m\bar u^\n\;.
\ee
Variational derivatives from products and ratios of other quantities are calculated my making use of the chain rule of differentiation and the above equations
\la{suq1as}\ba
\frac{\d\lef(\bar\r_{\rm q}\bar\m_{\rm q}\ri)}{\d\bar g_{\m\n}}&=&\bar\r_{\rm q}\bar\m_{\rm q}\bar u^\m\bar u^\n\;,\\
\frac{\d}{\d\bar g_{\m\n}}\lef(\frac{\bar\r_{\rm q}}{\bar\m_{\rm q}}\ri)&=&0\;,\\
\frac{\d\lef(\bar p_{\rm q}-\bar\e_{\rm q}\ri)}{\d\bar g_{\m\n}}&=&0\;.
\ea

\subsubsection{Variational derivatives of four-velocity of the Hubble flow}
Variational derivatives from four-velocity of the fluid are derived from the definition (\ref{pf8tn}) of the four-velocity given in terms of the potential $\bar\Phi$ or $\bar\Psi$ which are independent dynamic variables that do not depend on the metric tensor. Taking variational derivative from (\ref{pf8tn}) and making use either (\ref{pr5n}) or (\ref{pr51}) we obtain 
\ba
\frac{\d\bar u_\a}{\d\bar g_{\m\n}}&=&-\frac12\bar u_\a\bar u^\m\bar u^\n\;,\\
\frac{\d\bar u^\a}{\d\bar g_{\m\n}}&=&-\frac12\bar u^\a\bar u^\m\bar u^\n-\bar g^{\a(\m}\bar u^{\n)}\;,\\\la{n3c6j}
\frac{\d\lef(\bar u^\a\p_\a\ri)}{\d\bar g_{\m\n}}&=&-\p^{(\m}\bar u^{\n)}-\frac12\bar u^\m\bar u^\n\lef(\bar u^\a\p_\a\ri)\;,
\ea
where equation (\ref{n3c6j}) accounts for the fact that $\phi_\a$ is an independent variable that does not depend on the metric tensor.

\subsubsection{Variational derivatives of the metric tensor perturbations}
Variational derivatives from the metric tensor perturbations $l^{\a\b}$ are determined by taking into account that $l^{\a\b}=\mathfrak{h}^{\a\b}/\sqrt{\bar g}$ and $\mathfrak{h}^{\a\b}$ is an independent dynamic variable which does not depend on the metric tensor. Therefore, its variational derivative is nil, and we have 
\be\la{nq8c4}
\frac{\d l^{\a\b}}{\d\bar g_{\m\n}}=\frac{\d }{\d\bar g_{\m\n}}\left(\frac{\mathfrak{h}^{\a\b}}{\sqrt{\bar g}}\right)=\mathfrak{h}^{\a\b}\frac{\d}{\d\bar g_{\m\n}}\left(\frac{1}{\sqrt{\bar g}}\right)=-\frac12l^{\a\b}\bar g^{\m\n}\;.
\ee
Other variational derivatives are derived by making use of tensor operations of rising and lowering indices with the help of $\bar g_{\a\b}$ and applying from (\ref{nq8c4}). It gives
\ba
\frac{\d l_{\a\b}}{\d\bar g_{\m\n}}&=&-\frac12l_{\a\b}\bar g^{\m\n}+2l_\a{}^{(\m}\d^{\n)}_\b\;,\\
\frac{\d l}{\d\bar g_{\m\n}}&=&l^{\m\n}-\frac12l\bar g^{\m\n}\;,\\
\frac{\d\mathfrak{q}}{\d\bar g_{\m\n}}&=&-\mathfrak{q}\lef(\bar u^\m\bar u^\n+\frac12\bar g^{\m\n}\ri)+\frac12\lef(l^{\m\n}+l\bar u^\m\bar u^\n\ri)\;,\\
\frac{\d}{\d\bar g_{\m\n}}\lef(l^{\a\b}l_{\a\b}-\frac{l^2}2\ri)&=&2l^{\a(\m}l^{\n)}{}_\a-ll^{\m\n}-\bar g^{\m\n}\lef(l^{\a\b}l_{\a\b}-\frac{l^2}2\ri)\;.
\ea
\subsection{Variational derivatives with respect to matter variables}
\subsubsection{Variational derivatives of dark matter variables}\la{fff5v6}
The dark matter variables do not depend on the Clebsch potential $\bar\Phi$ directly but merely on its first derivatives $\bar\Phi_\a$. Therefore, any variational derivative of dark matter variable, say, ${\cal Q}={\cal Q}(\bar\Phi_\a)$, is reduced to a total divergence
\be\la{pop55}
\frac{\d {\cal Q}}{\d\bar\Phi}=-\frac{\pd}{\pd x^\a}\frac{\pd {\cal Q}}{\pd\bar\Phi_\a}\;.
\ee
We present a short summary of the partial derivatives with respect to $\bar\Phi_\a$.
\ba\la{subeq2a}
\frac{\pd\bar\m_{\rm m}}{\pd\bar\Phi_{\a}}&=&\bar u^\a\;,\\\la{subeq2b}
\frac{\pd\bar p_{\rm m}}{\pd\bar\Phi_{\a}}&=&\bar\r_{\rm m}\bar u^\a\;,\\\la{subeq2c}
\frac{\pd\bar\r_{\rm m}}{\pd\bar\Phi_{\a}}&=&\frac{c^2}{c^2_{\rm s}}\frac{\bar\r_{\rm m}}{\bar\m_{\rm m}}\bar u^\a\;,\\\la{subeq2d}
\frac{\pd\bar\e_{\rm m}}{\pd\bar\Phi_{\a}}&=&\frac{c^2}{c^2_{\rm s}}\bar\r_{\rm m}\bar u^\a\;,\\
\frac{\pd\lef(\bar\r_{\rm m}\bar\m_{\rm m}\ri)}{\pd\bar\Phi_{\a}}&=&\lef(1+\frac{c^2}{c^2_{\rm s}}\ri)\bar\r_{\rm m}\bar u^\a\;,\\
\frac{\pd}{\pd\bar\Phi_{\a}}\lef(\frac{\bar\r_{\rm m}}{\bar\m_{\rm m}}\ri)&=&-\lef(1-\frac{c^2}{c^2_{\rm s}}\ri)\frac{\bar\r_{\rm m}}{\bar\m_{\rm m}^2}\bar u^\a\;,\\
\frac{\pd\lef(\bar p_{\rm m}-\bar\e_{\rm m}\ri)}{\pd\bar\Phi_{\a}}&=&+\lef(1-\frac{c^2}{c^2_{\rm s}}\ri)\bar\r_{\rm m}\bar u^\a\;.
\ea
Partial derivatives of four velocity
\be\la{subeq5}
\frac{\pd\bar u_\a}{\pd\bar\Phi_{\b}}=-\frac{\bar P_\a{}^\b}{\bar\m_{\rm m}}\qquad,\qquad\frac{\pd\bar u^\a}{\pd\bar\Phi_{\b}}=-\frac{\bar P^{\a\b}}{\bar\m_{\rm m}}\;.
\ee
It allows us to deduce, for example,
\ba\la{gbu71}
\frac{\pd\lef(\bar u^\a\p_\a\ri)}{\pd\bar\Phi_{\b}}&=&-\frac1{\bar\m_{\rm m}}\bar P^{\a\b}\p_\b\;,\\
\frac{\pd\mathfrak{q}}{\pd\bar\Phi_{\a}}&=&-\frac2{\bar\m_{\rm m}}\bar P^\a{}_\m l^{\m\n}\bar u_\n\;.
\ea
\subsubsection{Variational derivatives of dark energy variables}
The dark energy variables depend on both the scalar potential $\bar\Psi$ and its first derivative $\bar\Psi_\a$ in the most generic situation. This is because there is a potential of the scalar field $W(\bar\Phi)$ that is absent in case of the dark matter. Therefore, variational derivative of the dark energy variable, say, ${\cal a}={\cal A}(\bar\Psi,\bar\Psi_\a)$, is 
\be\la{pop57}
\frac{\d {\cal A}}{\d\bar\Psi}=\frac{\pd {\cal A}}{\pd\bar\Psi}-\frac{\pd}{\pd x^\a}\frac{\pd {\cal A}}{\pd\bar\Psi_\a}\;.
\ee
Partial derivatives $\pd {\cal A}/\pd\bar\Psi=(\pd {\cal A}/\pd W)(\pd W/\pd\bar\Psi$, and their particular form depends on the shape of the potential $W$. As for the patial derivatives with respect to the derivatives of the field, they can be calulated explicitly for each variable, and 
we present a short summary of these partial derivatives below. More specifically,
\ba
\frac{\pd\bar\m_{\rm q}}{\pd\bar\Psi_{\a}}&=&\bar u^\a\;,\\
\frac{\pd\bar p_{\rm q}}{\pd\bar\Psi_{\a}}&=&\bar\r_{\rm q}\bar u^\a\;,\\
\frac{\pd\bar\r_{\rm q}}{\pd\bar\Psi_{\a}}&=&\bar u^\a\;,\\
\frac{\pd\bar\e_{\rm q}}{\pd\bar\Psi_{\a}}&=&\bar\r_{\rm q}\bar u^\a\;,\\
\frac{\pd\lef(\bar\r_{\rm q}\bar\m_{\rm q}\ri)}{\pd\bar\Psi_{\a}}&=&2\bar\r_{\rm q}\bar u^\a\;,\\
\frac{\pd}{\pd\bar\Psi_{\a}}\lef(\frac{\bar\r_{\rm q}}{\bar\m_{\rm q}}\ri)&=&0,\\
\frac{\pd\lef(\bar p_{\rm q}-\bar\e_{\rm q}\ri)}{\pd\bar\Psi_{\a}}&=&0\;.
\ea
Partial derivatives of four velocity
\be\la{subeq5rt}
\frac{\pd\bar u_\a}{\pd\bar\Psi_{\b}}=-\frac{\bar P_\a{}^\b}{\bar\m_{\rm q}}\qquad,\qquad\frac{\pd\bar u^\a}{\pd\bar\Psi_{\b}}=-\frac{\bar P^{\a\b}}{\bar\m_{\rm q}}\;.
\ee
It allows us to deduce, for example,
\ba\la{mkl49}
\frac{\pd\lef(\bar u^\a\psi_\a\ri)}{\pd\bar\Psi_{\b}}&=&-\frac1{\bar\m_{\rm q}}\bar P^{\a\b}\psi_\b\;,\\
\frac{\pd\mathfrak{q}}{\pd\bar\Psi_{\a}}&=&-\frac2{\bar\m_{\rm q}}\bar P^\a{}_\m l^{\m\n}\bar u_\n\;.
\ea

\end{document}